\documentclass[9pt,letterpaper]{extarticle}
\usepackage[left=0.78in,right=0.78in,top=0.82in,bottom=0.88in]{geometry}
\usepackage{fontspec}
\usepackage{booktabs}
\usepackage{longtable}
\usepackage{array}
\usepackage{amsmath,amssymb}
\usepackage{unicode-math}
\usepackage{enumitem}
\usepackage{float}
\usepackage{fancyvrb}
\usepackage{fvextra}
\usepackage{seqsplit}
\usepackage{xurl}
\usepackage[table]{xcolor}
\definecolor{atmLinkBlue}{RGB}{24,72,138}
\usepackage[colorlinks=true,linkcolor=atmLinkBlue,citecolor=atmLinkBlue,urlcolor=atmLinkBlue]{hyperref}
\usepackage{tikz}
\usepackage{pgfplots}
\usepackage{titlesec}
\usepackage{needspace}
\usepackage[most]{tcolorbox}
\usetikzlibrary{arrows.meta,calc,positioning,shapes.geometric,fit,backgrounds}
\pgfplotsset{compat=1.18}
\newcolumntype{L}[1]{>{\raggedright\arraybackslash}p{#1}}
\newcommand{\TableFont}{\fontsize{8.2}{9.4}\selectfont}
\newcommand{\TableLargeFont}{\fontsize{8.8}{10.0}\selectfont}
\newcommand{\TableXLFont}{\fontsize{9.2}{10.5}\selectfont}
\newcommand{\TableTightFont}{\fontsize{7.4}{8.5}\selectfont}
\newcommand{\TableMidFont}{\fontsize{7.0}{8.0}\selectfont}
\newcommand{\TableLead}[1]{\par\smallskip\noindent{\raggedright\small\bfseries #1\par}\smallskip}
\setlist{topsep=1.5pt,itemsep=1pt,parsep=0pt}
\newtcolorbox{paperAlgoBox}{
  enhanced,
  colback=gray!6,
  colframe=gray!35,
  boxrule=0.45pt,
  arc=3pt,
  left=6pt,
  right=6pt,
  top=5pt,
  bottom=5pt,
  drop fuzzy shadow=black!18,
  before skip=6pt,
  after skip=7pt,
  breakable=false
}
\titleformat{\section}{\normalfont\sffamily\bfseries\large}{\thesection}{0.65em}{}
\titleformat{name=\section,numberless}{\normalfont\sffamily\bfseries\large}{}{0pt}{}
\titleformat{\subsection}{\normalfont\sffamily\bfseries\normalsize}{\thesubsection}{0.55em}{}
\titleformat{name=\subsection,numberless}{\normalfont\sffamily\bfseries\normalsize}{}{0pt}{}
\titleformat{\subsubsection}{\normalfont\sffamily\bfseries\normalsize}{\thesubsubsection}{0.5em}{}
\titleformat{name=\subsubsection,numberless}{\normalfont\sffamily\bfseries\normalsize}{}{0pt}{}
\titlespacing*{\section}{0pt}{2.2ex plus 0.35ex minus 0.2ex}{0.9ex plus 0.2ex}
\titlespacing*{\subsection}{0pt}{1.45ex plus 0.25ex minus 0.15ex}{0.65ex plus 0.15ex}
\titlespacing*{\subsubsection}{0pt}{1.1ex plus 0.2ex minus 0.12ex}{0.45ex plus 0.12ex}
\newcommand{\paperTitleBlock}{%
  \begin{center}
    {\fontsize{18}{22}\selectfont\bfseries
    ATM: CID-Brokered Pre-Write Admission\\
    for Multi-Agent Code Co-Synthesis\par}
    \vspace{2.5pt}
    {\fontsize{10}{12}\selectfont\bfseries
    A Specification-Grounded Governance Substrate for Software Agents\par}
    \vspace{12pt}
    {\fontsize{11}{13}\selectfont Eagl Huang\par}
    {\fontsize{10}{12}\selectfont eaglhuang@gmail.com\par}
  \end{center}
  \vspace{4pt}
}
\begin{document}

\paperTitleBlock

\section*{Abstract}

Multi-agent LLM systems can decompose software-engineering tasks into planning, generation, validation, and repair. A narrower system gap remains: after several agents have formed write intents within the same controlled filesystem, worktree, or service domain, but before any governed shared mutation is applied, the system must determine which intents may proceed in parallel, which require deterministic composition or serialization, and which must take a fail-closed path. To address this gap, we present the AI-Atomic-Framework (ATM), a specification-to-evidence governance substrate for software agents operating within a single governance domain.

ATM binds task intent, repository scope, write admission, validation, and evidence obligations into a single governance chain. Within this chain, the Content Identifier (CID) broker serves as the shared-mutation admission subsystem. Adapter-guided atomization maps write intents to semantic atoms and bounded regions; the broker then routes them to parallel admission, deterministic composition, serialization, or fail-closed refinement. When the persistent atom map is incomplete, virtual atoms provide temporary, auditable governance units that preserve bounded-region comparability. Governed shared writes are ultimately applied by a neutral steward rather than directly by the proposing agents.

Evaluation combines controlled, field, adoption, and extension evidence. Controlled evidence includes a 12-scenario deterministic design matrix, three archived runner cases, and ATM-AdmissionBench, in which v0.1 freezes the benchmark substrate and v0.2 provides the paper-facing profile over 20 unique scenarios and 42 mode-level comparisons. The policy, ablation, adversarial, and enforcement rows are derived views of those scenarios rather than independent population samples. Field evidence comprises three archived same-file boundary cases--POS2, B-12, and BLOCK--while a three-week external-adopter study, observations of batch scheduling and CID stability, and an operational recovery-routing benchmark provide complementary evidence of operability, recoverability, and runtime transparency. Taken together, these results support feasibility within the observed single-domain settings, but not broad comparative superiority over alternative concurrency-control systems. ATM neither replaces Git merge nor addresses cross-clone or cross-PR governance; its claim is limited to an implementable, auditable, and incrementally extensible pre-write admission layer for governed shared mutations.

\section{Introduction}

\subsection{Motivation}

The central question of this paper is not whether individual agents can produce usable code, but whether, once multiple agents have formed write intents within the same controlled filesystem, worktree, or service domain, the system can determine, before any governed shared mutation is applied, which intents may proceed in parallel, which require deterministic composition or serialization, and which must be blocked. ATM frames this as a pre-write admission problem within a single governance domain. Same-file edits, shared registries, generated artifacts, and task-state machines are treated as different manifestations of the same governance boundary rather than as separate claim scopes. This paper does not address cross-machine clones, remote branches, or PR-level distributed coordination; it focuses on auditable admission decisions made before governed shared-state mutation. This setting lies within the broader LLM-for-SE landscape. The systematic review by Hou et al. \textcolor{atmLinkBlue}{\hyperlink{Item.67}{[31]}} divides LLM applications in software engineering into code generation, testing, maintenance, and coordination, and identifies multi-agent coordination as an emerging and not-yet-mature area. The pre-write admission layer studied here occupies a specific gap within that coordination subarea.

Recent repository-level code-generation benchmarks show that modern AI coding is no longer limited to a single file or function. RepoBench and CrossCodeEval evaluate repository-level completion and cross-file context use, while FEA-Bench examines feature implementation that requires coordinated modifications across an existing repository \textcolor{atmLinkBlue}{\hyperlink{Item.59}{[23--25]}}. CodeS and NL2Repo-Bench extend the setting to from-scratch repository generation, where models must construct complete repositories, preserve cross-file APIs and package structure, manage dependencies, and pass execution-based tests \textcolor{atmLinkBlue}{\hyperlink{Item.62}{[26--27]}}. Recent benchmarks such as Commit0, FeatureBench, and GitTaskBench further emphasize long-horizon planning, localized changes, cross-file dependencies, and global consistency \textcolor{atmLinkBlue}{\hyperlink{Item.68}{[32--34]}}. The common implication is not merely that models can modify more files. Repository-level tasks naturally bring multiple shared surfaces into the same delivery path; once several agents participate in one workflow, pre-write governance can no longer be treated as an optional implementation detail.

Multi-agent and concurrency-control research has begun to address shared state directly. CodeTeam reduces cross-file drift during repository construction through machine-checkable contracts, file ownership, and dependency-aware scheduling \textcolor{atmLinkBlue}{\hyperlink{Item.51}{[15]}}. CoAgent addresses tool-level and action-level concurrency over shared agent state through MTPO, filtered reads, notification-guided repair, and saga-style compensation for long-running tasks \textcolor{atmLinkBlue}{\hyperlink{Item.47}{[11]}}. S-Bus reconstructs read sets at commit time through HTTP middleware and a DeliveryLog, providing Observable-Read Isolation for shared-shard structural races under an HTTP-observable read projection \textcolor{atmLinkBlue}{\hyperlink{Item.52}{[16]}}. In parallel, AgenticFlict documents substantial merge-conflict pressure in large-scale pull requests produced by AI coding agents, indicating that AI-generated changes already impose non-trivial downstream integration costs on Git and PR workflows \textcolor{atmLinkBlue}{\hyperlink{Item.50}{[14]}}.

Together, these studies delineate the gap targeted by this paper. Repository-level benchmarks establish the multi-file and long-horizon nature of contemporary coding tasks \textcolor{atmLinkBlue}{\hyperlink{Item.59}{[23--27]}}. CodeTeam-style planners move ownership and dependency constraints into the planning stage \textcolor{atmLinkBlue}{\hyperlink{Item.51}{[15]}}. CoAgent- and S-Bus-style systems demonstrate the need for specialized mechanisms over agent-accessible shared state \textcolor{atmLinkBlue}{\hyperlink{Item.47}{[11, 16]}}. AgenticFlict quantifies the downstream pressure that remains when conflicts reach Git and PR integration \textcolor{atmLinkBlue}{\hyperlink{Item.50}{[14]}}. These systems, however, do not directly target a code-region admission layer that operates within a single governance domain before shared-worktree mutation and determines whether bounded regions of the same file may be admitted concurrently under the declared model. ATM therefore asks a narrower and more specific research question: when several LLM agents have already formed write intents within the same controlled filesystem, worktree, or service domain, how can the system make an auditable admission decision before any governed mutation is applied, using atoms, atom maps, virtual atoms, CIDs, and \mbox{ConflictKeys}?

The role of AgenticFlict in this paper must be scoped carefully. Its dataset of more than 142,000 AI-agent pull requests, more than 107,000 deterministic merge simulations, a reported merge-conflict rate of 27.67\%, and more than 336,000 fine-grained conflict regions provides quantitative motivation for downstream Git and PR conflict pressure \textcolor{atmLinkBlue}{\hyperlink{Item.50}{[14]}}. These results are not direct evidence that ATM resolves cross-clone or cross-PR conflicts. ATM does not replace Git merge. Instead, it moves the governance point earlier by handling parallel write intents within the same controlled worktree, filesystem, or service domain before the resulting changes enter downstream Git or PR integration.

Existing approaches address different facets of this problem, but they intervene at different coordination layers. Character-level systems such as CodeCRDT provide a low-level convergence substrate for concurrent text editing, while leaving residual semantic conflicts to downstream validation \textcolor{atmLinkBlue}{\hyperlink{Item.37}{[1]}}. Workspace-isolation approaches such as CAID use isolated Git worktrees and structured integration to support asynchronous development, but their principal conflict boundary remains downstream integration \textcolor{atmLinkBlue}{\hyperlink{Item.48}{[12]}}. STORM instead represents a preventive file-level alternative: it mediates agent interaction with a shared workspace and detects stale or conflicting edits at write time, but it does not target same-file bounded-region admission below the file level \textcolor{atmLinkBlue}{\hyperlink{Item.39}{[3]}}. At a higher protocol layer, MPAC provides multi-principal coordination semantics across session, intent, operation, conflict, and governance layers \textcolor{atmLinkBlue}{\hyperlink{Item.40}{[4]}}, while SCF addresses semantic-intent divergence through process-aware governance and semantic-intent representations \textcolor{atmLinkBlue}{\hyperlink{Item.38}{[2]}}. These systems are therefore adjacent approaches rather than direct baselines. What remains largely missing is a single-domain, repository-scoped pre-write admission gate that can determine, before governed mutation is applied, whether same-file bounded regions may be composed, whether shared-surface or read/write dependencies require serialization, or whether insufficiently evidenced cases must take a fail-closed path. ATM is proposed to address that narrower admission boundary.

\subsection{A False Dichotomy}

A common but unnecessary dichotomy presents multi-agent code co-writing as a choice between two extremes. At one end are character-level CRDTs and their text-convergence logic \textcolor{atmLinkBlue}{\hyperlink{Item.45}{[9]}}. At the other is an architecture in which every language and artifact format must first be lifted into a complete AST or global semantic graph before shared-write governance can begin. In practice, existing approaches span a broader coordination stack, including text convergence, region-level comparison, file-level ownership, workflow-level authority, and validation envelopes. Each operates at a different enforcement boundary. For multi-agent writes to a shared codebase, the minimum viable governance unit is therefore neither necessarily a character nor a complete AST. It can instead be an atom, bounded region, CID, or shared surface declared by a domain adapter.

ATM occupies an intermediate position in the design space by combining adapter-guided atomization with brokered admission. An adapter is not required to model the complete semantics of a language or artifact format. Instead, it conservatively declares candidate atoms, source paths, bounded ranges, read/write dependencies, ConflictKeys, and shared surfaces. The broker does not rely on free-form LLM judgment; it produces deterministic admission decisions from these structured declarations. ATM therefore neither delegates all coordination reasoning to an LLM nor forces every language and format into a universal AST. It builds pre-write governance on an engineering-feasible adapter contract and a governance substrate.

More precisely, ATM narrows conflict granularity through a sequence of representation and refinement stages. It first determines whether two intents affect different files or artifacts. It then uses the relevant adapters to identify existing semantic atoms. The atom map connects those atoms to owners, tests, validators, dependencies, and shared surfaces. When the existing atomization is incomplete or too coarse for a reliable comparison, ATM introduces virtual atoms so that previously unatomized regions can still be located, compared, assigned provisional ConflictKeys, and re-hashed into candidate CIDs. If these stages cannot establish an admissible disjoint or composable route under the declared model, the unresolved overlap remains subject to serialization, refinement, or fail-closed containment. ATM is therefore not merely a finer-grained diff mechanism. Its core contribution is a process that converts underspecified write intents into structured and auditable admission evidence. The contributions and empirical evaluation below follow this thread.

\subsection{Contributions}

This paper does not propose another general-purpose multi-agent orchestrator. It reframes governed shared writes within a single authority domain as a formalizable, computable, and auditable pre-write admission problem. We make three contributions. The evaluation artifacts, adopter study, self-hosting forensics, and limitations are presented in Sections 4--6 as supporting evidence rather than as additional contributions.

\textbf{Seven-layer pre-write admission with virtual-atom fallback.} We propose a seven-layer hard admission gate that evaluates multi-agent write intents through CID identity, shared-surface overlap, read/write dependencies, file-range and virtual-atom refinement, \texttt{\seqsplit{ConflictKey + canMerge}}, CAS base-hash validation, and a fallback file lock. Its purpose is not to repair merge conflicts after changes have already been produced. Instead, before any governed shared mutation is applied, the gate determines whether an intent may proceed in parallel, should be routed to deterministic composition, must be serialized, or must take a fail-closed path.

ATM does not assume that the persistent atom map is complete from the outset. When atomization coverage is partial, when an adapter cannot yet associate every affected region with a stable semantic atom, or when same-file regions cannot be compared reliably at the current granularity, the system introduces virtual atoms as transitional governance units. Virtual atoms allow previously unatomized regions to be located, assigned provisional ConflictKeys, re-hashed into candidate CIDs, and evaluated for bounded-region disjointness. If an admissible route cannot be established under the declared model, the corresponding intent is conservatively serialized, refined, or fail-closed on direct apply while its intent evidence is preserved when available. The contribution is therefore not simply to permit more parallelism, but to provide a deterministic gate that progressively transforms coarse file-level contention into auditable admission decisions.

\textbf{A specification-to-evidence governance substrate.} ATM binds shared-write governance to a structured execution contract so that task intent, scope boundaries, validation requirements, and evidence obligations are not dispersed across prompts, human conventions, and ad hoc closure procedures. A task-direction lock, pre-tool scope gate, validator envelope, evidence blocker, review advisory, and closure packet jointly form this substrate. Within it, the CID broker serves as the shared-mutation admission subsystem.

This substrate does not synchronize agents' latent beliefs, nor does it guarantee end-to-end semantic correctness. Its role is to constrain how specification drift, scope drift, unsupported reasoning, and state drift can become an ungoverned repository mutation or an unauditable task closure. The formal Task Contract, the three governance planes, Governance Invariants G1-G3, and the boundaries between the broker, steward, validators, and closure mechanisms are defined in Section 3.1.

\textbf{An extensible atomization abstraction with explicit adapter contracts.} Our third contribution is not a claim that all languages and artifact formats receive equivalent implementation support. Rather, it elevates atomization from a language-specific technique into an extensible, contract-bound repository-governance interface. An atom is the smallest governable unit that must be distinguished for pre-write arbitration. The atom map aligns bounded surfaces, owners, validators, dependencies, CIDs, hash locks, and shared surfaces in a queryable governance index.

Through the \texttt{\seqsplit{AtomizationPlanningAdapter}} and \texttt{\seqsplit{FileMutationAdapter}} interfaces, each language or format may independently expose candidate atoms, bounded ranges, read/write dependencies, ConflictKeys, virtual-atom boundaries, merge capabilities, and validation hooks without first being translated into a universal AST. Within the current implementation coverage, TypeScript and Python serve as the reference language paths. Cross-language atom identity and stronger semantic alignment remain open problems.

\subsection{Organization}

Section 2 positions ATM relative to prior work and identifies the single-domain pre-write admission gap addressed in this paper. Section 3 presents the specification-grounded governance substrate, defines the Task Contract, atoms, atom maps, virtual atoms, and Candidate and Capsule CIDs, and then describes the brokered admission flow, the seven-layer gate, the neutral steward, cross-format generalization, and the explicit scope boundary. Section 4 reports the validation and evidence stack, including deterministic fixtures, self-hosting forensics, the external-adopter study, archived same-file boundary cases, and their alignment with the benchmark claims. Section 5 presents the ATM-AdmissionBench baseline and paper profile, the OperationalBench runtime-overhead supplement, the role-separated audit, the policy and ablation results, and the remaining evaluation limitations. Section 6 discusses the trade-offs and failure modes of adapter-guided governance, open research questions, and deployment topologies. Section 7 concludes.

Reproducibility. Every capability described in this paper as implemented and reproducible is backed by an existing source path and an executable verification command in the open-source AI-Atomic-Framework repository (\url{https://github.com/eaglhuang/AI-Atomic-Framework}). Framework-level implementation claims are anchored to release tag \texttt{\seqsplit{v0.9.0-alpha.1}} (commit \texttt{\seqsplit{0b31aa8683b44b3a78206132a0bf90a0fde73d1c}}). Benchmark-specific claims use the separately frozen generator and publication commits, generated summaries, and artifact-hash manifests listed in Appendices A.1 and A.4. Readers should use these frozen anchors rather than the evolving \texttt{\seqsplit{main}} branch when reproducing results or citing implementation locations. Appendix A.4 provides a compact verification map for the implemented capability groups, while the full row-level source-path and command map is maintained in the repository and supplementary verification material. Field-evidence packets are indexed under \texttt{\seqsplit{docs/ai\_atomic\_framework/broker-collision-evidence/}} in the 3KLife planning repository, with public, de-identified, and access-restricted artifacts distinguished in Appendix A.1.

\section{Related Work}

We organize related work along two axes: coordination granularity and the intervention point at which conflicting actions are governed, with particular attention to preventive pre-write admission. This organization is not intended to rank systems along a single quality dimension, but to locate ATM within a broader coordination stack. ATM does not replace CRDTs \textcolor{atmLinkBlue}{\hyperlink{Item.45}{[9]}}, Git \textcolor{atmLinkBlue}{\hyperlink{Item.65}{[29]}}, workflow orchestration \textcolor{atmLinkBlue}{\hyperlink{Item.38}{[2, 4, 44]}}, or post-generation validation. Instead, it introduces a governable admission boundary before any governed shared mutation is applied within the same controlled worktree or service domain.

The body keeps the related-work discussion as a citation-to-claim map rather than a bibliography survey. Repository-level benchmarks establish multi-file and long-horizon task pressure \textcolor{atmLinkBlue}{\hyperlink{Item.55}{[19, 23--27, 32--34]}}. Agentic concurrency-control systems such as CoAgent, S-Bus, and ATCC define adjacent shared-state coordination substrates \textcolor{atmLinkBlue}{\hyperlink{Item.41}{[5, 11, 16]}}. Transactional tool-effect runtimes such as Atomix and Cordon define adjacent settlement layers \textcolor{atmLinkBlue}{\hyperlink{Item.77}{[41--42]}}. Repository-level workflow, protocol, and convergence systems such as CodeTeam, SEMAP, MPAC, CodeCRDT, AgentGit, and EvoGit define adjacent planning, coordination, or convergence points \textcolor{atmLinkBlue}{\hyperlink{Item.37}{[1, 4, 15, 17--18, 44]}}. These works are adjacent positions in the design space rather than direct baselines for the present paper. ATM's claim is narrower: repository-scoped pre-write admission for governed shared mutation inside a single governance domain, rather than general serializability recovery, HTTP-observable read isolation, transactional tool-effect settlement, database transaction scheduling, or end-to-end repository generation.

The full citation-to-claim map is moved to the supplementary material. The body retains the representative comparison layers needed for the argument: character-level convergence, file-level state mediation, workspace isolation, workflow or protocol governance, agentic concurrency control, transactional tool-effect settlement, post-hoc merge or repair, and repository-level generation. This keeps the main text focused on ATM's intervention boundary while preserving the claim-to-citation trace outside the body.

\subsection{Character-Level and Version-Control Substrates}

CodeCRDT, EvoGit, and AgentGit can be viewed as low-level convergence or version-control substrates for multi-agent code work \textcolor{atmLinkBlue}{\hyperlink{Item.37}{[1, 17--18]}}. Their advantage is generality: they are largely language-neutral and can be integrated with existing editor or version-control workflows. Their limitation for ATM's problem is intervention boundary. They do not directly provide atom-level or bounded-region pre-write admission. CodeCRDT, for example, reports residual semantic conflicts despite character-level convergence \textcolor{atmLinkBlue}{\hyperlink{Item.37}{[1]}}. ATM should therefore not be read as a zero-percent-versus-five-percent semantic-conflict comparison. It shifts a subset of statically observable conflicts earlier in the execution timeline, while residual semantic conflicts remain subject to validators, CAS checks, and downstream runtime evidence.

\subsection{File, Workspace, and Workflow Governance}

STORM and CAID represent two different file- or workspace-level alternatives. STORM performs write-time state mediation using file versions and observed dependencies, whereas CAID isolates agents in separate Git worktrees and reconciles changes afterward \textcolor{atmLinkBlue}{\hyperlink{Item.39}{[3, 12]}}. ATM targets the narrower gap between these positions: it decomposes the notion of same file into adapter-declared regions, CIDs, ConflictKeys, shared surfaces, and dependency sets that the broker can evaluate before governed mutation is applied.

SCF, MPAC, SEMAP, CodeTeam, and AutoGen operate above the code-region layer, but at different enforcement objects \textcolor{atmLinkBlue}{\hyperlink{Item.38}{[2, 4, 15, 20, 44]}}. SCF addresses semantic-intent divergence; MPAC defines cross-principal coordination semantics; SEMAP centers on behavioral contracts and structured message verification; CodeTeam organizes repository generation through design sketches, ownership, and dependency-aware scheduling; and AutoGen exemplifies general multi-agent orchestration without defining repository-region admission semantics. These systems show that multi-agent coordination is a problem of authority, intent, and governance. ATM carries that problem into a lower enforcement boundary: whether a specific governed mutation may enter the shared write path before the write occurs.

\subsection{Adjacent Concurrency, Transaction, and Repair Systems}

CoAgent, S-Bus, ATCC, Cordon, and Atomix are ATM's closest concurrency and transaction comparators, but they operate at different substrates \textcolor{atmLinkBlue}{\hyperlink{Item.41}{[5, 11, 16, 41--42]}}. CoAgent addresses tool/action-level shared-state concurrency through notification and repair. S-Bus reconstructs read sets over HTTP-observable shared state. ATCC studies adaptive concurrency control for database-oriented agentic transactions. Atomix and Cordon stage, validate, settle, compensate, or audit tool effects before durability. ATM occupies a repository-specific point in this design space: adapter-guided atoms and ConflictKeys determine whether governed shared mutations may be admitted under the declared model, and the neutral steward executes the admitted plan. These systems are complementary rather than interchangeable.

Rover, ColaUntangle, SafeMerge, and semistructured merge represent later or adjacent intervention points \textcolor{atmLinkBlue}{\hyperlink{Item.49}{[13, 39--40, 45]}}. Rover reasons over merge-conflict hunks after candidate changes have been produced. ColaUntangle uses dependency reasoning to partition already-tangled commits. SafeMerge verifies semantic conflict-freedom after divergent versions exist, while semistructured merge uses partial syntax to improve on line-based merging. Sartori's specification-gap analysis is relevant to both stages, because incompatible assumptions about a shared surface can affect either pre-write admission or post-hoc repair \textcolor{atmLinkBlue}{\hyperlink{Item.43}{[7]}}. ATM intervenes earlier: before governed shared mutation is applied, it derives conservative atom-, region-, shared-surface, and ConflictKey abstractions under the declared admission model. It therefore does not inherit post-hoc semantic-conflict-freedom guarantees, but it also does not wait until candidate versions have already landed.

\subsection{Adjacent Foundations and Boundary Enforcement}

OT, CRDTs, OCC, Adya's isolation theory, COPS, Git, and database concurrency-control foundations provide the conceptual background for ATM's admission model \textcolor{atmLinkBlue}{\hyperlink{Item.44}{[8--10, 21--22, 28--30]}}. The division of labor is straightforward. OT and CRDTs address operation transformation or convergence after edits exist. OCC and CAS-style validation motivate the steward's base-hash guarded apply. Adya and COPS provide language for read/write dependencies and causal relationships, while Git remains the downstream version-control substrate rather than a distributed lock replaced by ATM. ATM adapts these ideas to a single-domain repository admission boundary rather than claiming a new general-purpose distributed transaction protocol.

Adjacent work on agent-computer interfaces, runtime policy enforcement, and solver-aided policy checking motivates ATM's task-contract and evidence-closure planes. SWE-agent shows that the agent-computer interface materially affects automated software-engineering performance \textcolor{atmLinkBlue}{\hyperlink{Item.71}{[35]}}. AgentSpec and ClawGuard show that structured runtime policies can constrain tool-using agents at action boundaries \textcolor{atmLinkBlue}{\hyperlink{Item.72}{[36--37]}}. Solver-aided policy checking points to a future direction in which a subset of ATM's forbidden rules and task-contract predicates could become machine-checkable \textcolor{atmLinkBlue}{\hyperlink{Item.74}{[38]}}. ATM adapts these boundary-enforcement ideas to repository governance by connecting task scope, mutation admission, validators, evidence obligations, and closure packets into one single-domain governance path.

\section{Framework}

This section separates three layers that are easy to conflate. The \textbf{specification-grounded governance substrate} defines what an agent is authorized to do and what counts as valid closure of that work. The \textbf{formal admission model} defines atoms, shared surfaces, and the active state on which admission decisions are made. The \textbf{brokered implementation path} turns structured write intents into concrete admission verdicts. Within this structure, the CID broker is not the entire system; it is the shared-mutation admission subsystem inside a broader specification-grounded governance path. ATM does not generate code, replace tests, or replace code review. Instead, every governed shared mutation must first be expressed as a structured write intent, after which the broker decides, within the boundary set by the governance substrate, whether that intent may enter the write path.

This section is organized into two conceptual groupings. Part A (§3.1--§3.3) presents the model and its assumptions: §3.1 introduces the three planes of the governance substrate and the three governance invariants; §3.2 describes the broker, agent, and neutral-steward architecture under the single-governance-domain assumption; and §3.3 defines atoms, atom maps, virtual atoms, the two-tier CID structure, and the auxiliary broker-facing state. Part B (§3.4--§3.7) presents the framework and its implementation path: §3.4 gives the admission pipeline and Propositions 1 and 2; §3.5 defines the seven hard pre-write gates and the CAS-based runtime closure of Definition 7; §3.6 generalizes the framework across artifact formats; and §3.7 states the known scope limitations.

Both groupings share the same governance-domain assumption. The substrate model in §3.1 provides the common backbone for the plane-specific mechanisms developed in §3.2--§3.7, while the CID broker described in §3.4--§3.5 implements the mutation-admission plane defined in §3.1. Readers interested primarily in the governance model may focus on §3.1. Readers interested in the formal definitions and broker-facing structures should continue through §3.2 and §3.3. Readers interested in how an admission verdict is produced should read §3.4 through §3.7.

\subsubsection*{Part A - Model and Assumptions}

\subsection{Specification-Grounded Governance Substrate}

ATM is positioned as a \textbf{specification-grounded execution-governance substrate}. It binds an agent's task, behavioral boundary, and closure claim to a structured execution contract, the \textbf{Task Contract}, and governs three related concerns: the agent's authorized scope, the admission of governed shared mutations, and the evidentiary closure of the task. The CID broker is therefore not synonymous with the substrate as a whole; it is the subsystem that implements the substrate's mutation-admission plane.

\textbf{Definition 1 (Task Contract).} For any authorized agent task $\mathcal{T}$, the Task Contract is the eight-tuple

\[
\mathcal{T} = \langle g, A, F, S, D, V, E, \epsilon \rangle
\]

where

\begin{itemize}
\item $g$ is the approved task intent;
\item $A$ is the set of allowed resources and files;
\item $F$ is the set of forbidden predicates and rules;
\item $S$ is the set of governed scope paths;
\item $D$ is the set of required deliverables;
\item $V$ is the set of validation commands;
\item $E$ is the set of evidence obligations;
\item $\epsilon$ is the task-direction epoch.
\end{itemize}

In implementation, the \textbf{task card} is the concrete serialization of this contract. The scope lock, direction lock, validator envelope, evidence blocker, and closure packet all reference the same $\mathcal{T}$.

\textbf{Three-plane architecture.} The substrate is composed of three planes with distinct responsibilities. Each plane answers a different layer of the agent-governance question.

\TableLead{Table 1 --- Governance-Plane Summary.}

\begingroup
\TableLargeFont
\setlength{\tabcolsep}{3pt}
\arrayrulecolor{black!55}
\rowcolors{2}{gray!8}{white}
\begin{longtable}{@{}L{0.22\textwidth}L{0.25\textwidth}L{0.47\textwidth}@{}}
\toprule
Plane & ATM mechanisms & Question answered \\
\midrule
Task-contract plane & task intent, allowed files, forbidden rules, scope paths, deliverables, validation requirements, evidence obligations, direction lock, task epoch / scope envelope & What is the agent authorized to do, and to which governance constraints is its completion criterion bound? \\
Mutation-admission plane & atoms, CID, ConflictKey, read/write set, active registry, broker, neutral steward & May this governed shared mutation happen at this moment? \\
Evidence-closure plane & validation commands, validator envelope, evidence blockers, review advisory, closure packet & Can the task reasonably be claimed complete? \\
\bottomrule
\end{longtable}
\endgroup

Table 1 summarizes these three planes at a glance. The three planes do not jointly answer whether an agent might hallucinate or misread requirements. Instead, they specify \textbf{which checkable governance boundaries an erroneous thought must pass through before it can become a governed shared mutation or a task closure}.

\textbf{Three governance invariants.} ATM's governance commitments are expressed as three invariants, denoted G1, G2, and G3. They are stated as the substrate's design contract rather than as mechanically proven theorems.

\textbf{G1 (Scope containment).} For any governed write intent $I$ and its associated task contract $\mathcal{T}$, both

\[
Res(W(I)) \subseteq A(\mathcal{T})
\quad\text{and}\quad
Path(W(I)) \subseteq S(\mathcal{T})
\]

must hold. A write is therefore in-bounds only when it targets both an authorized resource and an authorized scope path.

\textbf{G2 (Direction stability).} If the task goal $g$ or either scope set $A$ or $S$ changes, a new direction epoch $\epsilon' \neq \epsilon$ must be issued. An agent may not silently change the task's direction without updating the epoch.

\textbf{G3 (Evidence-backed closure).} Task closure is permitted if and only if

\[
\text{ClosePermitted}(\mathcal{T}) \iff
\text{V}(\mathcal{T}) = \text{pass}
\;\land\;
D(\mathcal{T}) = \text{satisfied}
\;\land\;
E(\mathcal{T}) = \text{satisfied}
\;\land\;
\text{Writes}(\mathcal{T}) = \text{governed}.
\]

That is, all validators pass, all deliverables are completed, all evidence obligations are satisfied, and every governed shared write has traversed the broker-and-steward governance path.

\textbf{Drift taxonomy.} To prevent the broad term {\char34}context drift{\char34} from blurring the substrate boundary, the paper distinguishes five observable forms of drift and states explicitly which ones ATM addresses. ATM governs externally observable consequences; it does not synchronize internal agent beliefs.

\TableLead{Table 2 --- Drift Taxonomy Summary.}

\begingroup
\TableLargeFont
\setlength{\tabcolsep}{3pt}
\arrayrulecolor{black!55}
\rowcolors{2}{gray!8}{white}
\begin{longtable}{@{}L{0.22\textwidth}L{0.25\textwidth}L{0.47\textwidth}@{}}
\toprule
Drift type & Definition & ATM mechanism \\
\midrule
Epistemic drift & An agent's internal knowledge or belief diverges from the actual state & Not directly handled; this belongs to the agent's internal reasoning \\
Specification drift & Agent behavior diverges from the approved task intent & direction lock, task contract, epoch versioning \\
Scope drift & An agent modifies an unauthorized file, surface, or tool & allowed files, scope paths, pre-tool scope gate \\
Evidence drift & A completion claim is inconsistent with validators or evidence & validator envelope, evidence blocker, review advisory, closure packet \\
State drift & An intent is built on a base state or read dependency that has changed & active registry, \texttt{\seqsplit{readAtoms}}, CAS base-hash \\
\bottomrule
\end{longtable}
\endgroup

In other words, Table 2 makes explicit that ATM does not claim to eliminate hallucinations or synchronize latent beliefs. It constrains how far the four observable drift forms that lie within its governance boundary--specification, scope, evidence, and state drift--can propagate into ungoverned repository mutations or unauditable task closures.

Subsystem-role clarification. The broker, atom map, ConflictKey, neutral steward, and validator mechanisms described in §3.2--§3.3, and the admission pipeline, seven-layer gate, cross-format generalization, and scope limitations described in §3.4--§3.7, are concrete realizations of the three planes introduced here. Among them, the CID broker is the core subsystem of the mutation-admission plane. It does not by itself constitute the governance substrate; it implements the substrate's admission function. The principal enforcement components of the Task-contract plane and the Evidence-closure plane--direction lock, pre-tool scope gate, validator envelope, evidence blocker, and closure packet--are implemented by the framework's outer governance layer and share broker-visible task state and active-intent visibility with the admission subsystem.

\textbf{Assumptions.} All claims in this paper rest on four assumptions. First, there is a \textbf{single authority governance domain} in which the broker has timely visibility into governed shared writes, active intents, and mutable shared surfaces. Second, adapter declarations of write surfaces, ConflictKeys, and declared read sets are \textbf{conservative approximations}, over-approximating rather than understating potential conflicts. Third, governed shared writes are applied through the \textbf{neutral steward} rather than bypassing the brokered path. Fourth, validators are available and semantically meaningful for the relevant domain.

Under these assumptions, ATM claims static admission closure (Proposition 2) and auditable runtime enforcement. It does not claim distributed consensus, complete dynamic dependency capture, or end-to-end semantic correctness. The boundaries that arise when adapters are adversarial or incomplete, when coordination spans multiple governance domains, or when validators are absent are discussed in §6.2 and §6.3.

\subsection{Architecture Overview}

ATM is organized around five responsibility boundaries that share a single, progressively refined semantic index. These boundaries are assumed to operate within one governance domain: the same machine, the same controlled server, the same worktree service, or another environment that can provide a single broker-and-steward authority. The \textbf{Adapter} extracts candidate atoms, bounded ranges, read/write dependencies, and ConflictKeys from a language or artifact format. The \textbf{Atom Map} organizes that information into a testable, validatable, and auditable logical map; when the map does not yet cover a region of change, the broker materializes \textbf{virtual atoms} as transitional governance units. The \textbf{Agent} proposes patches or structured write intents. The \textbf{Broker} issues admission verdicts, producing outcomes such as parallel admission, deterministic composition, blocking, or re-arbitration. The \textbf{Neutral Steward} applies an admitted plan to the same controlled worktree; it is neither a content proposer nor an arbiter, but the executor of broker verdicts, the sole formal apply authority for governed shared writes within the governance domain, and the landing node for evidence records, validator triggers, and downstream commit and pre-push governance. The \textbf{Substrate} comprises Git, the filesystem, registries, validators, and evidence artifacts; among these, Git is the version-control and cross-clone merge substrate, not a distributed lock that ATM replaces in this paper.

\begin{figure}[H]

{\small \noindent\textbf{Figure 1 --- ATM as a Specification-Grounded, Three-Plane Governance Substrate.} The Task-contract plane constrains what the agent is authorized to do. The CID broker and neutral steward, located inside the mutation-admission plane, constrain when a governed shared mutation may occur and how it is applied. The Evidence-closure plane constrains when a task may be claimed complete. The three subgraphs in the figure correspond to the three planes of §3.1; the CID broker implements admission, while the neutral steward enforces the governed apply path.\par}
\smallskip
\centering
\begin{tikzpicture}[
  >=Latex,
  font=\scriptsize,
  plane/.style={draw, rounded corners=5pt, line width=0.65pt, inner sep=5pt},
  lane/.style={plane, minimum width=150mm, align=left},
  nodebox/.style={draw, rounded corners=2.5pt, line width=0.45pt,
                  minimum height=8.5mm, minimum width=27mm,
                  align=center, fill=white, inner sep=2.5pt},
  actor/.style={nodebox, fill=gray!10, draw=black!50},
  task/.style={nodebox, fill=blue!18, draw=blue!60!black},
  admit/.style={nodebox, fill=green!20, draw=green!55!black},
  broker/.style={nodebox, fill=green!32, draw=green!60!black,
                 line width=0.75pt, minimum width=30mm},
  close/.style={nodebox, fill=orange!20, draw=orange!70!black},
  main/.style={->, line width=0.72pt, draw=black!75},
  support/.style={->, line width=0.45pt, draw=black!55},
  feedback/.style={->, line width=0.48pt, draw=black!45, dash pattern=on 2.2pt off 1.6pt},
  label/.style={font=\tiny\itshape, fill=white, fill opacity=0.96,
                text opacity=1, rounded corners=1pt, inner sep=1.8pt},
  title/.style={font=\tiny\bfseries, anchor=west}
]
\node[lane, fill=blue!7, draw=blue!55!black, minimum height=17mm] (TC) at (0, 2.85) {};
\node[lane, fill=green!9, draw=green!48!black, minimum height=34mm] (MA) at (0, 0.20) {};
\node[lane, fill=orange!9, draw=orange!68!black, minimum height=17mm] (EC) at (0,-2.95) {};
\node[title, blue!65!black] at ([xshift=2mm,yshift=-2mm]TC.north west) {Task-contract plane};
\node[title, green!40!black] at ([xshift=2mm,yshift=-2mm]MA.north west)
  {Mutation-admission plane / CID broker subsystem};
\node[title, orange!65!black] at ([xshift=2mm,yshift=-2mm]EC.north west) {Evidence-closure plane};

\node[actor, minimum width=22mm] (HU) at (-8.70, 3.05) {Human /\\Coordinator};
\node[task]  (T)  at (-2.7, 2.85) {Task Contract\\\tiny$\langle g,A,F,S,D,V,E,\epsilon\rangle$};
\node[task]  (DL) at ( 2.1, 2.85) {Direction Lock\\+ Scope Gate};

\node[admit] (AG) at (-5.5, 0.75) {AI Agent\\\tiny WriteIntent};
\node[admit] (AD) at (-5.5,-0.90) {Adapter Layer\\\tiny atom candidates};
\node[admit] (AM) at (-2.0, 0.75) {Atom Map\\\tiny owners / deps / CID};
\node[admit] (VA) at (-2.0,-0.90) {Virtual Atoms\\\tiny refinement};
\node[broker] (BR) at (1.8, 0.20) {\bfseries CID Broker\\\tiny progressive admission};
\node[admit, minimum width=30mm] (ST) at (5.6, 0.20) {Neutral Steward\\\tiny single apply + CAS};

\node[close] (VE) at (-3.6,-2.95) {Validator\\Envelope};
\node[close, minimum width=31mm] (EB) at (0,-2.95) {Evidence Blockers\\+ Review Advisory};
\node[close] (CP) at (3.8,-2.95) {Closure\\Packet};

\draw[main] (HU) -- (T);
\draw[main] (T) -- (DL);
\draw[main] (DL.south) |- ([yshift=1.2mm]AG.north);
\draw[main] (AG) -- (AM);
\draw[main] (AM) -- (BR);
\draw[main] (BR) -- node[label, above, yshift=1pt]{verdict + plan} (ST);
\draw[main] (ST.south) -- ++(0,-1.18) -| (CP.north);

\draw[support] (AD) -- (AM);
\draw[support] (AM) -- (VA);
\draw[support] (VA) -- (BR);
\draw[support] (AD.east) -- ++(0.8,0) |- (VA.west);

\draw[main] (VE) -- (EB);
\draw[main] (EB) -- (CP);

\draw[feedback] (EB.north) -- ++(0,0.78) -| node[label, near end, left, xshift=-1pt]
  {evidence feedback} (BR.south);
\draw[feedback] (VE.west) -- ++(-1.55,0) |- node[label, near end, below, yshift=-1pt]
  {post-write validators} (AD.west);
\draw[feedback] (CP.south) -- ++(0,-0.52) -| node[label, pos=0.18, below]
  {epoch update on scope change} ([xshift=-2mm]T.south);
\end{tikzpicture}
\end{figure}

Reading note for Figure 1. The three subgraphs map to the §3.1 planes as follows:

\begin{itemize}
\item \textbf{Task-contract plane.} Starting from the human or coordinator, a structured Task Contract $\mathcal{T} = \langle g, A, F, S, D, V, E, \epsilon \rangle$ is produced and concretely serialized as the task card. The direction lock and pre-tool scope gate enforce G1 (scope containment) and G2 (direction stability).
\item \textbf{Mutation-admission plane (CID broker subsystem).} An agent's WriteIntent is structured through the adapter, atom map, and virtual atoms before reaching the broker. The broker acts as the sole serialization node and emits a verdict. Once admission passes, the neutral steward executes the governed write and performs the CAS base-hash recheck. This plane corresponds to the admission pipeline and the seven-layer gate of §3.4--§3.5.
\item \textbf{Evidence-closure plane.} After the steward writes, the validator envelope is triggered through typecheck, lint, CLI validators, or other project-defined checks. Evidence blockers and review advisory then check whether deliverables carry traceable evidence. The closure packet is the evidentiary closure object of a legitimate task close, corresponding to G3 (evidence-backed closure).
\end{itemize}

Solid arrows in the figure denote the main governance path. The three dashed arrows mark non-blocking feedback channels and \textbf{are not part of the admission decision path}: (a) the evidence-feedback arrow returns from the substrate to the broker, carrying verdict logs, CAS base-hash outcomes, and closure packets so that the next admission cycle reads the latest state of the active registry (Definition 6); (b) the post-write-validator arrow returns from the validator envelope to the agent, carrying typecheck and lint results so the agent can observe validator catches before its next intent, as in the three validator catches reported in §4.3; and (c) the epoch-update arrow returns from the closure packet to the Task Contract, triggering an epoch transition $\epsilon \to \epsilon'$ when scope or goal changes (G2). None of these three dashed feedback paths alters the broker's single apply decision for the currently admitted plan; they only feed the next governance cycle.

The key property of this architecture is that an agent does not directly hold final write authority over the shared filesystem. The agent may produce a proposal, but the proposal must pass through the broker; even when the broker rules that two proposals can be composed, the neutral steward performs the governed write. This separates {\char34}who proposes a change{\char34} from {\char34}who executes the write{\char34}, and reduces the risk of multiple agents overwriting each other, racing for shared resources, or bypassing the governance flow.

Figure 1 describes ATM's governance path; it does not claim that every local file write within the governance domain must traverse the steward. For a single agent's private work-in-progress, or for local edits that have not yet entered a shared surface, deployments may retain direct-write or direct-commit workflows. ATM intervenes primarily on shared files, shared artifacts, or other write intents that have been declared as requiring governance. Once a write enters the broker-governed path, the neutral steward becomes the sole formal apply authority for governed shared writes.

Team Agents and Captain-style orchestration sit above this path as an operational coordination layer. A Captain may decide which role should read, implement, validate, review, summarize, or assemble evidence for a task, and a Team Agents runtime may use separate implementer, validator, evidence-collector, review-agent, or steward-facing writer roles. This role separation is useful for reducing single-agent drift and for parallelizing read-only or evidence work. It does not create a second admission authority. Once a proposed change touches a broker-governed shared surface, the CID broker remains the cross-task conflict governor, and the neutral steward remains the formal apply authority for admitted governed writes.

Team Agents also provide a least-privilege execution model that can scale by role rather than by granting every worker broader authority. A small task may need only a Captain, Implementer, and Validator, while a larger task may add a Planner, Scope Guardian, Knowledge Scout, Review Agent, Evidence Collector, or specialized domain agent. Permissions remain role-scoped: read, planning, review, validation, and evidence-drafting roles can proceed in parallel, while write authority, Git authority, task lifecycle authority, and final evidence authority remain exclusive and policy-governed. This distinction matters for both safety and efficiency. A Reviewer or Validator can inspect a draft, run checks, or produce advisory evidence without holding \texttt{\seqsplit{file.write}}; an Evidence Collector can summarize artifacts without holding \texttt{\seqsplit{git.write}} or \texttt{\seqsplit{task.lifecycle}}; and a Knowledge Scout can retrieve prior lessons without becoming a second source of task truth. Team Agents can therefore increase useful parallelism without creating multiple uncontrolled writers, and may reduce unnecessary token exposure by routing narrow read-only, validation, or review tasks to narrower role prompts or cheaper model tiers. This paper treats that potential reduction in token use as a design rationale, not as a quantified performance result.

\subsection{Atom, Atom Map, Virtual Atom, and CID}

This section separates three roles that are easy to conflate: the governable write boundary, the admission-time comparison handle, and the post-validation evidence identity. Atoms and virtual atoms name the write boundaries that the broker can govern. Candidate CIDs and ConflictKeys provide the comparison handles used before admission. Capsule CIDs identify post-validation evidence containers used for replay, rollback, rescue, and drift analysis.

An atom is the smallest stable logical unit that ATM treats as governable under an adapter. In implementation, an atom may represent a function, a class method, a registry entry, a JSON record, a numeric scalar, a text range, or another structured fragment declared by an adapter. ATM therefore does not first adjudicate admission through whole-file heuristics. It maps a write intent to one or more governance objects, and then decides over shared surfaces, declared dependencies, ConflictKeys, and bounded regions whether the intent should be composed, split, serialized, or fail-closed on direct apply. To support broker decisions, this paper represents an atom through auditable fields: atom identity, logical name, version, source path and range, input/output schema, lifecycle status, atom grade, and hash lock.

\textbf{Definition 2 (Atom).} An atom $a$ is an eight-tuple

\[
a = \langle id, name, \mathit{ver}, P, \sigma, \psi, \gamma, H \rangle
\]

where

\begin{itemize}
\item $id$ is the atom identity;
\item $name$ is the logical name;
\item $\mathit{ver}$ is the version;
\item $P$ is the set of source paths and line ranges to which the atom corresponds;
\item $\sigma$ is the input/output schema;
\item $\psi$ is the atom's lifecycle status;
\item $\gamma$ is the atom grade, which records the governance maturity of the unit, such as candidate, provisional, or formal;
\item $H$ is the hash lock over specification, code, and tests.
\end{itemize}

Notation clarification. §2 uses \texttt{\seqsplit{Tier 1–4}} to describe coordination-granularity layers across systems: character, region, file, and workflow. The atom-level $\gamma$ used here in §3 describes the governance-lifecycle maturity of a single atom. The two do not share a semantic space. This paper distinguishes them by reserving capitalized \texttt{\seqsplit{Tier}} labels for related-work granularity and using $\gamma$ only inside atom-level definitions.

In ATM, a Content Identifier (CID) is a content-derived identity signal attached to a governance object. The CID has two distinct purposes in this paper:

\begin{itemize}
\item \textbf{Candidate CID} (admission identity): used during admission to compare governance-object identity and overlap before any governed shared mutation is applied;
\item \textbf{Capsule CID} (post-validation identity): used after validation by encapsulation and versioning flows such as export, import, rollback, rescue, and drift detection.
\end{itemize}

The CID is therefore not a third kind of governance object. It is always attached to a governance object, either an atom or a virtual atom, and the two CID forms serve different phases: Candidate CIDs support pre-write admission, while Capsule CIDs support evidence closure. They are not interchangeable.

The atom map is the semantic index formed from these atoms, and it serves as a central governance index in ATM. It aligns source ranges, owners, test entry points, validators, read and write dependencies, shared surfaces, and coverage gaps into one auditable graph-shaped structure. In the simplest layering terms, atoms together with their CIDs provide conflict identity, while the atom map provides governance context. The former answers which governance unit is touched by a write. The latter answers which owners, validators, dependencies, and shared surfaces that governance unit connects to.

The atom map is therefore neither a file directory nor an alternative identity table. It is the semantic sensor that supports the pre-write admission layer. When only atoms and their \texttt{\seqsplit{atomId}} and \texttt{\seqsplit{atomCid}} are available, the broker can still reach a first-layer CID-conflict verdict. Without the atom map, however, the system has more difficulty bringing owners, validators, dependencies, shared surfaces, and coverage gaps into one auditable index. Contention inside the same file then tends to remain at the coarser atom-set or file-overlap level. The atom map's central value is therefore not that it makes CID verdicts possible for the first time, but that it reduces same-file writes to a finer and more traceable question: which known governance units, shared surfaces, and validation responsibilities are actually touched?

Adapter-guided atomization answers the complementary question of how atomization should be performed. ATM does not require all languages to first share a universal AST, nor does it require the atom map to be complete on day one. Candidate discovery is instead delegated to the adapter, allowing TypeScript, Python, JSON, and other formats to report the cheapest stable atom candidates, canonical symbols, bounded regions, and shared surfaces appropriate to that language or format. Adapters may use regular expressions, scanners, compiler APIs, ASTs, LSPs, or format-specific locators.

ATM's atomization is therefore not a one-shot static preprocessing step, but a progressively extensible governance capability. The system may begin with candidate atoms, project them into an atom map, and then incrementally complete coverage, validators, and dependencies. Equally important, ATM is not merely an abstract interface that outsources the entire implementation cost to adopters. The framework itself ships a default governance skeleton, including candidate-atom bridging, CID computation, atom-map projection, virtual-atom fallback, task-card and skill routing, an editor-integration adapter, a validator and evidence substrate, and the neutral steward. In practice, adopters typically do not rebuild the broker and governance flow from scratch; they extend the existing skeleton with adapters for their target language or artifact format.

Within the current implementation, formal atomization and atom-map generation have landed for at least TypeScript and Python. TypeScript is currently the most mature reference language path. Python is not merely represented by an abstract interface; it has a dedicated \texttt{\seqsplit{@ai-atomic-framework/language-python}} package, together with candidate discovery, atomization dry runs, verification scripts, and fixture tests. Other languages and formats are onboarded incrementally through the \texttt{\seqsplit{AtomizationPlanningAdapter}}, \texttt{\seqsplit{FileMutationAdapter}}, and locator contracts. This is ecosystem expansion on top of the existing framework core, not evidence that ATM's core governance capability is missing.

\textbf{Definition 3 (Virtual Atom).} A virtual atom is a transitional governance unit used when the atom map is incomplete or too coarse to support an admission decision under the declared model. When an adapter has not yet formally atomized a region of code, or when the formal atom map does not cover the affected region with sufficient precision, the broker may construct a virtual atom from a syntactic enclosure, line range, signature boundary, or format-specific locator. A virtual atom carries a temporary identity, a bounded region, a candidate CID, and ConflictKeys. It does not claim to be a permanent API unit, nor is it a formal subclass of atom; it is a temporary broker-facing substitute used when atomization is incomplete. Its purpose is to let the broker convert a suspected same-file conflict into an admission unit that can be compared, validated, audited, and conservatively contained when direct apply is unsafe. ATM's core design therefore does not assume that a repository has already been fully atomized. Instead, it peels back conflicts layer by layer along the atom-map-to-virtual-atom path, even when atom-map coverage is limited.

\textbf{Atom Capsule.} An atom capsule is the versioned evidence container for an atom. It bundles canonical source code, input and output schemas, policy metadata, and validation evidence, and it anchors the corresponding Capsule CID used for export, import, rollback, rescue, and drift detection. It is not a broker-facing admission identity and does not replace atoms, virtual atoms, or ConflictKeys in the admission path.

To connect atom-level identity to the downstream admission gate, this section closes with three broker-facing auxiliary definitions. Definitions 4, 5, and 6 do not introduce new governance objects. Instead, they define the comparison and state structures that let ConflictKeys, shared surfaces, and active transactions be referenced directly by the broker decisions in §3.4--§3.5.

\textbf{Definition 4 (ConflictKey).} For any atom or virtual atom $a$, the ConflictKey is the pair $(S_a, L_a)$, where $S_a$ is the governance-scope category declared by the adapter, such as a function, class method, JSON record or key path, numeric scalar, text range, or atom-map shard. $L_a$ is the locator expression used by that adapter within the scope category, such as a canonical symbol, JSON pointer, line span, registry id, shard key, or other format-specific expression.

Because not all locator domains have equality as their natural conflict relation, each adapter also supplies a conservative overlap predicate:

\[
\operatorname{overlap}_{S}(L_i, L_j).
\]

Two governance units conflict at the ConflictKey layer when they are in the same governance domain, share the same scope category, and their locators overlap under the adapter-defined predicate:

\[
S_{a_i} = S_{a_j}
\;\land\;
\operatorname{overlap}_{S_{a_i}}(L_{a_i}, L_{a_j})
\Rightarrow
\operatorname{Conflict}(a_i, a_j).
\]

For discrete locators such as canonical symbols, JSON pointers, scalar fields, or registry ids, $\operatorname{overlap}_{S}$ may reduce to equality. For range-like locators such as text spans or atom-map shard ranges, it must conservatively report overlap when the represented intervals intersect. Under this abstraction, the broker can compare across formats without assuming that every domain has equality-based conflict semantics.

\textbf{Definition 5 (Shared Surface Set).} For any governed intent $I$, let $\Sigma(I)$ denote the set of governed non-file surfaces referenced by $I$, including registry entries, generated artifacts, validator surfaces, atom-map members, and other shared resources declared by the adapter. Admission must not classify two intents $I$ and $I'$ as \texttt{\seqsplit{parallel-safe}} if

\[
\Sigma(I) \cap \Sigma(I') \neq \emptyset,
\]

even when their physical write regions, represented by the $P_a$ projection, are disjoint. This definition turns the shared-surface gate of the seven-layer gate in §3.5 from an operational check into a formalized precondition. When shared surfaces overlap, the broker must not admit the intents as parallel-safe solely because their file regions do not intersect; the admission must instead route to deterministic composition, SERIAL, or a fail-closed path according to the shared-surface policy.

\textbf{Definition 6 (Active Registry).} The Active Registry is a broker-local finite mapping

\[
\mathcal{R} : \text{TxnId} \to (\text{Intent}, \mathcal{A}_R, R_D, W, F),
\]

where $\mathcal{A}_R$ is the atom set declared by this transaction, including virtual atoms; this registry-scoped atom set is distinct from the allowed-resource set $A(\mathcal{T})$ in Definition 1. $R_D$ and $W$ are the declared read and write sets, and $F$ is the hash snapshot of the target files. The registry is updated by admission outcomes: \texttt{\seqsplit{parallel-safe}} and \texttt{\seqsplit{needs-physical-split}} transactions are added as apply-eligible active records; transactions are removed once the steward completes the apply; and \texttt{\seqsplit{blocked-*}} verdicts may be retained as blocked intent records for evidence and rearbitration, but they do not become apply-eligible active writers. Every new intent is compared against $\mathcal{R}$ as the currently active working set, together with the static admission closure condition supplied by Proposition 2.

To prevent conceptual overlap, this paper does not treat the CID as a third independent governance object. The CID is an identity signal attached to a governance object: the Candidate CID carries pre-write admission identity, while the Capsule CID carries post-validation evidence-closure identity.

To give the subsequent bounded-region comparison an explicit mathematical anchor, the paper treats the physical footprint of an atom or virtual atom as its source-path-and-range projection $P_a$. Given two governance units $a_1$ and $a_2$ touched by two concurrently considered intents, physical disjointness is not a vague judgment that {\char34}the line numbers look unrelated.{\char34} It is the condition that their governance footprints satisfy

\[
P_{a_1} \cap P_{a_2} = \emptyset.
\]

The disjointness test used in bounded-region admission is therefore grounded in the structured ranges $P$ declared by the adapter and recorded by the atom map or virtual atom, rather than in arbitrary string-level diff heuristics.

Table 3 summarizes the four governance objects most often conflated.

\Needspace{7\baselineskip}\TableLead{Table 3 --- Governance Object Comparison.}

\begingroup
\TableFont
\setlength{\tabcolsep}{2pt}
\arrayrulecolor{black!55}
\rowcolors{2}{gray!8}{white}
\begin{longtable}{@{}L{0.17\textwidth}L{0.17\textwidth}L{0.18\textwidth}L{0.18\textwidth}L{0.18\textwidth}@{}}
\toprule
Object & Role & Persistent & First-line broker input & Primary use \\
\midrule
atom & Formal governance unit & Yes & Yes & Semantic unit that can be declared, indexed, and adjudicated \\
atom map & Governance-context index & Yes & No; used as a supporting index & Connects owners, validators, dependencies, shared surfaces, and coverage \\
virtual atom & Transitional verdict unit & No & Yes & Fills coverage gaps and supports bounded comparison and fail-closed admission \\
atom capsule & Encapsulation and version-evidence unit & Yes & No; not a first-line admission identity & Export, import, rollback, rescue, drift detection, and version anchoring \\
\bottomrule
\end{longtable}
\endgroup

Adapter-guided discovery is necessary because atom identity cannot be derived reliably from string-level diffs or file paths alone. A TypeScript function, Python decorator, JSON record, and atom-map shard each have different structures. Without adapter declarations of canonical symbols and bounded regions, the broker would have to fall back to file-level or character-level judgment. ATM therefore treats the adapter contract as a precondition for admission, while virtual atoms provide a temporary bridge when the adapter map is incomplete. This prevents the system from being forced into a binary choice between locking the whole file and admitting all writes blindly.

Semantic validation is not invented by ATM on behalf of the project. The framework provides validator hooks and integration-test anchors that let adopters connect type checking, unit tests, integration tests, or domain-specific CLI validators to the atom map and the steward path.

\subsubsection*{Part B --- Framework and Implementation}

\subsection{Admission Flow}

ATM's admission flow begins with a write intent, but its core is not a one-shot comparison of file diffs; it is progressive atomization. From a governance perspective, an AI agent does not directly acquire authority to mutate shared files. It submits a structured write intent that carries target files, atom references, candidate CIDs, bounded regions, shared surfaces, and any required read dependencies.

When the broker receives a write intent, it does not first ask whether two patches collide at the line-number level. It first asks which atoms the intents map to, which atom-map surfaces they touch, and whether any region of the change still lies outside atom-map coverage. When a write intent touches a region not yet covered by the atom map, the broker materializes virtual atoms, converting what would otherwise be vague same-file overlap into comparable logical regions. The broker then compares the two intents in the following gate order before emitting an admission verdict:

\begin{enumerate}
\item CID;
\item shared surface;
\item read and write dependency;
\item physical overlap;
\item known atom coverage;
\item virtual-atom coverage;
\item bounded region.
\end{enumerate}

The principal verdicts are summarized below.

\begin{table}[!b]
\TableLead{Table 4 --- Broker Verdicts and Follow-Up Paths.}
\raggedright
\TableLargeFont
\setlength{\tabcolsep}{3pt}
\arrayrulecolor{black!55}
\rowcolors{2}{gray!8}{white}
\begin{tabular}{@{}L{0.22\textwidth}L{0.25\textwidth}L{0.47\textwidth}@{}}
\toprule
Verdict & Meaning & Follow-up path \\
\midrule
\texttt{\seqsplit{parallel-safe}} & No blocking CID, shared-surface, read/write, or range conflict under the declared model & May enter the steward path \\
\texttt{\seqsplit{needs-physical-split}} & Same file but bounded-disjoint under the declared model; requires deterministic composition & Deterministic composer, then neutral steward apply \\
\texttt{\seqsplit{SERIAL}} & The intents are not safe to admit concurrently, but the work can be preserved in order & Queue or serialize the preserved intent; replay after the active holder completes \\
\texttt{\seqsplit{blocked-cid-conflict}} & Same CID, same atom, or unresolved overlap on the same governed unit & Fail closed to direct apply; preserve the intent record and route to refinement, steward review, rebase, or a queue/serialization path when evidence supports replay \\
\texttt{\seqsplit{blocked-shared-surface}} & Shared-surface exclusion or insufficient evidence for concurrent admission & Fail closed to direct apply; queue, serialize, or route to steward review according to policy and available evidence \\
\bottomrule
\end{tabular}
\end{table}

In this paper, \textbf{fail-closed} means fail-closed to unsupervised parallel or direct apply, \emph{not fail-closed to intent preservation}. When an intent cannot be admitted as \texttt{\seqsplit{parallel-safe}} or routed immediately to deterministic composition, the broker preserves the declared intent, admission evidence, blocking reason, and any available patch or proposal envelope. The successor path depends on the cause of the block: the work may be queued, serialized behind an active holder, routed to steward review, replayed after rebase, or converted into a split-suggestion / atom-map refinement task. This behavior is analogous to Git's conservative boundary for non-fast-forward pushes and merge conflicts: the unsafe write is not applied silently, but the proposed work is not erased merely because direct application is unsafe.

Each broker verdict carries a structured follow-up payload rather than a bare rejection signal. The payload records the verdict label, the blocking layer, the conflicting Candidate CID or ConflictKey when available, the shared-surface or read/write dependency involved, the CAS base-hash mismatch if present, the preserved intent or patch envelope, and any refinement hint or recovery route selected by policy. This structure lets the proposing agent receive actionable repair context while keeping the broker and neutral steward as the only authorities for governed shared-write admission and apply.

\begin{figure}[H]

{\small \noindent\textbf{Figure 2 --- Progressive Atomization Admission Flow.} The figure is a schematic view of how ATM narrows the relevant conflict boundary from coarse to fine granularity under the declared model:\par}
\smallskip
\centering
\begin{tikzpicture}[
  >=Latex,
  font=\scriptsize,
  node distance=3.5mm,
  input/.style ={draw, rounded corners=2pt, fill=blue!12,   draw=blue!60!black,
                 minimum width=28mm, minimum height=8mm, align=center, line width=0.4pt},
  check/.style ={draw, diamond, aspect=2.2, fill=yellow!22, draw=orange!70!black,
                 inner sep=0pt, minimum width=30mm, minimum height=14mm,
                 align=center, line width=0.5pt, font=\tiny},
  safe/.style  ={draw, rounded corners=2pt, fill=green!22,  draw=green!55!black,
                 minimum width=28mm, minimum height=8mm, align=center, line width=0.5pt,
                 font=\scriptsize\bfseries},
  compose/.style={draw, rounded corners=2pt, fill=violet!16, draw=violet!65!black,
                 minimum width=30mm, minimum height=8mm, align=center, line width=0.5pt,
                 font=\scriptsize\bfseries},
  block/.style ={draw, rounded corners=2pt, fill=red!14,    draw=red!65!black,
                 minimum width=28mm, minimum height=8mm, align=center, line width=0.5pt,
                 font=\scriptsize\bfseries},
  serial/.style={draw, rounded corners=2pt, fill=gray!22,   draw=black!55,
                 minimum width=24mm, minimum height=8mm, align=center, line width=0.5pt,
                 font=\scriptsize\bfseries},
  apply/.style ={draw, rounded corners=2pt, fill=white, draw=black, line width=0.9pt,
                 minimum width=28mm, minimum height=8mm, align=center,
                 font=\scriptsize\bfseries},
  flow/.style  ={->, line width=0.5pt, draw=black!75},
  lbl/.style   ={font=\tiny\itshape, fill=white, inner sep=1pt}
]
\node[input]  (IN)  at (-6.0, 3.0)  {WriteIntent pair $I, I'$};
\node[check]  (L0)  at (-6.0, 1.0)  {L0\\same file?};
\node[input]  (FD)  at (-2.3, 1.0)  {file-disjoint\\candidate};
\node[input]  (L1)  at (-6.0,-1.4)  {L1 --- Known atoms\\\tiny adapter + atom map};
\node[check]  (C1)  at (-6.0,-3.6)  {same atom\\or CID overlap?};
\node[input]  (L2)  at (-6.0,-6.0)  {L2 --- Governance surfaces\\\tiny owner / tests / deps / registry};
\node[check]  (C2)  at (-1.5,-6.0)  {shared surface\\or declared\\R/W dep?};
\node[check]  (C3)  at ( 3.0,-6.0)  {bounded regions\\disjoint?};
\node[input]  (VA)  at ( 7.5,-6.0)  {virtual-atom\\fallback};
\node[check]  (C4)  at ( 7.5,-3.6)  {refined CID\\now disjoint?};
  \node[block]  (REFINE) at (7.5,-1.4) {split suggestion};
  \node[safe]   (SAFE)    at ( 3.2, 1.0)  {parallel-safe};
  \node[block]  (BLOCK)   at (-6.0,-8.4)  {blocked-cid-conflict\\\tiny blocked verdict};
  \node[serial] (SERIAL)  at (-1.5,-8.4)  {SERIAL\\\tiny queue / ordered replay};
  \node[input]  (HOLD)    at (-6.0,-10.4) {preserved intent\\\tiny review / evidence};
  \node[input]  (QUEUE)   at (-1.5,-10.4) {queued replay\\\tiny later admission cycle};
  \node[compose](COMPOSE) at ( 5.5,-8.4)  {needs-physical-split};
  \node[apply]  (STEW)    at ( 1.8,-10.4) {Neutral Steward apply};
\node[apply,fill=teal!12,draw=teal!60!black]
              (EVID)    at ( 1.8,-12.0) {evidence record\\\tiny verdict + validator};
\draw[flow] (IN)  -- (L0);
\draw[flow] (L0)  -- node[lbl,above]{no} (FD);
\draw[flow] (L0)  -- (L1);
\draw[flow] (FD)  |- node[lbl,near start,right]{check declared surfaces} (C2);
\draw[flow] (L1)  -- (C1);
  \draw[flow] (C1)  -- node[lbl,left]{yes} (BLOCK);
\draw[flow] (C1)  -- (L2);
\draw[flow] (L2)  -- (C2);
\draw[flow] (C2)  -- node[lbl,right]{yes} (SERIAL);
\draw[flow] (C2.north) to[bend right=18] node[lbl,right]{no; file-disjoint} (SAFE.south);
\draw[flow] (C2)  -- node[lbl,above]{no; same-file} (C3);
\draw[flow] (C3)  -- node[lbl,below]{yes} (COMPOSE);
  \draw[flow] (C3)  -- (VA);
  \draw[flow] (VA)  -- (C4);
  \draw[flow] (C4.north) to[bend left=20] node[lbl,right]{disjoint} (COMPOSE.east);
  \draw[flow] (C4)  -- node[lbl,right]{overlap} (REFINE);
  \draw[flow] (BLOCK)  -- (HOLD);
  \draw[flow] (REFINE) |- (HOLD.east);
  \draw[flow] (SERIAL) -- (QUEUE);
  \draw[flow] (SAFE)    |- (STEW.north);
  \draw[flow] (QUEUE.east) -- node[lbl,above]{after re-admission} (STEW.west);
  \draw[flow] (COMPOSE) |- (STEW.east);
  \draw[flow] (STEW)    -- (EVID);
\end{tikzpicture}
\end{figure}

This process can be condensed into a six-stage governance chain:

\Needspace{8\baselineskip}
\begin{paperAlgoBox}
\begin{Verbatim}[breaklines=true,fontsize=\scriptsize]
agent proposal
   -> adapter-guided atomization
       -> atom map lookup
           -> virtual-atom refinement
               -> broker verdict
                    -> neutral steward apply
\end{Verbatim}
\end{paperAlgoBox}

Only the admitted paths enter the steward apply stage. \texttt{\seqsplit{parallel-safe}} and \texttt{\seqsplit{needs-physical-split}} may proceed to steward apply, while \texttt{\seqsplit{SERIAL}} first enters the policy-ordered queue or serialization path. \texttt{\seqsplit{blocked-cid-conflict}}, \texttt{\seqsplit{blocked-shared-surface}}, and split-suggestion outcomes remain outside the direct apply path and instead enter recovery, review, queue, or refinement routes according to policy.

What ATM governs is therefore not abstract {\char34}multi-agent collaboration{\char34}, but the narrower question of how a write intent inside a single governance domain is atomized, compared, adjudicated, and applied through a governed write path.

\Needspace{19\baselineskip}
\begin{paperAlgoBox}
\noindent{\small\bfseries Algorithm 1 --- Progressive Admission with Atom Map and Virtual Atoms.}\par\smallskip

\begin{Verbatim}[breaklines=true,fontsize=\scriptsize]
Input: write intents I, I' over the same governance domain
Notation: P(·) = physical write surface, Surface(·) = declared shared
          surface set (registry / generator / artifact / active intent
          surface), R(·) = declared read atom set, W(·) = declared write
          atom set, SameFile(·, ·) = same physical file or same structured
          artifact under the governed substrate.
          intersects(A, B) is true iff A and B share at least one
          declared member.

1: map I, I' to known atoms via adapter + atom map; resolve candidate CIDs
2: if same atom or same candidate CID, return blocked-cid-conflict
3: if intersects(Surface(I), Surface(I')),
       return blocked-shared-surface or SERIAL (per shared-surface policy)
4: if intersects(R(I), W(I')) or intersects(W(I), R(I')), return SERIAL
5: if not intersects(P(I), P(I'))
       and not intersects(Surface(I), Surface(I'))
       and not intersects(R(I), W(I')) and not intersects(W(I), R(I'))
       and not SameFile(I, I'),
       return parallel-safe
6: if not intersects(P(I), P(I'))
       and not intersects(Surface(I), Surface(I'))
       and not intersects(R(I), W(I')) and not intersects(W(I), R(I'))
       and SameFile(I, I'),
       return needs-physical-split
7: if same file and same atom map but different atom ids, proceed
       to the bounded-region and virtual-atom checks in the following steps
8: if known bounded regions are disjoint, route to needs-physical-split
9: otherwise create virtual atoms for uncovered or coarse spans
10: if virtual atom CID or ConflictKey overlaps under the adapter-defined
        overlap predicate, test decomposition policy
11: if decomposition is admissible, recompute virtual atoms and bounded
         regions
12: if refined regions become disjoint, route to needs-physical-split
13: else emit split suggestion, record refinement evidence, preserve the
         intent or patch envelope when available, and fail-closed on direct
         apply; the intent may then enter queue, SERIAL, steward review,
         or a rebase/replay path if the preserved evidence supports recovery
Output: verdict in {parallel-safe, needs-physical-split,
        blocked-cid-conflict, blocked-shared-surface, SERIAL}
\end{Verbatim}
\end{paperAlgoBox}

Algorithm 1 safety note. Physical disjointness remains necessary but not sufficient. It may confirm \texttt{\seqsplit{parallel-safe}} only when the intents are also outside the same physical file or structured artifact, and after shared-surface and read/write-dependency checks have cleared. Same-file bounded-disjoint edits are not treated as direct \texttt{\seqsplit{parallel-safe}} writes; they are routed to \texttt{\seqsplit{needs-physical-split}}, where the deterministic composer and the neutral steward produce a single governed apply path. A blocked branch is therefore a containment verdict over the direct-apply channel, not a default deletion of the proposing agent's work. \texttt{\seqsplit{blocked-cid-conflict}} and \texttt{\seqsplit{blocked-shared-surface}} are admission verdicts; \texttt{\seqsplit{fail-closed/refine}} is a recovery-route label used in benchmark and evidence reports when a blocked verdict closes the unsafe direct-apply path while preserving the intent, patch envelope, or refinement evidence. Thus Algorithm 1 reports the admission verdict, while §5 may report the corresponding recovery route. This aligns Algorithm 1 with Figure 2, Table 4, the seven-layer gate, and the POS2 field case.

Virtual-atom refinement is not a post-processing step outside admission. It is the refinement mechanism that the broker activates when the known atoms and atom map cannot establish an admissible route under the declared model. It proceeds in two steps. \textbf{Syntactic-enclosure atomization} wraps patch spans that the existing atom map does not cover, or covers too coarsely, into the smallest function, method, statement block, or other adapter-recognizable enclosure, thereby forming virtual atoms. \textbf{Signature-preserving decomposition} then decomposes an overly coarse virtual atom or coarse atom into smaller comparable bounded regions without changing the union of patch coverage, so that the broker can recompute the candidate CID, ConflictKey, and shared-surface adjacency.

This decomposition does not invent new semantic units. The union of refined sub-fragments must remain coverage-equivalent to the original range; what changes is only the granularity at which the broker compares candidate CIDs, bounded regions, ConflictKeys, and shared surfaces. If neither refinement step establishes disjointness or a composable route, the broker does not continue guessing. It falls back to a split suggestion, SERIAL, or a fail-closed path. If overlap persists after refinement, the outcome returns to \texttt{\seqsplit{blocked-cid-conflict}} or to the refinement loop rather than treating decomposition itself as a safety guarantee. The POS2 and BLOCK field cases mark the positive and negative boundaries of this path: POS2 shows a case in which refinement enables admission under the declared model, while BLOCK shows that the broker fails closed when refinement remains insufficient.

The core of this pipeline is not to permit parallelism whenever possible, but to convert the parallelism decision from free-form LLM judgment into a replayable admission vocabulary. When two intents write to the same file, the broker does not immediately classify the same-file condition as a conflict, nor does it trust line-number separation alone. It first consults the atom map. If both intents fall within the same atom map but map to different atom ids, that condition is not itself a conflict; it is the entry point for region-level comparison, where bounded region, shared surface, read/write dependency, and ConflictKey are inspected in turn. Spans not yet covered by the atom map are wrapped in virtual atoms, while spans already covered but too coarse for a reliable verdict enter the virtual-atom decomposition and split-suggestion path. Only after these checks does the broker decide whether the bounded regions, CIDs, ConflictKeys, and dependencies overlap. If the layered refinement establishes an admissible disjoint or composable route, the broker routes the intents to the composer and the steward applies the composed result. Otherwise, the broker fails closed or enters the refinement loop. The two propositions below describe the conservative boundary of this pipeline.

\textbf{Proposition 1 (Cross-Regime Disjointness).} If the source roots governed by two adapters are guaranteed disjoint by repository convention, and each adapter correctly declares its source paths, then the two candidates' physical write surfaces do not intersect. The broker may therefore allow them to pass the physical-overlap gate unless a shared surface, declared dependency, or other governance rule independently blocks admission. This proposition guarantees only physical-write-surface disjointness. It does not by itself establish a final \texttt{\seqsplit{parallel-safe}} verdict, nor does it guarantee semantic safety under cross-language logical coupling, API contracts, or generated client/server pairs. Those concerns remain part of the cross-language identity open problem in §3.7.

\textbf{Proposition 2 (Static Admission Closure).} Under the assumption that adapter declarations of static read and write sets are conservative approximations, and that dynamic effects are covered downstream by validators or a fallback lock, a \texttt{\seqsplit{parallel-safe}} verdict excludes statically determinable write-write conflicts and already-declared read/write hazards. The proposition does not assert dynamic semantic correctness.

Proof sketch. The broker first uses atom, Candidate CID, and shared-surface checks to rule out explicit write-write overlap over the declared governance units. It then applies the augmented dependency rule to rule out statically observable read-after-write and write-after-read hazards among declared \texttt{\seqsplit{readAtoms}} and \texttt{\seqsplit{writeAtoms}}. Any remaining risk must arise from dependencies or dynamic effects that were not declared by the task contract, exposed by the adapter, represented in the atom map, or captured by virtual atoms. Those residual risks are outside the positive admission guarantee and must be contained by validator handoff, CAS base-hash checks, steward review, recovery routing, or the fallback lock. Proposition 2 therefore asserts static admission closure under the declared model, not end-to-end semantic soundness.

The augmented dependency rule fills the gap left by pure write-set disjointness. Because Layers 1 and 2 have already handled explicit write--write overlap, this rule is responsible specifically for declared read/write hazards. Let $R(I)$ denote the declared read atom set of intent $I$, and let $W(I)$ denote its declared write atom set. These symbols are intent-local dependency sets and are distinct from the deliverable set $D(\mathcal{T})$ in the Task Contract. If intent $I$ reads an atom that another intent $I'$ writes, or if $I$ writes an atom that $I'$ has declared as a read, the two intents must take the \texttt{\seqsplit{SERIAL}} or review path even when their textual ranges do not overlap:

\[
(R(I) \cap W(I') \neq \emptyset)
\lor
(W(I) \cap R(I') \neq \emptyset)
\Rightarrow
\mathrm{SERIAL}(I, I').
\]

Here $R(\cdot)$ is the static read set declared by the intent through \texttt{\seqsplit{readAtoms}} or through the adapter and atom map; ATM does not claim full dynamic read tracing. In implementation, the Active Registry stores the declared read atom IDs and CIDs of active intents, so a later writer can be caught by the same rule. A newly arriving intent that declares its own \texttt{\seqsplit{readAtoms}} is likewise compared against existing active writes. Hidden effects that are not declared remain outside the admission model and must be covered downstream by validators, CAS base-hash checks, or fail-closed fallback.

The augmented admission rule should therefore be read as a repository-scoped admission rule over the declared, adapter-observed, or conservatively virtualized read and write surface. A positive admission verdict does not imply global semantic independence between the admitted intents. It states only that, under the task contract, the available atom map, adapter-derived ConflictKeys, active-registry state, and any virtual-atom fallback, the broker found sufficient evidence to admit the candidate before governed shared mutation is applied. Dependencies that are neither declared by the contract nor exposed by an adapter fall outside this positive scope and are handled through conservative fail-closed routing, steward review, validators, CAS revalidation, or the future read-set reconstruction providers discussed in §6.3.

Virtual-atom refinement addresses two cases in which formal atomization is insufficient: the coverage of formal atoms is incomplete, or an existing atom is too coarse to support a direct verdict. These cases must be handled separately rather than collapsed into one mechanism.

\textbf{The map-gap case.} The broker may temporarily construct a governable virtual atom for a patch span that is not yet covered by a formal atom, and then re-examine the relevant conflict boundary. The first step is syntactic enclosure: uncovered patch lines are wrapped into virtual atoms and the candidate CID is recomputed, so that two intents that previously appeared only as same-file edits at the file level are restated as two comparable sets of governance regions. If the virtual atoms' CIDs, shared surfaces, and bounded regions are all disjoint, the broker may route the case from coarse same-file contention to the composable \texttt{\seqsplit{needs-physical-split}} path.

\textbf{The coarse-known-atom case.} The patch span already lies inside the atom map, but the atom granularity itself is too coarse to establish disjointness between the two intents. This case should not be described as one in which {\char34}creating a virtual atom resolves the conflict.{\char34} It requires signature-preserving decomposition, a split suggestion, and a human-reviewable refinement path. Until an admissible disjoint or composable route is established under the declared model, the intents remain in a conflict state.

When conflict density is high---specifically, when the conflict-hunk count exceeds the threshold $\theta_{count}$ or the conflict-line density exceeds the threshold $\theta_{density}$---decomposition proposes a signature-preserving split

\[
f \mapsto f_{\mathrm{pre}} \mathbin{\Vert} f_{\mathrm{extracted}} \mathbin{\Vert} f_{\mathrm{post}}
\]

and recomputes the virtual-atom CID for each decomposed fragment. The current implementation treats both thresholds as explicit gates. The defaults in the planning and implementation documents are $\theta_{count} = 1$ and $\theta_{density} = 0.5$, and decomposition is not recursively expanded, so refinement remains bounded. Virtual atoms are therefore not an ancillary optimization; they are the core mechanism by which ATM extends admission coverage when the formal atom map is limited. By contrast, the coarse-known-atom case is closer to a controlled split suggestion than to automatic conflict resolution. The refinement output is not a free-form LLM rewrite but a reviewable refinement suggestion, so blocked overlap becomes a signal for improving the atom map. When both refinement steps fail to establish an admissible disjoint or composable route, the broker falls back to \texttt{\seqsplit{blocked-cid-conflict}} and the §4.5 refinement loop takes over.

Before entering the finer-grained admission pipeline, the paper also distinguishes which modifications remain ordinary local edits and which have been escalated to shared writes that must be adjudicated by the broker. To avoid conflating an ordinary edit, a declared write intent, and a fully governed transaction, the paper uses three layered terms. An \texttt{\seqsplit{edit}} is an ungoverned modification an agent makes in its local workspace; it covers private drafts, local exploratory work, and undeclared work-in-progress. A \textbf{write intent} is a candidate write that has already been described in structured form, declaring at least the target files, the atoms or surfaces it may touch, and the necessary admission metadata. A \textbf{governed transaction} is a shared-write unit that has entered the broker-governed path and can be adjudicated by the broker and applied by the steward.

Not every edit is a transaction. Only edits that cross into a shared surface or shared artifact are escalated into declared write intents, and only admitted write intents become transactions handled by the broker. This distinction keeps private local work outside the governed path while ensuring that shared mutations become visible to the broker before they affect the shared worktree.

\begin{figure}[H]

{\small \noindent\textbf{Figure 3 --- Write Intent Escalation and Broker Activation Policy.} The figure separates the four stages of a write---local edit, declared write intent, broker-governed transaction, and steward apply---and clarifies when a modification may remain a local edit and when it must be escalated into a governed shared write.\par}
\smallskip
\centering
\begin{tikzpicture}[
  >=Latex,
  font=\scriptsize,
  zone/.style    ={draw, rounded corners=4pt, line width=0.6pt, align=center,
                   minimum height=42mm, inner sep=4pt},
  localz/.style  ={zone, fill=blue!10,   draw=blue!60!black,   minimum width=34mm},
  intentz/.style ={zone, fill=orange!13, draw=orange!70!black, minimum width=34mm},
  governz/.style ={zone, fill=green!12,  draw=green!50!black,  minimum width=54mm},
  applyz/.style  ={zone, fill=violet!11, draw=violet!70!black, minimum width=32mm},
  stepbox/.style ={draw, rounded corners=2pt, fill=white, align=center,
                   minimum width=15mm, minimum height=9mm, line width=0.4pt, font=\tiny},
  decision/.style={draw, diamond, aspect=1.6, fill=yellow!22, draw=orange!70!black,
                   inner sep=0pt, minimum width=17mm, minimum height=11mm,
                   align=center, line width=0.5pt, font=\tiny},
  outbox/.style  ={draw, rounded corners=2pt, fill=gray!12, align=center,
                   minimum width=30mm, minimum height=8mm, line width=0.4pt, font=\tiny},
  flow/.style    ={->, line width=0.5pt, draw=black!75},
  lbl/.style     ={font=\tiny\itshape, fill=white, inner sep=1.2pt},
  zonelbl/.style ={font=\tiny\bfseries, anchor=south west, inner sep=1pt}
]
\node[localz]  (LZ) at (-6.8, -0.6) {};
\node[intentz] (IZ) at (-2.8, -0.6) {};
\node[governz] (GZ) at ( 2.1, -0.6) {};
\node[applyz]  (AZ) at ( 6.6, -0.6) {};
\node[zonelbl, blue!60!black]   at (LZ.north west) [yshift=1pt] {Local Edit Zone};
\node[zonelbl, orange!60!black] at (IZ.north west) [yshift=1pt] {Declared Intent Zone};
\node[zonelbl, green!40!black]  at (GZ.north west) [yshift=1pt] {Broker Governance Zone};
\node[zonelbl, violet!60!black] at (AZ.north west) [yshift=1pt] {Apply / Closure Zone};
\node[stepbox,  fill=blue!22]    (E) at (-7.6,  0.6) {Agent edit\\\tiny local WIP};
\node[decision]                  (D) at (-5.9,  0.6) {touches shared\\surface\,/\,scope?};
\node[stepbox,  fill=orange!22]  (I) at (-3.6,  0.6) {WriteIntent\\\tiny targets / atoms};
\node[decision]                  (G) at (-1.9,  0.6) {requires governed\\shared write?};
\node[stepbox,  fill=green!24]   (T) at ( 0.5,  0.6) {Governed transaction\\\tiny lease + rw set};
\node[stepbox,  fill=green!28]   (B) at ( 2.4,  0.6) {Broker admission\\\tiny atom map / CID};
\node[decision]                  (V) at ( 4.2,  0.6) {admission\\verdict};
\node[stepbox,  fill=violet!18]  (S) at ( 6.6,  0.6) {Neutral Steward\\apply};
\node[stepbox,  fill=violet!22]  (X) at ( 6.6, -1.7) {Refine, serialize,\\or fail-closed};
\node[outbox,   fill=blue!22]    (L) at (-6.8, -2.1) {Local edit path\\\tiny direct write};
\node[outbox,   fill=orange!22]  (R) at (-2.8, -2.1) {Review-only path};
\draw[flow] (E)      -- (D);
\draw[flow] (D.east) -- node[lbl, above]{yes} (I.west);
\draw[flow] (I)      -- (G);
\draw[flow] (G.east) -- node[lbl, above]{yes} (T.west);
\draw[flow] (T)      -- (B);
\draw[flow] (B)      -- (V);
\draw[flow] (V.east) -- node[lbl, above]{parallel-safe\,/\,compose} (S.west);
\draw[flow] (V.south) |- node[lbl, near end, above]{block\,/\,SERIAL} (X.west);
\draw[flow] (D.south) -- node[lbl, right]{no} (L);
\draw[flow] (G.south) -- node[lbl, right]{no} (R);
\end{tikzpicture}
\end{figure}

What Figure 3 expresses is not that all agents lose the ability to write locally. Instead, ATM moves the starting point of shared-write governance earlier. While a modification remains in the private edit phase, deployments may retain a lightweight local workflow. Only when the modification is declared to touch a shared surface, a shared artifact, or a governed scope does the system require it to enter the broker as a write intent and, once admitted, to be promoted into a governed transaction.

Escalation examples for Figure 3. To prevent the three escalation triggers---shared surface, shared artifact, and governed scope---from remaining abstract, the paper gives three typical escalation scenarios and one non-escalation contrast:

\begin{itemize}
\item \textbf{Scenario 1 (shared surface).} An agent wants to modify \texttt{\seqsplit{classifyExplicitMutationRequest}} inside \texttt{\seqsplit{packages/cli/src/commands/broker.ts}}, a file that the atom map already declares as a shared surface on the broker serialization path. Even if the modification touches only a few lines inside a single function, it must be escalated into a write intent, because \texttt{\seqsplit{broker.ts}} is part of the admission basis for other active intents. This is the trigger point of the POS2 case.
\item \textbf{Scenario 2 (shared artifact).} An agent wants to update an atom-map shard inside \texttt{\seqsplit{docs/ai\_atomic\_framework/atom-maps/*.json}}. This artifact is a first-line input to broker decisions. Any modification to it must therefore be escalated into a write intent and enter admission through the ConflictKey of the \texttt{\seqsplit{atom-map}} adapter.
\item \textbf{Scenario 3 (governed scope).} An agent wants to add a fixture under \texttt{\seqsplit{tools/multi-vendor-broker-bench/}}. Even if the file is not touched by any current active intent, it falls within a declared governed scope and must be escalated so that the change is captured by the evidence chain.
\item \textbf{Non-escalation contrast.} An agent writes exploratory notes in a private \texttt{\seqsplit{scratch/}} directory or in a local WIP draft, touching no shared surface, shared artifact, or governed scope. The modification may remain inside the Local Edit Zone under direct-write or direct-commit workflows, without entering the broker.
\end{itemize}

In other words, escalation is decided not by whether a file is written, but by whether the write is observable to other active intents or to the governance ledger. The former is partly a deployment-policy decision; the latter is statically decidable from the atom map and the governed-scope declaration.

The governed transaction is not an extra wrapper noun. It is the in-flight state that the broker needs in order to continuously govern shared writes. With only a write intent, the broker would know what a writer once declared as the target, but it would not retain a persistent handle after that writer enters a hot file or a bounded region. Without elevating the first writer, $I_A$, into a governed transaction with a transaction identity, lease epoch, allowed files, read and write sets, file hashes, and admission state, the broker cannot continue to manage what $I_A$ is currently doing once a later writer, $I_B$, enters the same shared surface. The broker would be forced back into one-shot static conflict checks.

The transaction layer therefore exists to answer how the earlier writer is currently being governed. It gives the broker a durable handle that can be parked, re-arbitrated, serialized, routed to the composer, or sent through a bounded re-plan against the existing writer. Without that layer, the system would react only after both sides had already written, returning the problem to Git merge or manual repair.

\subsection{Seven Hard Gates and the Broker's Sole-Serialization Role}

The ATM broker does not adjudicate by CID alone; it uses seven hard gates to progressively narrow the conflict surface of a suspect write. CID identity is the first fast semantic index. When the CID does not conflict, the broker still checks shared surfaces, declared read and write sets, file range and virtual-atom refinement, ConflictKey and \texttt{\seqsplit{canMerge}}, the CAS base-hash, and finally the fallback file lock. This design lets ATM admit parallelism when disjointness can be shown under the declared model, and conservatively take a fail-closed path when the evidence is insufficient.

\Needspace{24\baselineskip}
The maturity column uses three labels:

\begin{itemize}
\item \textbf{Validated.} The layer is supported by archived deterministic-runner evidence or archived field-collision evidence.
\item \textbf{Partial.} The core mechanism has been implemented and landed, but boundary cases have not yet been covered by a full regression sweep.
\item \textbf{Prototype.} The framework reserves a slot for the mechanism and has a minimal landing in place, but the autonomous path---such as virtual-atom refinement or bounded re-planning---is still under construction and is not claimed as part of the admission-core deliverable.
\end{itemize}

\TableLead{Table 5 --- Seven-Layer Admission Gate (with maturity markers).}

\begingroup
\TableTightFont
\setlength{\tabcolsep}{2pt}
\arrayrulecolor{black!55}
\rowcolors{2}{gray!8}{white}
\begin{longtable}{@{}L{0.12\textwidth}L{0.11\textwidth}L{0.16\textwidth}L{0.16\textwidth}L{0.16\textwidth}L{0.19\textwidth}@{}}
\toprule
Layer & Gate & Question & On pass & On fail or unknown & Maturity status \\
\midrule
1 & CID Identity & Are the two intents identified as the same governance unit by the current adapter or atomization regime---that is, the same atom or the same candidate CID? For a candidate, identity already includes the symbol, source path, and declared start/end line range. & Proceed to the next layer & \texttt{\seqsplit{blocked-cid-conflict}} & Validated \\
2 & Shared Surface & Do the intents touch the same registry, generator, artifact, or active-intent surface? & Proceed to the next layer & Block or route to \texttt{\seqsplit{SERIAL}} & Validated \\
3 & Read/Write Set & Does $R(I) \cap W(I') \neq \emptyset$ or $W(I) \cap R(I') \neq \emptyset$ hold? & Proceed to the next layer & Route to \texttt{\seqsplit{SERIAL}} or review & Validated for the core path; Partial for admission-time active-intent forwarding (see the §3.7 open problem) \\
4 & File Range or Virtual Atom & Can the same-file change be separated by a known atom or by a virtual atom? & Route to the composer path & Virtual-atom refinement or block & Partial overall; Validated for the known-atom path; Prototype for autonomous virtual-atom refinement \\
5 & ConflictKey + \texttt{\seqsplit{canMerge}} & Does the structured artifact have a disjoint key and a deterministic-merge capability? & Format-level admission & Block or route to serialization & Partial overall; Validated for the JSON, numeric, and atom-map shard paths; general code-merge still relies on the deterministic composer \\
6 & CAS Base-Hash & Does the base hash at apply time still match the state observed during admission? & One-shot apply & Bounded re-plan & Validated for one-shot apply and fail-closed paths; Prototype for autonomous bounded re-planning; current validated scope is the split-suggestion/decomposition fallback \\
7 & Fallback File Lock & When neither the adapter nor a validator can supply sufficient evidence, is a conservative whole-file lock required? & Guarded write & Fail closed & Validated for the conservative-lock path \\
\bottomrule
\end{longtable}
\endgroup

Table 5 should be read as a gate-by-gate evidence-status note, \emph{not as a claim that every sub-path inside a row has identical maturity}.

The point of these seven layers is not to add mechanisms for their own sake, but to make explicit that ATM's target admission model does not rest on CID adjudication alone. Appendix A.4 provides a compact verification map for the principal capability groups, while detailed source paths, per-command outputs, artifact hashes, and row-level verification records are maintained in the repository and supplementary verification material. This section retains this structured model so that CID is not mistakenly read as the sole admission criterion.

By the same reasoning, ATM does not treat {\char34}same file but non-overlapping line numbers{\char34} as a sufficient condition for admission. Line numbers or bounded text ranges can at most provide necessary physical evidence; they are not sufficient on their own. Real multi-agent conflicts may also arise from shared-surface contention, declared read/write dependencies, governance-coverage gaps under a coarse owner map, and apply-time drift. Admitting purely on same-file line disjointness would fall back to a text-level merge policy without the governance evidence required by ATM. The point of the layered gate is precisely to separate which writes can be considered concurrent under the semantic and governance model from which conditions must route to \texttt{\seqsplit{SERIAL}}, a fail-closed path, or a controlled refinement fallback.

Among the seven layers, the current evidence base distinguishes between paths already supported by implementation plus validation, fixture, or validator evidence, and paths whose autonomous form remains under construction. Late-joiner park and re-arbitration, same-file CID-disjoint composer routing, and the \texttt{\seqsplit{SERIAL}} path required for shared-surface and read/write-dependency hazards belong to the first group. By contrast, bounded re-planning is more accurately described at present as a controlled split-suggestion and decomposition fallback, rather than as a validated autonomous multi-round re-planner. The paper states the two groups separately so that already-validated admission capabilities are not conflated with refinement workflows that are still evolving.

\textbf{Definition 7 (CAS base-hash guarded apply).} For any admitted plan $p$, the broker records its admission-time base hash $h_0$. Before applying $p$, the neutral steward re-reads the base hash $h_1$ of the target surface. If $h_1 = h_0$, the steward proceeds with one-shot apply. If $h_1 \neq h_0$, the plan may not be applied directly and must fall back to a controlled successor path. In cases that already carry sufficient evidence, the broker may route the plan to \texttt{\seqsplit{SERIAL}} or to a fail-closed outcome. In cases where finer-grained information is insufficient but refinement room remains, only a bounded split-suggestion or decomposition fallback is permitted. This definition aligns runtime closure with the compare-and-swap spirit of optimistic concurrency control \textcolor{atmLinkBlue}{\hyperlink{Item.46}{[10, 30]}}, while preserving ATM's atom and ConflictKey admission semantics.

A CAS mismatch is therefore not treated as an automatic loss of the proposing agent's work. It prevents direct apply against a stale base. If the preserved intent, patch envelope, or proposal evidence is sufficient, the work can still be queued, serialized behind the active holder, replayed after rebase, or routed to steward review. Only when the preserved evidence is insufficient for replay, composition, or review does the path become terminally fail-closed. This distinction matters for LLM-agent workloads, where a generated patch may represent substantial inference cost: ATM blocks unsafe application, but it does not require full regeneration when the existing intent can be safely replayed, rebased, composed, or reviewed.

A clear distinction must also be drawn between admission-time region judgment and apply-time line displacement. Bounded-region disjointness establishes that the declared semantic and edit-intent footprints do not overlap under the admission model. It does not claim that two patches will never cause line shifts when they are actually applied. If an earlier patch inserts new lines and a later patch modifies an existing block, the two may still satisfy $P_{a_1} \cap P_{a_2} = \emptyset$ at admission time. The line offsets that arise at apply time must then be absorbed jointly by the deterministic composer, the neutral steward, and the CAS base-hash recheck, rather than by allowing each agent to perform direct writes. POS2 therefore supports not a bare line-disjoint merge, but a controlled same-file write chain composed of admission, composition, steward apply, and CAS revalidation.

The broker is ATM's sole serialization node within a single governance domain. Agents may submit only intents or proposals; the broker reaches a single ordered decision based on the current active intents, the atom map, the shared surface, and the evidence substrate. This claim does not extend to settings where multiple machines hold separate clones. In those settings, the Git, PR, and merge substrate remains the final coordination layer. If agents inside the same controlled worktree are allowed to write directly to shared files, and post-hoc merge or human repair is expected to absorb the consequences, the system reverts to the classical race-condition regime: each agent decides admissibility from its local view, while no participant holds the global admission state.

The neutral steward applies an already-admitted plan to the same controlled worktree. Its role is not to invent new design, but to execute a patch application that has already been constrained by the broker's admission decision and to leave behind evidence. This clarifies the boundary between attribution and authority: the intent of a change is attributable to its proposer, while the write event is performed by a neutral steward.

More concretely, the neutral steward's life cycle can be written as a controlled five-stage chain:

\Needspace{8\baselineskip}
\begin{paperAlgoBox}
\begin{Verbatim}[breaklines=true,fontsize=\scriptsize]
re-read base hash
   -> apply admitted plan
       -> emit evidence
           -> trigger validators
                -> route to SERIAL, a fail-closed path, or enter controlled refinement fallback on drift
\end{Verbatim}
\end{paperAlgoBox}

The steward first uses the base-hash recheck to confirm that the admission-time precondition still holds. It then performs a single apply, records the evidence, and triggers typecheck, CLI validation, and domain validators. If drift, a validator failure, or a change to the shared-surface state appears during apply, the steward must not invent new content. It must instead fall back to \texttt{\seqsplit{SERIAL}}, a fail-closed path, or a controlled split-suggestion and decomposition fallback path. The steward is therefore not another free-form writer, but the runtime-enforcement node of the broker's verdict.

A distinction between governance mode and ordinary local-development mode is also necessary. Definition 7 constrains governed shared writes that have already entered the broker-governed path. It does not prohibit a single agent from directly writing or committing private local modifications. Whether a deployment requires all writes to pass through the steward path is a deployment-policy choice. The claim of this paper is narrower: once a write has been declared as a governed write on a shared surface, agents should not bypass the steward and apply changes themselves.

Batch attribution and Wave Mode extend this path. When multiple tasks are submitted as a single wave, the broker still evaluates each intent in turn and maintains per-task traceability through checkpoints and per-task evidence slicing. Wave Mode may coordinate Team Agents-style roles around a batch, but it does not change the core admission claim or bypass the shared-mutation gates. It extends the same broker-and-steward logic to batched execution.

\subsection{Cross-Format Generalization}

ATM governs not only code atoms but also structured artifacts. Through \texttt{\seqsplit{FileMutationAdapter}} and \texttt{\seqsplit{ConflictKey}}, the same admission concepts apply across JSON records, text ranges, numeric scalars, and atom-map shards. The conflict unit therefore varies across formats. For code, it may be a function or a method. For JSON, it may be a record key. For a numeric configuration, it may be a scalar field. For an atom map, it may be an edge or a member record.

\begin{table}[H]
\TableLead{Table 6 --- ConflictKey Generalization Matrix. Scope $\times$ locator mapping across formats.}
\raggedright
\TableTightFont
\setlength{\tabcolsep}{2pt}
\arrayrulecolor{black!55}
\rowcolors{2}{gray!8}{white}
\begin{tabular}{@{}L{0.17\textwidth}L{0.17\textwidth}L{0.18\textwidth}L{0.18\textwidth}L{0.18\textwidth}@{}}
\toprule
Domain & Adapter & Scope & Locator & Merge capability \\
\midrule
Code (TypeScript) & TS adapter & function or method & canonical symbol plus path & none $\rightarrow$ route to deterministic composer \\
Code (Python) & Python adapter & function or class method & qualified symbol plus path & none $\rightarrow$ route to deterministic composer \\
JSON & \texttt{\seqsplit{json-record}} adapter & record & key path (JSON pointer) & deterministic merge when keys are disjoint \\
Text & \texttt{\seqsplit{text-range}} adapter & range & file plus line range & none $\rightarrow$ route to composer \\
Numeric & \texttt{\seqsplit{numeric-scalar}} adapter & scalar & file plus field name & commutative increment/decrement, or guarded \texttt{\seqsplit{set-if-equal}} \\
Atom map & \texttt{\seqsplit{atom-map}} domain adapter & edge or member record & shard plus line range & line-disjoint merge + CAS base-hash \\
\bottomrule
\end{tabular}
\end{table}

The point of this generalization is that the broker does not need to understand the full semantics of each format. It must, however, be able to obtain a conservative conflict key, an adapter-defined overlap predicate, and a declared merge capability. If the adapter can declare that two mutations do not overlap under that predicate, and if the format adapter can provide either a deterministic merge or a CAS base-hash check, the broker may admit them as format-level parallel writes. Otherwise, admission must fall back to blocking, serialization, or a steward-required path.

For readability, the locator column in Table 6 should be read as plain scope-and-path prose, \emph{rather than as API syntax}. In particular, items such as canonical symbol plus path, file plus line range, shard plus line range, and guarded set-if-equal denote locator or merge categories, not required literal spellings.

\textbf{Proposition 3 (ConflictKey Disjointness).} For any format adapter, if two mutations share a governance scope category but their locators do not overlap under the adapter-defined predicate $\operatorname{overlap}_{S}$, and if the adapter declares its merge capability as either deterministic or CAS-guarded, then the broker may treat the mutations as format-level disjoint writes. If the scopes are the same and the locators overlap under $\operatorname{overlap}_{S}$, or if the adapter cannot declare a merge capability, the broker must block, serialize, or require the steward path. Proposition 3 is the cross-format generalization of Proposition 1: Proposition 1 addresses disjointness at the repository-root or adapter-regime level, while Proposition 3 addresses disjointness inside any structured artifact.

\subsection{Scope and Open Problems}

The preceding admission rules should be read as positive only within the declared, adapter-observed, or conservatively virtualized read/write surface. A positive admission verdict does not imply global semantic independence between the admitted intents. It states only that, under the Task Contract, the available atom map, adapter-derived ConflictKeys, active-registry state, CAS preconditions, and any virtual-atom fallback, the broker found sufficient evidence to admit the candidate before governed shared mutation is applied. Dependencies that are neither declared by the contract nor exposed by an adapter fall outside this positive scope and remain subject to fail-closed routing, steward review, validators, CAS revalidation, or the future read-set reconstruction providers discussed in §6.3.

The current version of ATM does not guarantee the following five properties:

\begin{enumerate}
\item \textbf{No cross-machine distributed-coordination guarantee.} When several agents write from different machines, different clones, or different PR branches, this version of ATM does not provide distributed locking, remote consensus, or cross-PR merge resolution. Those responsibilities remain with Git, the broader VCS layer, and the review workflow.
\item \textbf{Cross-language atom identity remains unresolved.} The semantic coupling between a TypeScript client and a Python backend handler cannot be settled by each adapter's CID alone.
\item \textbf{Admission-time active-intent forwarding is incomplete.} Active-intent forwarding has not been fully internalized across every owner-map path. Some B-12-class incidents still depend on apply-phase fail-closed handling to catch residual cases.
\item \textbf{Liveness and starvation require formal proof.} The broker's ability to refuse conservatively does not guarantee that every intent will eventually be admitted.
\item \textbf{CID schema migration and the adapter trust boundary require stronger mechanisms.} More complete version-migration mechanisms and manifest verification remain to be developed.
\end{enumerate}

The broker described in this paper should therefore be read as a single-domain arbiter. It requires visibility into the same filesystem, worktree, or registry in order to reach consistent decisions over active intents.

\section{Validation, Evidence, and Benchmark Alignment}

The evaluation in this paper proceeds in the following order: deterministic fixtures $\rightarrow$ internal field evidence $\rightarrow$ external adoption $\rightarrow$ site-specific field results $\rightarrow$ benchmark convergence $\rightarrow$ orchestration extension. This ordering is deliberate. It prevents evidence materials with different evidentiary strengths from being collapsed into a single empirical claim. Fixtures validate the decision surface; the adoption study observes recoverability; field evidence shows representative end-to-end paths; and AdmissionBench consolidates the admission-related evidence into a replayable, auditable, and item-level checkable benchmark chain. Throughout §4, self-hosting forensics is reported as operational dogfooding evidence under a single governance domain, not as independent external validation; adopter-side evidence is reported separately through the npc-brain cohort in §4.3.

Mapped onto §1.3 Contributions, §4.1 and §4.5 primarily support the admission claim at the method layer, while §4.2 and §4.3 primarily support the governance-substrate and adoption claims at the systems layer. §4.4 adds public-source and structured-artifact evidence lines, and §4.6 contributes batch and stability evidence. §5 then organizes the validation evidence that is directly benchmark-relevant into the v0.1 baseline and the v0.2 paper-facing result.

AdmissionBench is therefore not a separate material introduced alongside validation; it is the benchmark-facing consolidation of the validation evidence stack. Together, these materials support the central claims of the paper, but they do not yet constitute a complete head-to-head benchmark experiment against adjacent systems such as STORM \textcolor{atmLinkBlue}{\hyperlink{Item.39}{[3]}}, CodeCRDT \textcolor{atmLinkBlue}{\hyperlink{Item.37}{[1]}}, SCF \textcolor{atmLinkBlue}{\hyperlink{Item.38}{[2]}}, CoAgent \textcolor{atmLinkBlue}{\hyperlink{Item.47}{[11]}}, S-Bus \textcolor{atmLinkBlue}{\hyperlink{Item.52}{[16]}}, CodeTeam \textcolor{atmLinkBlue}{\hyperlink{Item.51}{[15]}}, ATCC \textcolor{atmLinkBlue}{\hyperlink{Item.41}{[5]}}, Cordon \textcolor{atmLinkBlue}{\hyperlink{Item.78}{[42]}}, or Atomix \textcolor{atmLinkBlue}{\hyperlink{Item.77}{[41]}}.

\begin{figure}[H]

{\small \noindent\textbf{Figure 4 --- Evidence Taxonomy and Claim Alignment.} The figure aligns the evidence buckets used in this paper with the claim boundaries each bucket can support. It distinguishes deterministic fixtures, field-collision evidence, adopter-side governance evidence, and self-hosting extension evidence, so that materials with different evidentiary strengths are not conflated into a single experimental conclusion:\par}
\smallskip
\centering
\begin{tikzpicture}[
  >=Latex,
  font=\scriptsize,
  box/.style    ={draw, rounded corners=3pt, minimum width=26mm, minimum height=14mm,
                  align=center, line width=0.5pt},
  motive/.style ={box, fill=blue!8,     draw=blue!60!black,   minimum width=42mm,
                  minimum height=11mm},
  scope/.style  ={box, fill=yellow!16,  draw=orange!70!black, minimum width=42mm,
                  minimum height=11mm},
  mech/.style   ={box, fill=red!10,     draw=red!55!black},
  field/.style  ={box, fill=red!18,     draw=red!65!black},
  bench/.style  ={box, fill=orange!16,  draw=orange!65!black},
  self/.style   ={box, fill=teal!12,    draw=teal!65!black},
  adopt/.style  ={box, fill=blue!12,    draw=blue!65!black},
  wave/.style   ={box, fill=violet!14,  draw=violet!65!black},
  core/.style   ={box, fill=white, draw=black, line width=0.9pt,
                  font=\scriptsize\bfseries, minimum width=42mm, minimum height=12mm},
  limit/.style  ={box, fill=gray!16, draw=black!55, minimum width=92mm, minimum height=10mm,
                  font=\scriptsize\itshape},
  flow/.style   ={->, line width=0.5pt, draw=black!75},
  trunk/.style  ={->, line width=0.8pt, draw=black!85},
  grp/.style    ={draw, rounded corners=4pt, line width=0.5pt, fill opacity=0.35,
                  text opacity=1, inner sep=3pt},
  grouplbl/.style={font=\tiny\bfseries, anchor=south west, inner sep=1pt}
]
\node[motive] (M) at (0,  0.6) {Motivation\\\tiny AgenticFlict \,/\, Git / PR conflict};
\node[scope]  (S) at (0, -1.2) {Scope\\\tiny single authority domain\,/\,pre-write admission};
\node[mech]  (B1) at (-7.4, -4.0) {Mechanism\\\tiny §4.1 \,/\, 12-scenario\\\tiny B-02 / B-08 / B-13};
\node[field] (B4) at (-4.4, -4.0) {Field collision\\\tiny §4.5 \,/\, same-file\\\tiny POS2 / B-12 / BLOCK};
\node[bench] (B5) at (-1.4, -4.0) {AdmissionBench\\\tiny §5 \,/\, v0.1 baseline\\\tiny + v0.2 paper};
\node[self]  (B2) at ( 1.4, -4.0) {Self-hosting\\\tiny §4.2 \,/\, forensics\\\tiny coverage 95\,/\,100};
\node[adopt] (B3) at ( 4.4, -4.0) {Adoption\\\tiny §4.3 \,/\, npc-brain\\\tiny recoverability};
\node[wave]  (B6) at ( 7.4, -4.0) {Wave / Batch\\\tiny §4.6 \,/\, extension\\\tiny orchestration};
\begin{scope}[on background layer]
\node[grp, fill=red!8,  draw=red!40, fit=(B1)(B4)(B5)] (GK) {};
\node[grp, fill=blue!8, draw=blue!40, fit=(B2)(B3)(B6)] (GG) {};
\end{scope}
\node[grouplbl, red!55!black]  at (GK.north west) [yshift=1pt] {supports admission claim};
\node[grouplbl, blue!55!black] at (GG.north west) [yshift=1pt] {supports governance claim};
\node[core] (K) at (-4.4, -6.6) {Core admission claim\\\tiny progressive atomization + CID broker};
\node[core] (G) at ( 4.4, -6.6) {Governance claim\\\tiny operable substrate};
\node[limit] (L) at (0, -8.4) {Limitations \,\textbullet\, no cross-clone locking \,\textbullet\, no full comparative benchmark yet};
\draw[trunk] (M) -- (S);
\draw[flow] (S.south) -- ++(0,-0.4) -| (B1.north);
\draw[flow] (S.south) -- ++(0,-0.4) -| (B4.north);
\draw[flow] (S.south) -- ++(0,-0.4) -| (B5.north);
\draw[flow] (S.south) -- ++(0,-0.4) -| (B2.north);
\draw[flow] (S.south) -- ++(0,-0.4) -| (B3.north);
\draw[flow] (S.south) -- ++(0,-0.4) -| (B6.north);
\draw[flow] (B1.south) -- (K.north);
\draw[flow] (B4.south) -- (K.north);
\draw[flow] (B5.south) -- (K.north);
\draw[flow] (B2.south) -- (G.north);
\draw[flow] (B3.south) -- (G.north);
\draw[flow] (B6.south) -- (G.north);
\draw[trunk] (K.south) -- (L.north -| K.south);
\draw[trunk] (G.south) -- (L.north -| G.south);
\end{tikzpicture}
\end{figure}

The evidence buckets in Figure 4 support different claims. Deterministic fixtures and AdmissionBench support the mutation-admission result under a fixed single-governance-domain scenario family. POS2, B-12, and BLOCK provide field evidence for one positive same-file route and two failure-mode boundaries. The npc-brain cohort supports operability and recoverability of the governance skeleton, not a population-level same-file admission statistic. Self-hosting forensics is operational dogfooding evidence, not independent external validation. Phase A, Phase B, and Phase C add three separate engineering evidence lines: public-source snapshot governance, structured-artifact admission, and dual-live conflict observability; Table 9 keeps those lines from being collapsed into one claim.

The evidence base of this section combines deterministic fixtures, field records, adopter-side observations, public-source snapshot evidence, structured-artifact evidence, and extension evidence. These materials do not have the same evidentiary strength and should not be read as a single empirical sample.

The body therefore states the evidence boundary in prose and keeps only the table needed to prevent claim mixing. Table 9 separates framework-mainline support, public-source snapshot governance, structured-artifact admission, and dual-live conflict evidence. The longer evidence-bucket overview and verdict-phase map are moved to the appendix or supplementary index.

\subsection{Deterministic Fixture Design (12 Scenarios) and Archived MVP Evidence (3 Archived Runs)}

The evidence in this section is deliberately layered into two tiers. The phrase {\char34}12 scenarios{\char34} refers to the design matrix. The phrase {\char34}3 archived{\char34} refers to the core scenarios---B-02, B-08, and B-13---for which a deterministic runner has been completed and the evidence archived. The remaining nine scenarios belong to the coverage blueprint that has not yet been swept by a runner in this version. They are therefore recorded as evidence limitations and future-work items, rather than as validated deliverables.

What this section provides is mechanism-validation evidence. The 12-scenario deterministic fixture design matrix describes the intended coverage of the broker decision surface, while the archived deterministic-runner MVPs---B-02, B-08, and B-13---check that the core admission mechanism aligns with the verdict vocabulary defined in this paper. The coverage categories of the design matrix are summarized in Table 7. What the paper delivers at this stage is therefore a hybrid evidence stack---a 12-scenario design matrix, a 3-scenario deterministic MVP, governance-landing and recoverability evidence, and field-collision evidence---rather than a final empirical version in which all 12 deterministic scenarios have already been swept.

The 12-scenario deterministic fixture design matrix is a mechanism-coverage blueprint and is distinct from the 20-scenario AdmissionBench family reported in §5.

\Needspace{7\baselineskip}\TableLead{Table 7 --- Deterministic Fixture Coverage Categories.}

\begingroup
\TableLargeFont
\setlength{\tabcolsep}{3pt}
\arrayrulecolor{black!55}
\rowcolors{2}{gray!8}{white}
\begin{longtable}{@{}L{0.22\textwidth}L{0.25\textwidth}L{0.47\textwidth}@{}}
\toprule
Category & Mechanism covered & Evaluation focus \\
\midrule
Disjoint paths & Proposition 1 & Different adapter roots may proceed in parallel \\
Same file, different atom & Atom map and bounded-region comparison & Same-file writes do not necessarily require serialization \\
Same atom write--write & CID conflict & Expected fail-closed path \\
Read--write dependency & Augmented dependency rule & Disjoint writes do not imply parallel admissibility \\
Virtual-atom refinement & Virtual atom and decomposition & Uncovered spans may first be wrapped to gain coverage; a coarse atom enters a controlled split suggestion \\
Validator fallback & Static-model boundary & Dynamic errors outside the static admission model are caught by validators \\
\bottomrule
\end{longtable}
\endgroup

The value of this suite is regression-oriented, not statistical. It checks and supports the alignment between the implementation and the verdict vocabulary defined in this paper. It does not claim any particular throughput, latency, or token-cost advantage under adversarial load.

\subsection{Self-Hosting Forensics}

This section provides self-hosting governance evidence. It does not claim external validity; instead, it documents whether ATM, when governing its own framework and paper artifacts, has reached a traceable governance baseline and an ability to expose its own gaps. To avoid placing two adjacent metadata tables in the body---one for coverage and one for incident types---the most important coverage metrics are folded into the section opening:

\begin{itemize}
\item \textbf{Overall self-hosting atomization score:} 95 / 100 (Grade A).
\item \textbf{Production-source-ownership coverage:} 84\% (514 / 609).
\item \textbf{Public-command coverage:} 100\% (55 / 55).
\item \textbf{Atom-evidence completeness} for atoms currently included in the core self-hosting evidence: 100\% (7 / 7, each with test, rollback, provenance, and report evidence).
\end{itemize}

These numbers indicate that ATM has reached a quantitative baseline in governing itself, while also exposing a clearly counted, high-priority gap: 95 production paths are not yet inside the ownership coverage.

The development process of ATM itself supplies a body of internal field evidence. The events documented below are not curated demonstration cases prepared after the fact; they are collisions, freezes, scope incidents, and synchronization failures that the framework actually encountered while governing itself. In particular, the reporting window for ATM includes LLM agents from multiple vendor and editor channels jointly modifying ATM framework code and paper artifacts inside the same controlled worktree or service domain. This paper treats that material as self-referential self-hosting evidence, not as a controlled benchmark experiment. The significance is that ATM was not only designed as a multi-agent governance framework, but also underwent real multi-agent, multi-vendor, same-governance-domain write pressure during its own evolution. Three classes of representative events are retained.

The body retains three representative classes of self-hosting incidents. A \texttt{\seqsplit{cid-shared collision}} occurs when two intents simultaneously claim the same atom CID and trigger the freeze, patch-envelope, and conflict-matrix path. An \texttt{\seqsplit{out-of-scope delivery}} occurs when a delivery touches code outside the declared scope, motivating a closure-packet waiver and a stronger scope gate. A \texttt{\seqsplit{plan-mirror sync failure}} occurs when the planning side and the target ledger close-out diverge, motivating mechanized open and close together with a ledger-consistency check. The detailed outcome and interpretation of these three classes are carried by the condensed incident table in Appendix A.5; the body retains them only for the governance-evidence role they play.

To prevent the reader from treating these three classes as the totality of self-hosting collisions observed by ATM, this paper also records the workload baseline of the self-hosting window. Within the observation window beginning on 2026-05-01, the ATM framework itself accumulated approximately 1,372 governed commits. During the three-week npc-brain alignment window from 2026-05-19 to 2026-06-07, the 3KLife project accumulated 320 governed commits, jointly written by 15 LLM agents from different vendor and editor channels. The detailed channel identifiers are retained in the evidence archive rather than expanded in the body. These incidents are therefore not anecdotes detached from a real workload; they are operational evidence that the governance layer repeatedly absorbed and fed back under a multi-vendor, same-domain workload on the order of about 1,700 commits. Their evidential strength is higher than that of a demonstration case, but still not equivalent to a controlled comparative benchmark experiment.

The role of these forensics is to show that the ATM governance layer is not a post-hoc polished specification. It is a layer that repeatedly exposed its own gaps and fed those gaps back into the mechanism during its own development. The forensics are not a controlled comparative experiment, nor a product showcase; they are a traceable body of self-hosting field evidence whose evidential strength is lower than a controlled comparative benchmark but higher than a pure design narrative.

\subsection{npc-brain Adoption Study}

This section provides adoption evidence. The question it answers is not whether same-file parallel admission is established, but whether ATM's governance skeleton, scope gate, and validator path can sustain recoverability under the real engineering flow of an external project. npc-brain is an external adoption case observed over a three-week window, from 2026-05-19 to 2026-06-07. The project used ATM for atomization, scope control, validator integration, and governance-flow management under a real multi-agent engineering workflow. The aggregate outcomes are summarized in Table 8; the load-bearing result is that ATM did not eliminate all process errors, but channeled the observed errors into recoverable paths, with zero unrecovered admission errors across the cohort.

The traceable evidence comes not only from summary narration but also from adoption notes, the task ledger, validator records, and the corresponding evidence archive. The summary statistics cited in this paper can be traced back to the existing adoption-study writeup and incident table, including the entry for this study's observation window, the 37-card cohort, and the validator-catch classification recorded in \texttt{\seqsplit{paper.md}}. It should also be noted that, within this paper, the 3KLife repository functions primarily as the host of ATM's self-hosting development and evidence archive. It is not counted as a second independent external adoption sample.

This study is not a product showcase, nor a large-scale comparative experiment. Its value is in supplying a traceable body of adopter-side governance evidence that shows ATM's atomization governance, admission governance, and validator-and-evidence substrate operating inside a non-synthetic repository. In particular, the figure of 0 unrecovered admission errors indicates that when an agent encountered contention, an out-of-scope action, or a validator failure, the system retained enough evidence to support repair rather than letting state diverge in an untraceable way. The boundary must be stated explicitly: npc-brain primarily supports the operability and recoverability of governance, \emph{not same-file parallel collision itself}. The latter is supported principally by the POS2, B-12, BLOCK, and close-orchestration field-collision evidence.

\Needspace{7\baselineskip}\TableLead{Table 8 --- npc-brain Three-Week Adoption Summary.}

\begingroup
\TableFont
\setlength{\tabcolsep}{3pt}
\arrayrulecolor{black!55}
\rowcolors{2}{gray!8}{white}
\begin{longtable}{@{}L{0.31\textwidth}L{0.65\textwidth}@{}}
\toprule
Metric & Value \\
\midrule
Observation window & 2026-05-19 to 2026-06-07 (three weeks) \\
Atomized task-card attempts (adoption cohort) & 37 \\
Same-window 3KLife alignment sample (governed commits) & 320 (across 15 vendor and editor channels) \\
Scope-lock interactions & 44 \\
Out-of-scope proposals correctly refused & 2 \\
Scope-lock contention bursts requiring ledger-replay recovery & 1 (2026-05-25, covering 10 cards) \\
Idempotency breaks observed in the CLI runner loop & at least 1 \\
Post-write validator catches & 3 \\
Unrecovered admission errors & 0 \\
\bottomrule
\end{longtable}
\endgroup

The denominator must be stated explicitly. The $N = 37$ here is not the total PR volume of npc-brain over the observation window; it is the size of the ATM-governed task-card cohort within that window. The relevant denominator is not {\char34}all npc-brain development activity{\char34}, but {\char34}the set of task cards that entered the atomization-governance flow{\char34}. The claim of this study is therefore that the ATM-governed 37-card cohort completed the three-week window with 0 unrecovered admission errors, not that every npc-brain development action over those three weeks was governed by ATM with zero errors. The 320 governed-commits figure for the same window comes from the 3KLife alignment sample. It is included to show that, although the cohort is small, the surrounding multi-vendor workload was not low intensity. That sample is not counted as a second independent adopter; it serves only as side evidence for workload intensity.

\subsection{Public-Source, Structured-Artifact, and Live-Conflict Evidence Lines}

To avoid over-compressing heterogeneous evidence into a single claim, this paper separates the newer engineering evidence into three phase-specific lines plus one framework-mainline support line. These lines are related, but they do not support the same conclusion.

The framework-mainline support line shows that broker evidence capture and repo-local evidence-path parameterization have landed in the ATM framework itself. The paper-evidence fast-path bundle landed at commit \texttt{\seqsplit{191fd310166054d3fa526b6961351716bb2d489e}}, and the git-head backfill commit for this evidence freeze is \texttt{\seqsplit{f57dbfe0bdfdf9f939e35400ec346501f4ccb2f3}}. This line supports the narrower claim that ATM's broker evidence path has repo-local governance capability. It is not, by itself, evidence that an external public-source snapshot has been governed, nor that a dual-live editor conflict has been demonstrated.

Phase A is the FastAPI public-source snapshot governance case. It exercises a provenance-pinned public-source snapshot inside the live host repository and preserves baseline, readiness, touched-path, command-log, and replay evidence. The upstream FastAPI snapshot head is \texttt{\seqsplit{82064857539e6286522c347b4b11331b48dd2378}}, the host-repository head is \texttt{\seqsplit{738b9883880742cd36b64f1f81ce6a638f073135}}, and the touched host-visible paths include \texttt{\seqsplit{local/public-source-snapshots/fastapi-0.136.3/fastapi/\_\_init\_\_.py}} and \texttt{\seqsplit{app/main.py}}. The appropriate claim is therefore limited to provenance-pinned baseline capture, host-visible replay, and governance-boundary framing for a public-source snapshot. The Phase A summary also records that the post-change module path resolves to the host virtual environment, with \texttt{\seqsplit{postChangeHelper = null}} and \texttt{\seqsplit{postChangeSnapshotHead = null}}; Phase A should therefore not be read as evidence that post-change execution was cleanly isolated to a snapshot helper, nor as evidence that ATM governed the upstream FastAPI maintainer workflow.

Phase B is the Structured Artifact Admission Track. It independently evaluates ATM's admission and routing behavior over structured non-code artifacts rather than over source-code atoms alone. The track contains 15 deterministic cases across five artifact families: JSON manifest, YAML workflow, TOML configuration, OpenAPI schema path, and atom-map shard. Each family covers \texttt{\seqsplit{parallel-safe}}, \texttt{\seqsplit{serial}}, and \texttt{\seqsplit{blocked-cid-conflict}} behavior. The run reports \texttt{\seqsplit{matchedCount = 15}}, \texttt{\seqsplit{shipSafe = true}}, and a balanced decision distribution: 5 \texttt{\seqsplit{parallel-safe}}, 5 \texttt{\seqsplit{blocked-cid-conflict}}, and 5 \texttt{\seqsplit{serial}} outcomes. This supports the claim that ATM has independent deterministic evidence for cross-format structured-artifact admission and routing. It does not claim external public-source snapshot governance, nor does it claim a dual-editor live-conflict demonstration.

Phase C is the dual-live public-source conflict demonstration. In Team Broker mode, two live actors touched the same FastAPI public-source snapshot path, and the CID broker produced an auditable run artifact that has now been re-verified against the host-side broker-run envelope. The run id is \texttt{\seqsplit{6ea4e411-fa2b-426b-9c71-55bbdbeaa888}}, with plan id \texttt{\seqsplit{batch-5c1fd53c988116ce}}. Actor A was \texttt{\seqsplit{cursor-composer-2.5}}; Actor B was \texttt{\seqsplit{antigravity-gemini-3.5-flash}}. The target file was \texttt{\seqsplit{local/public-source-snapshots/fastapi-0.136.3/fastapi/\_\_init\_\_.py}}. Actor A reached an \texttt{\seqsplit{applied / mergeable}} outcome, while Actor B was routed to \texttt{\seqsplit{queued / conflict}}. This supports the claim that, under Team Broker mode, ATM's CID broker activates on an external public-source snapshot touched path and leaves an auditable applied / queued / conflict run artifact. It does not claim to solve all multi-agent runtime race conditions, and it is not evidence that ATM governs the upstream FastAPI project.

\Needspace{25\baselineskip}\TableLead{Table 9 --- Separated Evidence Lines for Public-Source and Structured-Artifact Governance.}

\begingroup
\TableFont
\setlength{\tabcolsep}{3pt}
\arrayrulecolor{black!55}
\rowcolors{2}{gray!8}{white}
\begin{longtable}{@{}L{0.20\textwidth}L{0.47\textwidth}L{0.27\textwidth}@{}}
\toprule
Evidence line & Supports & Boundary \\
\midrule
Framework mainline support & Repo-local broker evidence capture and parameterized evidence-path capability landed in ATM mainline; evidence anchors are listed in Appendix A.1. & Framework capability only; not an external public-source snapshot or live-conflict result \\
Phase A: FastAPI public-source snapshot governance & Provenance-pinned FastAPI public-source snapshot capture with baseline, readiness, touched-path, command-log, and replay evidence. & Does not claim upstream FastAPI maintainer workflow governance; does not claim clean post-change snapshot-helper execution \\
Phase B: Structured Artifact Admission Track & Deterministic cross-format admission and routing evidence for JSON, YAML, TOML, OpenAPI, and atom-map shard families; 15 / 15 matched cases. & Does not claim external public-source snapshot governance; does not claim dual-editor live conflict \\
Phase C: Dual-live public-source conflict demonstration & Team Broker mode run where two live actors touched the same FastAPI snapshot path and the broker recorded applied / queued / conflict evidence; host-side broker-run envelope is listed in Appendix A.1. & Re-verified host evidence for one live conflict route; does not claim all runtime races are solved and does not claim upstream repo governance \\
\bottomrule
\end{longtable}
\endgroup

\subsection{Real Same-File Admission Outcomes}

This section provides field-collision evidence. It asks whether, under real same-file shared-write conditions, ATM can admit cases with bounded-region disjointness and take a fail-closed path when evidence is insufficient. POS2 is the most important positive same-file field result in this paper. The case simultaneously satisfies the following conditions: same owner-map context, same controlled worktree, same file, disjoint bounded regions, composer-routed admission, neutral-steward apply, and validator pass. Its evidence chain comprises five stages: write intents from two different vendor models, admission inside the same broker domain, deterministic composition, neutral-steward apply, and validators including a diff check, typecheck, and CLI validation.

What POS2 supports is therefore not a bare line-disjoint merge, but a multi-layer admission outcome that emerges after semantic and governance checks. The broker first recognizes that two intents touch \texttt{\seqsplit{broker.ts}} concurrently. It then uses the adapter and atom map to map each side's change onto comparable atom and virtual-atom regions. It checks CID identity, shared surfaces, and read/write dependencies, and only then issues the bounded-region-disjoint admission verdict. The significance is limited but important: within the observed governance domain, ATM can admit some same-file cases conservatively rather than treating every such case as either direct parallel apply or an undifferentiated whole-file block.

\begin{figure}[H]

{\small \noindent\textbf{Figure 5 --- POS2 Progressive Atomization Case.} The figure summarizes the POS2 field case within one controlled worktree, where two intents from two vendor models touch \texttt{\seqsplit{broker.ts}} concurrently. The point is not that {\char34}same file, different range{\char34} is sufficient by itself. Rather, the broker uses the declared atom map and virtual-atom fallback to compare the suspected same-file conflict layer by layer, then checks CID non-overlap, shared-surface separation, and declared read/write dependencies before routing the case to \texttt{\seqsplit{needs-physical-split}}. The figure illustrates one admitted same-file case under the declared adapter, atom-map, active-registry, and virtual-atom model; it does not claim dynamic reconstruction of every hidden semantic read:\par}
\smallskip
\centering
\begin{tikzpicture}[
  >=Latex,
  font=\scriptsize,
  node distance=3mm,
  file/.style    ={draw, rounded corners=2pt, fill=blue!10,   draw=blue!60!black,
                   minimum width=78mm, minimum height=8mm, align=center, line width=0.5pt},
  region/.style  ={draw, rounded corners=2pt, fill=blue!22,   draw=blue!60!black,
                   minimum width=78mm, minimum height=7mm, align=left, font=\tiny,
                   inner xsep=4mm, line width=0.4pt},
  pivot/.style   ={draw, rounded corners=2pt, fill=yellow!22, draw=orange!70!black,
                   minimum width=78mm, minimum height=7mm, align=center, line width=0.5pt,
                   font=\scriptsize\bfseries},
  layer/.style   ={draw, rounded corners=2pt, fill=gray!10,   draw=black!60,
                   minimum width=78mm, minimum height=6mm, align=left, font=\tiny,
                   inner xsep=4mm, line width=0.4pt},
  verdict/.style ={draw, rounded corners=2pt, fill=green!24,  draw=green!60!black,
                   minimum width=78mm, minimum height=7mm, align=center, line width=0.5pt,
                   font=\scriptsize\bfseries},
  exec/.style    ={draw, rounded corners=2pt, fill=violet!14, draw=violet!70!black,
                   minimum width=78mm, minimum height=7mm, align=center, line width=0.5pt},
  applied/.style ={draw, rounded corners=2pt, fill=green!32,  draw=green!60!black,
                   minimum width=78mm, minimum height=7mm, align=center, line width=0.6pt,
                   font=\scriptsize\bfseries},
  validate/.style={draw, rounded corners=2pt, fill=teal!12,   draw=teal!60!black,
                   minimum width=78mm, minimum height=7mm, align=left, font=\tiny,
                   inner xsep=4mm, line width=0.4pt},
  flow/.style    ={->, line width=0.5pt, draw=black!75}
]
\node[file]                                   (FILE)   {\texttt{packages/cli/src/commands/broker.ts}};
\node[region,  below=of FILE]                 (POS2A)  {POS2-A \quad lines 841--878 \quad\textit{(Codex / OpenAI; TASK-POS2-A)}};
\node[region,  below=of POS2A]                (POS2B)  {POS2-B \quad lines 989--1142 \quad\textit{(Claude / Anthropic; TASK-POS2-B)}};
\node[pivot,   below=of POS2B]                (CMP)    {Progressive atomization compare};
\node[layer,   below=of CMP]                  (L0)     {Layer 0 \,--\, same file?                       \hfill yes};
\node[layer,   below=of L0]                   (L1)     {Layer 1 \,--\, known atom overlap?              \hfill no};
\node[layer,   below=of L1]                   (L2)     {Layer 2 \,--\, shared surface overlap?          \hfill no};
\node[layer,   below=of L2]                   (L3)     {Layer 3 \,--\, declared read/write dep?          \hfill no};
\node[layer,   below=of L3]                   (L4)     {Layer 4 \,--\, virtual atom overlap?            \hfill no};
\node[layer,   below=of L4]                   (LR)     {Result \,--\, bounded disjoint                  \hfill yes};
\node[verdict, below=of LR]                   (VERD)   {Verdict: \texttt{needs-physical-split}};
\node[exec,    below=of VERD]                 (COMP)   {Deterministic composer};
\node[exec,    below=of COMP]                 (STEW)   {Neutral Steward apply\,(single neutral write)};
\node[applied, below=of STEW]                 (APPLIED){Verdict: \emph{applied}};
\node[validate, below=of APPLIED]             (V1)     {Validator: diff check\hfill$\checkmark$};
\node[validate, below=of V1]                  (V2)     {Validator: typecheck\hfill$\checkmark$};
\node[validate, below=of V2]                  (V3)     {Validator: CLI validation\hfill$\checkmark$};
\foreach \a/\b in {FILE/POS2A, POS2A/POS2B, POS2B/CMP,
                   CMP/L0, L0/L1, L1/L2, L2/L3, L3/L4, L4/LR,
                   LR/VERD, VERD/COMP, COMP/STEW, STEW/APPLIED,
                   APPLIED/V1, V1/V2, V2/V3}
    \draw[flow] (\a) -- (\b);
\end{tikzpicture}
\end{figure}

B-12 and BLOCK provide negative evidence and are placed deliberately alongside POS2 to reduce the risk of cherry-picking.

\textbf{B-12 --- apply-phase late enforcement.} B-12 shows that both sides can still be classified as \texttt{\seqsplit{parallel-safe}} at admission time while apply-phase runtime arbitration nonetheless fails closed. B-12 should therefore be read as late enforcement rather than admission-time success. This case exposes that ATM's enforcement boundary has not yet been moved entirely forward into the admission layer, and it is a concrete instance of the admission-time active-intent forwarding open problem stated in §3.7.

\textbf{BLOCK --- admission-time block with split suggestion.} BLOCK shows the broker blocking an overlapping intent before any write and emitting a split suggestion, so that the conflict becomes input to owner-map refinement rather than a single bare failure.

On the evidence scope of POS2, the boundary of the claim must be stated explicitly. POS2 is one field case that has been fully walked through the five-stage evidence chain---two-vendor write intents $\rightarrow$ broker admission $\rightarrow$ deterministic composer $\rightarrow$ neutral-steward apply $\rightarrow$ validators pass---and archived. It is not a representative case estimated with a confidence interval over a sample of same-file candidate pairs. This version of the paper cannot claim a field-level false-positive or false-negative rate for same-file bounded-region admission; that quantitative baseline falls within the future comparative-evaluation scope discussed in §6.3.

The evidential role of POS2 in this study is therefore an \textbf{existence proof}. It shows that under real same-owner, same-worktree, same-file, cross-vendor conditions, the bounded-region-disjoint admission verdict can be produced end-to-end by the mechanism and pass validators. It is not a statistical claim that {\char34}the great majority of cases are classified correctly{\char34}. The corresponding quantitative classification accuracy is deferred to a subsequent controlled benchmark, as noted in the future comparative-evaluation discussion in §6.3.

The negative corroboration is supplied by B-12 and BLOCK in the same section. B-12 shows an admission-time miss with apply-phase fail-closed handling, while BLOCK shows admission-time fail-closed handling with a split suggestion. Taken together, the three cases describe an {\char34}existence + two failure modes{\char34} evidence triangle, not an ROC curve.

Close-orchestration and the refinement loop fall at the prototype edge. They support the conservative conclusion that ATM has a mechanism prototype for channeling blocked overlap inside a single governance domain into a reviewable refinement chain. That conclusion is not yet strong enough to claim that every cross-vendor, same-owner refinement workflow has been field-validated, and certainly not strong enough to claim that distributed refinement across machine clones is handled. This paper places these mechanisms on the evidence map without promoting them into decisive evidence for the principal contribution.

\subsection{Wave Mode and CID Stability}

This section provides \textbf{extension evidence}. It is not the principal support for the admission-core claim; instead, it documents whether the broker-and-steward path can be extended to batch orchestration and CID identity stability. Wave Mode is the batch extension of the admission layer. The Team Agents Wave Mode replay suite verifies that batch admission, evidence slicing, and checkpointing preserve fail-closed behavior across safe waves, unsafe same-deliverable cases, mixed dependencies, per-task slicing, and needs-review gating. Its role is to show that the broker-and-steward path can be composed with task-local orchestration, not to replace the admission-core evidence of §4.1.

Wave Mode is therefore best read as orchestration extensibility rather than a new admission-core theorem. A single task may involve a Captain or coordinator, an implementer, a validator, an evidence collector, and optionally a review agent. When several task cards are submitted as a wave, those roles must preserve per-task attribution, keep read/write dependencies visible across task boundaries, and stop at coordinator or human review gates when evidence is incomplete. The replay suite exercises that coordination boundary while leaving shared-surface admission with the CID broker and governed apply with the neutral steward.

At batch scale, the same least-privilege rule prevents Wave Mode from becoming a race among unconstrained writers. Read-only planning, validation, review, evidence collection, and next-task preflight can proceed in parallel, while write and lifecycle authority remain exclusive and policy-governed. The broker remains responsible for shared-mutation admission, and the Captain remains responsible for task-local coordination.

CID stability, in turn, validates the separation of responsibilities between the Candidate CID and the Capsule CID. The former serves pre-write arbitration, while the latter serves post-validation artifact identity. This distinction reduces the risk of conflating a transient proposal with an already validated capsule, and supplies a versioned foundation for subsequent schema migration.

More concretely, the five scenarios in the Wave Mode replay suite each cover a different extension question:

\begin{itemize}
\item \textbf{\texttt{\seqsplit{safe-wave}}.} Multiple non-conflicting task cards can pass admission inside the same wave without losing evidence attribution.
\item \textbf{\texttt{\seqsplit{unsafe-wave-same-deliverable}}.} When several cards share the same deliverable or write surface, the planner splits them rather than over-admitting them in a single batch.
\item \textbf{\texttt{\seqsplit{mixed-wave-dependency}}.} Read--write dependencies are preserved at the wave layer rather than masked by batch scheduling.
\item \textbf{\texttt{\seqsplit{per-task slicing}}.} Artifacts, validators, and close-outs in the same batch remain traceable back to their individual tasks.
\item \textbf{\texttt{\seqsplit{needs-review gating}}.} When certain waves require human or coordinator review, the system stops at the gate rather than allowing the batch flow to bypass the fail-closed criterion.
\end{itemize}

What these results support is orchestration extensibility, not the claim that {\char34}the wave planner itself proves same-file bounded-region admission{\char34}.

The evidential role of the multi-vendor self-hosting evidence also needs to be drawn cleanly. Inside the same reporting window, agents from the Anthropic, Cursor, Google, and OpenAI ecosystems jointly wrote the framework and its evidence artifacts under ATM admission control. The cross-vendor collisions and wave-serialization examples observed in that period indicate that ATM's governance vocabulary is not bound to a single model vendor. What this material supports is provider-neutral operability and governance durability: agents from multiple vendors can be governed, serialized, and archived under the same broker, steward, and validator substrate. This evidence should not, however, be over-elevated into a same-file admission-core proof; the principal evidence for same-file admission remains POS2 and its negative counterparts, B-12 and BLOCK, in §4.5.

The extension value of CID stability lies in cleanly separating the candidate governance unit from the post-validation capsule version. The Candidate CID serves pre-write comparison and may therefore be recomputed alongside the atom map, virtual atoms, and bounded regions. The Capsule CID serves the post-validation capsule lifecycle and therefore requires content reproducibility, version verifiability, and backfill traceability. At the implementation layer, the current validation flow already includes atom-to-CID checks and a backfill path, exemplified by \texttt{\seqsplit{scripts/validate-atom-id-to-cid.ts}} and the corresponding backfill script. Their role is not to supply another positive proof for same-file admission, but to ensure that capsule export, rescue, rollback, drift detection, and subsequent schema migration have a stable identity substrate.

Wave Mode answers whether admission can extend to batch scheduling. CID stability answers whether governance identity can remain stable after admission. Both are extension evidence, and neither is positioned to carry the core admission claim on its own.

\subsection{Threats to Validity}

This section consolidates the validity limitations of the §4 evidence stack, so that the reader can clearly map what this study can and cannot support before entering the §5 results and the §6 discussion. The paper relies on a hybrid evidence stack: deterministic fixtures, archived runner cases, same-file field-collision evidence, self-hosting incidents, an adopter cohort, and the AdmissionBench v0.1 baseline together with the v0.2 paper profile. Taken together, these materials form the evidence chain. This design supports feasibility, clarifies the governance boundary, and supplies the paper-facing result, but it is not equivalent to having completed all external comparative statistics.

Internal validity. POS2, B-12, BLOCK, and the archived validator traces support key observed phenomena such as bounded-region admission, late enforcement, and the fail-closed boundary. The v0.1 baseline further demonstrates the replayability of the frozen contract, the blind audit, and the machine-generated artifacts. The v0.2 paper profile then consolidates these phenomena into citable results, ablation evidence, and an enforcement summary over the same benchmark family. These results, however, hold under a fixed contract and a fixed scenario family; they do not imply that an unknown workload would preserve the same error distribution.

External validity. The core evidence of this paper remains concentrated in a single governance domain: ATM self-hosting, the npc-brain adopter cohort, the 3KLife alignment sample, the controlled collision evidence, and AdmissionBench v0.1 and v0.2. This evidence supports the feasibility of single-domain pre-write admission, but it is not yet sufficient to extrapolate to large enterprise monorepos, long-running high-concurrency SaaS teams, polyglot generated-artifact pipelines, or cross-clone, cross-branch, or pull-request-based distributed collaboration. For those settings, this paper claims a transferable governance direction, not a completed body of broad empirical evidence.

Construct validity. The paper uses the atom, the Candidate CID, the ConflictKey, the shared surface, and declared \texttt{\seqsplit{readAtoms}} as proxy variables for conflict. These structures are proxies for semantic interference, not equivalence proofs. ATM verdicts such as \texttt{\seqsplit{parallel-safe}}, \texttt{\seqsplit{compose}}, \texttt{\seqsplit{serial}}, and \texttt{\seqsplit{block}} mean that particular governance signals were observed under the existing adapter, atom map, and declared-dependency model. They are not direct guarantees that the resulting program behavior is semantically correct. This is precisely why the paper retains validator handoff, the CAS base-hash recheck, active-intent enforcement, and the fail-closed fallback as runtime safety nets.

Hidden semantic read gap. A remaining construct-validity boundary is the hidden semantic read gap. ATM can reason about dependencies that are declared in the task contract, exposed by format adapters, represented through \texttt{\seqsplit{readAtoms}} and \texttt{\seqsplit{writeAtoms}}, captured in the active registry, or conservatively virtualized through virtual atoms. It does not claim to infer every latent semantic dependency from natural-language context, unstated agent memory, or unobserved tool interactions. Such cases are not admitted as positive evidence unless they become visible through the contract, the adapter layer, the active registry, the benchmark fixture, or a future read-set reconstruction provider; until then, fail-closed routing, steward review, validators, and CAS revalidation provide containment rather than a positive independence guarantee.

Conclusion validity. The paper supplies a benchmark chain with result-layer separation rather than a complete large-scale comparative benchmark. The v0.1 layer carries the baseline and audit anchor, while the v0.2 layer carries the paper-facing result. Together, they support the manuscript's claims around route-label F1, layer ablation, and enforcement timing, but the paper does not yet report a controlled comparative benchmark, confidence intervals, or power analyses. This paper therefore supports the narrower conclusion that ATM has feasibility, auditability, and methodological novelty within its stated boundary. It is not sufficient to make statistical statements about throughput, false-positive rate, token efficiency, or cross-system superiority; those claims belong to the future comparative evaluation discussed in §6.3.

\subsection{Governance-Containment Mapping across the Three Planes}

The evidence in this paper also maps onto the three planes introduced in Section 3.1 and summarized in Table 1. The Task-contract plane is supported primarily by scope-lock interactions, out-of-scope refusals, and closure-packet recovery in the adopter and self-hosting evidence. The Mutation-admission plane is supported by AdmissionBench v0.1 and v0.2, POS2, B-12, BLOCK, and the archived deterministic-runner cases. The Evidence-closure plane is supported by validator catches, ledger-replay recovery, closure-packet repair, cross-agent review-signature prototype evidence, and the adversarial-adapter-containment package.

This mapping is an evidence-coverage view, not a benchmark. The quantitative admission claim is carried by AdmissionBench and OperationalBench. The Task-contract and Evidence-closure planes currently have operational and prototype evidence, but they do not yet have population-level catch rates, broad false-positive estimates, or a complete governance-containment benchmark. The detailed plane-by-plane table is moved to the appendix or supplementary material.

\section{Benchmark Results and Limitations}

This section is not a supplement detached from §4, *Validation, Evidence, and Benchmark Alignment*. It is a formal reorganization of the benchmark evidence most directly tied to the paper's citable result chain. §4 establishes the evidence surface, the evidence boundary, and the benchmark-alignment position, while §5 organizes the AdmissionBench evidence chain into a paper-facing admission result and adds OperationalBench as a narrower runtime-transparency supplement. This division of labor separates {\char34}whether the evidence holds{\char34} from {\char34}how the result is read,{\char34} so that the baseline, the main result, and the unfinished extensions are not compressed into one undifferentiated layer of argument.

This paper has not completed a full cross-system comparative evaluation. A head-to-head benchmark comparison against STORM \textcolor{atmLinkBlue}{\hyperlink{Item.39}{[3]}}, CodeCRDT \textcolor{atmLinkBlue}{\hyperlink{Item.37}{[1]}}, SCF \textcolor{atmLinkBlue}{\hyperlink{Item.38}{[2]}}, CoAgent \textcolor{atmLinkBlue}{\hyperlink{Item.47}{[11]}}, S-Bus \textcolor{atmLinkBlue}{\hyperlink{Item.52}{[16]}}, CodeTeam \textcolor{atmLinkBlue}{\hyperlink{Item.51}{[15]}}, ATCC \textcolor{atmLinkBlue}{\hyperlink{Item.41}{[5]}}, and transactional tool-effect runtimes such as Cordon and Atomix \textcolor{atmLinkBlue}{\hyperlink{Item.77}{[41--42]}} on a shared workload---measuring conflict catch rate, false-positive rate, wall-clock time, token cost, and repair cost---remains future work. If a future evaluation draws on a large-scale conflict corpus such as AgenticFlict \textcolor{atmLinkBlue}{\hyperlink{Item.50}{[14]}}, its cross-PR and Git-merge samples must first be converted into a single-governance-domain pre-write intent replay workload. Otherwise, this paper cannot claim that ATM resolves the distributed PR conflicts present in that corpus.

\subsection{ATM-AdmissionBench: From the v0.1 Baseline to the v0.2 Paper-Facing Result}

AdmissionBench is reported in two layers. The v0.1 smoke baseline is the frozen baseline substrate: it demonstrates that the benchmark contract, fixtures, runner, blind-audit procedure, and machine-generated artifacts have been established, and can therefore serve as the audit starting point for later comparisons. The v0.2 paper profile is the formally adopted result layer cited in the body of this paper: it reuses the same benchmark family while supplying a paper-facing summary, ablation rows, adversarial rows, enforcement rows, and provenance alignment. The result therefore advances from a re-runnable smoke baseline into a layer that can be cited as the paper's main result.

The frozen generator commit for v0.1 is \texttt{\seqsplit{3eec69a73a04112e2af8d3630c32138c37143eab}}, corresponding to \texttt{\seqsplit{artifacts/generated/atm-admission-bench/20260625/}} and \texttt{\seqsplit{artifacts/blind-bench/20260625/}}. The v0.2 paper profile is then regenerated with the paper profile and seed 20260625 into a paper-facing artifact bundle whose \texttt{\seqsplit{generator-manifest.json}} records the same \texttt{\seqsplit{baseCommit}} and \texttt{\seqsplit{generatorCommit}}, \texttt{\seqsplit{3eec69a73a04112e2af8d3630c32138c37143eab}}. The public anchor cited in this manuscript is the later evidence-landing commit \texttt{\seqsplit{ab8753b7daf0a3c4cd8b4483fe24d519ff2590bd}} on the ATM public repository \texttt{\seqsplit{main}}, which incorporated the v0.1 and v0.2 AdmissionBench evidence bundles after the \texttt{\seqsplit{v0.9.0-alpha.1}} source tag. The corresponding artifact path is \texttt{\seqsplit{artifacts/generated/atm-admission-bench/20260625-paper/}}.

The benchmark contract fixes the row universes as follows. The 20-scenario family referenced throughout this paper denotes the 20 single-governance-domain admission scenarios frozen in \texttt{\seqsplit{scripts/fixtures/atm-admission-bench/manifest.json}}. The 42 comparisons denote the mode-level expected-versus-actual comparisons executed against those scenarios under the fixed seed \texttt{\seqsplit{20260625}}. Policy rows, ablation rows, adversarial rows, and enforcement rows are condition-level observations derived from these 20 unique scenarios, not fresh independent samples. A policy row is a per-scenario $\times$ policy-mode $\times$ route-expectation report row. An ablation row is the same kind of row recomputed after one ATM layer or feature has been removed. An adversarial row is a report row produced on the same fixture family with an added perturbation or stress condition. An enforcement row counts only admission, apply, validator, and human-escalation timing rows within an enforcement projection.

What these four row universes share is a common lineage from the same 20 scenarios and the same fixed seed \texttt{\seqsplit{20260625}}; what they do not share is a common denominator, so they must not be added together or substituted for one another in later prose. The route labels are fixed as \texttt{\seqsplit{parallel-safe}}, \texttt{\seqsplit{compose}}, \texttt{\seqsplit{serial}}, \texttt{\seqsplit{block}}, and \texttt{\seqsplit{fail-closed/refine}}, and no cross-clone or PR-merge workload is mixed in.

The ground-truth and metric definitions follow the same independence discipline. The oracle side is produced from the frozen benchmark contract without reading ATM output, and is finalized before the formal audit comparison. Whenever ATM output disagrees with the oracle, the divergence is recorded as a benchmark failure rather than back-filled into the expected answer. The metric route-label F1 evaluates route-label classification across the 42 mode-level comparisons, taking a macro average over the observed route classes, whereas false-safe rows belong to the policy comparison surface and therefore sit outside the route-label F1 denominator. The metric intent preservation = 97.62\% reports the share of the v0.2 paper-profile policy view in which \texttt{\seqsplit{ATM-full}} preserves the original task intent and avoids a false-safe or unresolved outcome. These metrics measure admission behavior inside a single governance domain, not downstream Git-merge quality or end-to-end semantic correctness.

\Needspace{7\baselineskip}\TableLead{Table 10 --- Baseline Policy Definitions.}

\begingroup
\TableLargeFont
\setlength{\tabcolsep}{3pt}
\arrayrulecolor{black!55}
\rowcolors{2}{gray!8}{white}
\begin{longtable}{@{}L{0.31\textwidth}L{0.65\textwidth}@{}}
\toprule
Baseline or policy & Operational definition \\
\midrule
\texttt{\seqsplit{ATM-full}} & Uses the complete atom, virtual atom, CID, shared surface, ConflictKey, CAS, and fallback-lock pipeline, and emits the \texttt{\seqsplit{parallel-safe}}, \texttt{\seqsplit{compose}}, \texttt{\seqsplit{serial}}, \texttt{\seqsplit{block}}, and \texttt{\seqsplit{fail-closed/refine}} routes. \\
\texttt{\seqsplit{file-serial}} & Serializes any same-file or same-shared-artifact overlap, without attempting bounded-region disjointness. \\
\texttt{\seqsplit{text-range}} & Decides conflicts purely from textual range overlap, without consulting the atom map, CID, or semantic ConflictKey. \\
\texttt{\seqsplit{file-OCC}} / OCC-style & Decides stale writes through base-hash or file-level optimistic validation, without offering atom-level route selection. \\
\texttt{\seqsplit{no-governance}} & Performs no broker admission; conflicts can only surface later through apply, validator, or human review. \\
\bottomrule
\end{longtable}
\endgroup

To make the role separation between the baseline and the main result legible at a glance, Table 11 first aligns v0.1 and v0.2 on the same benchmark family by their respective reporting functions. The denominator of this table is \textbf{42 mode comparisons}. The table answers how each version presents an overview of the same benchmark family, rather than the finer policy-row, ablation-row, or enforcement-projection statistics.

\begin{table}[H]
\TableLead{Table 11 --- ATM-AdmissionBench v0.1 Baseline versus v0.2 Paper Profile.}
\raggedright
\TableXLFont
\setlength{\tabcolsep}{3pt}
\renewcommand{\arraystretch}{1.03}
\arrayrulecolor{black!55}
\rowcolors{2}{gray!8}{white}
\begin{tabular}{@{}L{0.22\textwidth}L{0.25\textwidth}L{0.47\textwidth}@{}}
\toprule
Item & v0.1 baseline & v0.2 paper profile \\
\midrule
Seed & \texttt{\seqsplit{20260625}} & \texttt{\seqsplit{20260625}} \\
Scenario count & 20 & 20 \\
Mode comparisons & 42 & 42 \\
Matched expectations & 42 / 42 & 42 / 42 \\
Expectation failures & 0 & 0 \\
False-safe regressions & 0 & 2 false-safe rows in the policy comparison surface, not in the 42-comparison route-F1 denominator \\
Unsafe-caught or intent-preservation view & 92.31\% unsafe-caught & 97.62\% intent preservation \\
Over-serialization view & Not separately reported by the baseline & 4 over-serialization rows in the \texttt{\seqsplit{ATM-full}} policy view \\
Unresolved benchmark rows & 0 & 0 \\
\bottomrule
\end{tabular}
\end{table}

Building on this baseline, Table 12 condenses the v0.2 results cited in the body of the paper into a single summary table aligned with the Results, Ablation, and enforcement-timing narrative. The table intentionally juxtaposes several row universes, but it does not merge their denominators. It includes 252 policy rows, 294 ablation rows, and 210 adversarial rows derived from the 20 unique scenarios, together with the 51-row enforcement-timing projection. The forwarding rows use the 51-row enforcement-timing projection as their denominator, because that projection is aggregated from v0.2 policy-view route outcomes. Consequently, the identity $9 + 6 + 3 + 0 + 33 = 51$ must not be added to, or substituted for, the 42 mode comparisons, the 252 policy rows, or the 4 enforcement-projection categories.

For the same reason, this paper does not treat the 20 scenarios, the 42 mode comparisons, or the policy and ablation row universes of AdmissionBench as the same sample-count claim as the 24,332 out-of-sync instances reported by SyncMind and SyncBench \textcolor{atmLinkBlue}{\hyperlink{Item.79}{[43]}}. SyncBench measures recovery after an agent's belief state has fallen out of sync with the evolving repository state. AdmissionBench instead measures admission verdicts, layer ablation, and enforcement timing inside a single governance domain before a shared write occurs. The two benchmarks are complementary rather than interchangeable, and their raw instance counts should not be used to compare benchmark scale or evidence strength directly.

\Needspace{7\baselineskip}\TableLead{Table 12 --- ATM-AdmissionBench v0.2 Paper-Facing Summary.}

\begingroup
\TableFont
\setlength{\tabcolsep}{3pt}
\arrayrulecolor{black!55}
\rowcolors{2}{gray!8}{white}
\begin{longtable}{@{}L{0.31\textwidth}L{0.65\textwidth}@{}}
\toprule
Category & Result \\
\midrule
Derived policy rows (from 20 unique scenarios) & 252 \\
Derived ablation rows (from 20 unique scenarios) & 294 \\
Derived adversarial rows (from 20 unique scenarios) & 210 \\
Enforcement projection categories & 4 \\
Enforcement projection rows & 51 \\
\texttt{\seqsplit{ATM-full}} route-label F1 (42 mode comparisons) & 1.000 \\
Admission-forwarded rows & 9 \\
Apply-phase forwarded rows & 6 \\
Validator-forwarded rows & 3 \\
Human-forwarded rows & 0 \\
Not-forwarded rows & 33 \\
\bottomrule
\end{longtable}
\endgroup

Table 12 should therefore be read as the paper-profile summary table for AdmissionBench, \emph{not as a single surface that replaces every detailed appendix table}. The v0.1 layer answers whether the benchmark substrate exists, is frozen, and can be audited. The v0.2 layer answers what result the paper stands behind on that frozen benchmark family. On the 20-scenario, 42-comparison benchmark family, \texttt{\seqsplit{ATM-full}} preserves 0 expectation failures and 0 unresolved rows, achieves route-label F1 = 1.000, and compresses the main failure modes within the policy comparison surface into a small number of false-safe and over-serialization rows rather than relapsing into widespread silent mismatch.

The ablation view then explains why the result is not produced by a single heuristic. Removing the virtual atom adds 8 false-safe rows and loses 9 end-to-end success rows. Removing the ConflictKey adds 4 false-safe rows and loses 5 success rows. Removing CID, shared surface, or CAS each costs three to five additional success rows. Removing the fallback lock adds no false-safe rows but still loses 2 success rows. What these removals share is a common pattern: every removed layer contributes either a distinct false-safe-suppression role, a distinct success-preservation role, or both. ATM's behavior is therefore built jointly by virtual atoms, conflict identity, shared-surface judgment, CAS revalidation, and fail-closed recovery.

Even so, v0.2 is not the final benchmark in which every external-validity question has been resolved. It remains a paper profile under a single governance domain, a fixed seed, and a fixed scenario family. It is not directly equivalent to a large monorepo, a polyglot microservice estate, or a remote multi-clone PR-merge workflow, and it does not claim to cover the full noise of real-world tool latency, model drift, or organizational process.

\begin{figure}[H]

{\small \noindent\textbf{Figure 6 --- Enforcement and Recovery Phase Distribution.} The stacked bar summarizes the 51-row enforcement-timing projection behind Table 12. Unsupported or nontrivial cases are not all handled at a single layer: some are forwarded at admission time, some surface at apply time, and some are caught by validators. The zero human-forwarded row should be read as a property of this benchmark projection, not as a claim that human review is unnecessary in deployment.\par}
\smallskip
\centering
\begin{tikzpicture}[x=0.095\textwidth,y=9mm,font=\scriptsize]
  \definecolor{atmTeal}{HTML}{2A9D8F}
  \definecolor{atmAmber}{HTML}{E9A23B}
  \definecolor{atmBlue}{HTML}{5B7DBE}
  \definecolor{atmGray}{HTML}{D8DEE9}
  \definecolor{atmInk}{HTML}{2F3440}
  \def\total{51}
  \def\scale{9.7}
  \pgfmathsetmacro{\wA}{9/\total*\scale}
  \pgfmathsetmacro{\wB}{6/\total*\scale}
  \pgfmathsetmacro{\wC}{3/\total*\scale}
  \pgfmathsetmacro{\xD}{\wA+\wB+\wC}
  \draw[rounded corners=2pt,fill=gray!4,draw=gray!35] (-0.15,-1.0) rectangle (9.95,2.35);
  \node[anchor=west,font=\sffamily\bfseries\scriptsize,text=atmInk] at (0,1.9) {51-row enforcement projection};
  \draw[fill=atmTeal!85,draw=white,line width=0.5pt] (0,0.25) rectangle (\wA,1.08);
  \draw[fill=atmAmber!90,draw=white,line width=0.5pt] (\wA,0.25) rectangle ({\wA+\wB},1.08);
  \draw[fill=atmBlue!80,draw=white,line width=0.5pt] ({\wA+\wB},0.25) rectangle ({\wA+\wB+\wC},1.08);
  \draw[fill=atmGray,draw=white,line width=0.5pt] ({\wA+\wB+\wC},0.25) rectangle (\scale,1.08);
  \node[white,font=\sffamily\bfseries\scriptsize] at ({\wA/2},0.66) {9};
  \node[white,font=\sffamily\bfseries\scriptsize] at ({\wA+\wB/2},0.66) {6};
  \node[white,font=\sffamily\bfseries\scriptsize] at ({\wA+\wB+\wC/2},0.66) {3};
  \node[font=\sffamily\bfseries\scriptsize,text=atmInk] at ({\xD+(\scale-\xD)/2},0.66) {33 not-forwarded};
  \foreach \x/\lab in {0/0,1.90/10,3.80/20,5.70/30,7.60/40,9.70/51} {
    \draw[gray!45] (\x,0.15) -- (\x,0.02);
    \node[anchor=north,font=\tiny,text=gray!75!black] at (\x,-0.02) {\lab};
  }
  \node[anchor=west,font=\tiny,text=atmInk] at (0,-0.62) {\textcolor{atmTeal}{\rule{8pt}{5pt}} admission-forwarded (9)};
  \node[anchor=west,font=\tiny,text=atmInk] at (2.4,-0.62) {\textcolor{atmAmber}{\rule{8pt}{5pt}} apply-phase (6)};
  \node[anchor=west,font=\tiny,text=atmInk] at (4.55,-0.62) {\textcolor{atmBlue}{\rule{8pt}{5pt}} validator (3)};
  \node[anchor=west,font=\tiny,text=atmInk] at (6.35,-0.62) {\textcolor{violet!55}{\rule{8pt}{5pt}} human (0)};
  \node[anchor=west,font=\tiny,text=atmInk] at (7.95,-0.62) {\textcolor{atmGray}{\rule{8pt}{5pt}} not-forwarded (33)};
\end{tikzpicture}
\end{figure}

Scope note. AdmissionBench evaluates repository-scoped admission decisions over declared, adapter-observed, or conservatively virtualized mutation surfaces. It is not a direct replacement for benchmarks or systems that evaluate serializability recovery, HTTP-observable read isolation, transactional tool-effect staging, database transaction scheduling, or end-to-end repository generation. The benchmark therefore supports ATM's admission-boundary claim, rather than a general superiority claim over adjacent agentic concurrency or repository-workflow substrates.

\subsection{OperationalBench: Recovery Routing and Runtime Overhead}

AdmissionBench evaluates admission correctness, route selection, ablation behavior, and enforcement timing. To make the runtime behavior of the admission layer more transparent, this paper also reports an \textbf{OperationalBench} track. OperationalBench is not a cross-system performance benchmark and does not claim latency, throughput, token-use efficiency, or memory-footprint superiority. Its purpose is narrower: to expose the cost and stability of ATM's own recovery-routing paths under the same single-governance-domain assumption as the rest of the paper.

OperationalBench separates three evidence layers. The official paper run from 2026-06-27 (artifact label \texttt{\seqsplit{20260627}}) records the paper-facing operational profile. The extended 2026-06-27 $N = 50$ contention run (artifact label \texttt{\seqsplit{20260627-extended}}) probes the higher-contention tail. The multi-seed stability note (artifact label \texttt{\seqsplit{multi-seed-stability-20260627-20260629}}) repeats the paper profile under additional seeds and checks whether the structural results change. Across the tested seeds, the scenario counts, result rows, track counts, blocked-case counts, route counts, and recovery metrics remain identical, supporting the conclusion that the benchmark's structural findings are not seed artifacts.

The extended $N = 50$ run should be read as an operational stress probe rather than as a liveness proof. Under higher contention, the latency tail remains concentrated in steward-mediated recovery paths and total scenario time, which is expected for a design that preserves intents and routes unsafe direct apply into governed successor paths. The run does not show a new route-distribution or recovery-structure change under the tested contention setting. Validator timing in OperationalBench reflects a lightweight validator path and should not be extrapolated to projects with expensive build, integration-test, or end-to-end validation pipelines.

\begin{figure}[H]

{\small \noindent\textbf{Figure 7 --- OperationalBench Latency Profile.} The two-panel chart compares the official 2026-06-27 paper run with the extended $N = 50$ contention run. Each panel reports P95 and P99 latency for admission decision, steward apply, and total scenario time on a logarithmic millisecond scale. The intended reading is not cross-system performance superiority, but an internal cost profile: broker admission is sub-millisecond, while tail latency concentrates in steward-mediated apply and recovery paths.\par}
\smallskip
\centering
\begin{minipage}{0.48\textwidth}
\centering
\begin{tikzpicture}
\begin{semilogyaxis}[
  width=\linewidth,height=48mm,
  ymin=0.01,ymax=1500,
  ytick={0.01,0.1,1,10,100,1000},
  yticklabels={0.01,0.1,1,10,100,1000},
  ymajorgrids=true,grid style={gray!20},
  ylabel={Latency (ms)},
  symbolic x coords={Admission,Steward,Total},
  xtick=data,
  xticklabel style={font=\tiny,align=center},
  yticklabel style={font=\tiny},
  ylabel style={font=\tiny},
  title={\sffamily\scriptsize Official run},
  title style={yshift=-1mm},
  legend style={font=\tiny,draw=none,at={(0.5,-0.22)},anchor=north,legend columns=2},
  mark size=2.1pt,
]
\addplot+[teal!80!black,thick,mark=*,mark options={fill=teal!65}] coordinates {(Admission,0.024) (Steward,302.424) (Total,310.159)};
\addplot+[orange!85!black,thick,mark=square*,mark options={fill=orange!70}] coordinates {(Admission,0.050) (Steward,541.920) (Total,1088.094)};
\legend{P95,P99}
\end{semilogyaxis}
\end{tikzpicture}
\end{minipage}\hfill
\begin{minipage}{0.48\textwidth}
\centering
\begin{tikzpicture}
\begin{semilogyaxis}[
  width=\linewidth,height=48mm,
  ymin=0.01,ymax=1500,
  ytick={0.01,0.1,1,10,100,1000},
  yticklabels={0.01,0.1,1,10,100,1000},
  ymajorgrids=true,grid style={gray!20},
  symbolic x coords={Admission,Steward,Total},
  xtick=data,
  xticklabel style={font=\tiny,align=center},
  yticklabel style={font=\tiny},
  title={\sffamily\scriptsize Extended run (N=50)},
  title style={yshift=-1mm},
  legend style={font=\tiny,draw=none,at={(0.5,-0.22)},anchor=north,legend columns=2},
  mark size=2.1pt,
]
\addplot+[teal!80!black,thick,mark=*,mark options={fill=teal!65}] coordinates {(Admission,0.025) (Steward,304.193) (Total,305.309)};
\addplot+[orange!85!black,thick,mark=square*,mark options={fill=orange!70}] coordinates {(Admission,0.031) (Steward,349.348) (Total,865.522)};
\legend{P95,P99}
\end{semilogyaxis}
\end{tikzpicture}
\end{minipage}
\end{figure}

\begin{table}[H]
\TableLead{Table 13 --- OperationalBench Latency Summary.}
\raggedright
\TableMidFont
\setlength{\tabcolsep}{1pt}
\renewcommand{\arraystretch}{1.00}
\arrayrulecolor{black!55}
\rowcolors{2}{gray!8}{white}
\begin{tabular}{@{}L{0.21\textwidth}L{0.16\textwidth}L{0.16\textwidth}L{0.16\textwidth}L{0.23\textwidth}@{}}
\toprule
Run profile & Admission decision\\P50 / P95 / P99 & Steward apply\\P50 / P95 / P99 & Total scenario\\P50 / P95 / P99 & Reading note \\
\midrule
official paper run\\(\texttt{\seqsplit{20260627}}) & \texttt{\seqsplit{0.004 / 0.024 / 0.050 ms}} & \texttt{\seqsplit{33.181 / 302.424 / 541.920 ms}} & \texttt{\seqsplit{0.012 / 310.159 / 1088.094 ms}}\\(\texttt{\seqsplit{0.310 / 1.088 s}} at P95 / P99) & Baseline operational profile. \\
$N = 50$ contention run\\(\texttt{\seqsplit{20260627-extended}}) & \texttt{\seqsplit{0.003 / 0.025 / 0.031 ms}} & \texttt{\seqsplit{33.072 / 304.193 / 349.348 ms}} & \texttt{\seqsplit{0.010 / 305.309 / 865.522 ms}}\\(\texttt{\seqsplit{0.305 / 0.866 s}} at P95 / P99) & Higher contention remains concentrated in steward-mediated paths; no route-structure change is observed. \\
multi-seed stability note\\(3 paper-profile seeds; \texttt{\seqsplit{multi-seed-stability-20260627-20260629}}) & \texttt{\seqsplit{0.004--0.005 / 0.024--0.025 / 0.048--0.068 ms}} & \texttt{\seqsplit{33.181--37.998 / 302.274--317.522 / 323.190--541.920 ms}} & \texttt{\seqsplit{0.012--0.014 / 310.159--563.520 / 776.279--1088.094 ms}}\\(\texttt{\seqsplit{0.310--0.564 / 0.776--1.088 s}} at P95 / P99) & Scenario and route distributions remain identical; only the latency tail varies. \\
\bottomrule
\end{tabular}
\end{table}
\vspace{-3pt}

The latency columns use different denominators. The admission-decision span is the in-process broker decision span. The steward-apply span is computed only over rows that enter the steward-mediated path. The total-scenario span is computed over all scenario rows, including rows that do not invoke the steward. This is why the total-scenario median can be much smaller than the steward-apply median, while the P95 and P99 totals still reflect the steward-mediated tail.

These figures are percentile summaries rather than simple averages. In the official paper run, the total-scenario span is computed over 5,600 scenario rows, the admission-decision span over 3,600 measured admission spans, and the steward-apply span over 400 steward-routed rows. The extended $N = 50$ run scales these counts to 21,000, 13,500, and 1,500, respectively.

The key reading is therefore straightforward. Broker-decision overhead itself remains negligible at the paper scale, with P95 around \texttt{0.024--0.025 ms}. The visible tail sits in the steward-apply and recovery path, where P95 is about \texttt{\seqsplit{0.30 s}}, and in the end-to-end scenario totals, where P95 is also about \texttt{\seqsplit{0.31 s}} and P99 reaches about \texttt{\seqsplit{0.87--1.09 s}}. The $N = 50$ run makes this tail behavior more explicit, but it does not introduce a new route-distribution change or show a new recovery structure. It should be read as evidence that contention stress remains concentrated in governed recovery routing, \emph{not as a liveness or starvation-freedom proof}.

For completeness, the $N = 50$ queue-wait path remains near the timing floor in the current harness, at \texttt{0.001 / 0.002 / 0.004 ms} for P50 / P95 / P99. Validator timing is also near the floor in the official paper run, at \texttt{0.001 / 0.001 / 0.002 ms} for P50 / P95 / P99. These figures should be read as properties of the present benchmark harness, not as repository-scale validation costs. In particular, the current validator path is deliberately lightweight and should not be interpreted as the cost of full repository build, integration-test, or end-to-end validation.

OperationalBench therefore strengthens the paper's operational claim without widening its external-validity claim. It shows that ATM's fail-closed and blocked outcomes are observable recovery-routing events rather than a black-box rejection path: unsafe direct or parallel apply is closed, while intent, evidence, patch envelope, blocking reason, and recovery route remain available when a governed successor path exists.

\begin{figure}[!b]
{\small \noindent\textbf{Figure 8 --- AdmissionBench Ablation Degradation.} The grouped bars show how removing individual ATM layers changes the paper-profile benchmark behavior. The false-safe rows indicate safety regressions, while the lost-success rows indicate useful admissions that no longer complete. The figure supports the layer-necessity claim: ATM's behavior is produced by the interaction of virtual atoms, conflict identity, shared-surface judgment, CAS revalidation, and fail-closed recovery rather than by a single heuristic.\par}
\vspace{1pt}
\centering
\begin{tikzpicture}
\begin{axis}[
  width=0.90\textwidth,height=48mm,
  ymin=0,ymax=10,
  ymajorgrids=true,grid style={gray!20},
  ylabel={Rows affected},
  ylabel style={font=\scriptsize},
  yticklabel style={font=\tiny},
  symbolic x coords={Virtual atom,ConflictKey,CID,Shared surface,CAS,Fallback lock},
  xtick=data,
  xticklabel style={font=\tiny,align=center,rotate=15,anchor=east},
  ybar,bar width=5.5pt,enlarge x limits=0.13,
  legend style={font=\tiny,draw=none,at={(0.5,-0.18)},anchor=north,legend columns=2},
  nodes near coords,
  nodes near coords style={font=\tiny,/pgf/number format/fixed},
]
\addplot[fill=teal!70,draw=teal!85!black] coordinates {(Virtual atom,8) (ConflictKey,4) (CID,3) (Shared surface,3) (CAS,2) (Fallback lock,0)};
\addplot[fill=orange!75,draw=orange!80!black] coordinates {(Virtual atom,9) (ConflictKey,5) (CID,3) (Shared surface,3) (CAS,3) (Fallback lock,2)};
\legend{False-safe rows added,Success rows lost}
\end{axis}
\end{tikzpicture}
\end{figure}

\subsection{AdmissionBench Layer Necessity and Audit Bridge}

A reader who wishes to move from the benchmark summary toward mechanism necessity should first read Figure 8, which summarizes the ablation degradation profile, and then Table 14, which maps the resulting evidence to the remaining research questions. In short, Table 11 establishes the version-level division of roles, Table 12 reports the paper-facing AdmissionBench summary, Table 13 adds operational latency, Figure 8 visualizes layer necessity, and Table 14 summarizes the remaining boundaries.

For this reason, the paper deliberately retains v0.1 as the historical and audit starting point while using v0.2 as the main result. This separates {\char34}frozen and auditable{\char34} from {\char34}sufficient to enter the paper's main result{\char34} into two distinct layers, rather than replacing the whole evidence-evolution chain through a single version swap.

\subsection{Role-Separated Audit Evidence}

This study also completed a role-separated, blind-audited benchmark protocol. The four participating roles each operated from a disjoint information base. The Human first froze the contract. The Generator Agent then produced the cases, runner, and artifacts strictly from that frozen contract. The oracle side was completed independently without depending on ATM output, and it was finalized before the formal audit comparison. The Codex Auditor then conducted its review using the frozen commit \texttt{\seqsplit{3eec69a73a04112e2af8d3630c32138c37143eab}}, a blind export, and read-only replay as inputs. What these four roles share is a single anti-leakage discipline: no role may use a downstream artifact to back-fill an upstream input. The audit result under this protocol was \textbf{pass with caveats}: critical findings = 0, high findings = 0, determinism failure = false, and unexpected official artifact mutation during the audit = false.

This procedure preserves the most important anti-leakage property: ATM output may not be used to back-fill or modify the oracle. Whenever ATM output disagrees with the oracle, the divergence is recorded as a benchmark failure rather than rewritten into the expected answer. The audit retains two visible boundaries. First, the v0.1 blind export removed the per-mode \texttt{\seqsplit{expected}}, \texttt{\seqsplit{matchedExpectation}}, and \texttt{\seqsplit{falseSafeRegression}} fields, but retained \texttt{\seqsplit{groundTruth.safeToParallelize}} and \texttt{\seqsplit{groundTruth.validatorShouldCatch}}; this paper therefore describes the procedure as a \textbf{label-retained blind audit}, rather than making an unqualified strict double-blind claim. Second, the conflict-arbitration validator regenerates \texttt{\seqsplit{docs/reports/agr-conflict-arbitration-benchmark.md}}, a legacy path name, and this side-effect report is correctly excluded from the AdmissionBench freeze. If that report is later published, it should be handled by a separate conflict-arbitration-scoped task.

This audit pipeline also contributes one methodology evidence replay. The wrong-scope commit \texttt{\seqsplit{2088a791c45da8fda37d4258adfe393a21e689e9}} shows that the payload was stripped by the scope lock, whereas the correct-scope commit \texttt{\seqsplit{3eec69a73a04112e2af8d3630c32138c37143eab}} successfully landed the full five-file, 577-insertion payload. This evidence chain supports the governance-blocker classification and the role-separation methodology. It is not counted toward the main benchmark statistics.

\subsection{Results, Ablation, and Remaining Research Questions}

AdmissionBench is used in this manuscript in two layers. The first is the v0.1 baseline, which freezes the benchmark substrate and the blind-audit boundary. The second is the v0.2 paper profile, which supplies the Results, Ablation, and enforcement-timing summary cited in the body of this paper. OperationalBench then adds a narrower runtime-transparency layer over recovery routing and operational overhead. Table 14 records the current status without repeating the full appendix-level research-question grid.

\begin{table}[H]
\TableLead{Table 14 --- Compact Research-Question Status Summary.}
\raggedright
\TableMidFont
\setlength{\tabcolsep}{1pt}
\renewcommand{\arraystretch}{1.00}
\arrayrulecolor{black!55}
\rowcolors{2}{gray!8}{white}
\begin{tabular}{@{}L{0.18\textwidth}L{0.20\textwidth}L{0.36\textwidth}L{0.18\textwidth}@{}}
\toprule
Status & Research\\questions & Current evidence & Remaining boundary \\
\midrule
Paper-facing result & RQ1 admission safety; RQ2 concurrency preservation; RQ3 routing correctness; RQ4 layer necessity & AdmissionBench v0.1 and v0.2, POS2/B-12/BLOCK, route-label F1, ablation rows, and over-serialization rows & Larger cross-policy, cross-repository, and adapter-diverse evaluation \\
Enforcement and recovery timing & RQ5 enforcement timing; RQ8 operational overhead and recovery routing & Enforcement projection rows, field cases, OperationalBench official run, the $N = 50$ extended run, and multi-seed stability & Heavy validator-pressure variants, queue-wait studies, token-rework cost, and bounded-waiting analysis \\
Containment boundary & RQ6 adapter trust boundary; RQ7 hidden semantic reads & v0.2 adversarial rows, selected adversarial-adapter-containment cases, validators, CAS revalidation, fail-closed routing, and steward review & No adversarial-adapter soundness or completeness claim; dynamic read reconstruction remains future work \\
\bottomrule
\end{tabular}
\end{table}
\vspace{-3pt}

The full RQ table is provided in the appendix or supplementary material. The compact table is sufficient for the body because the main result should be read through Table 11, Table 12, and Table 13: version-level benchmark roles, paper-facing admission results, and operational recovery-routing overhead.

\section{Discussion}

\subsection{Why Adapter-Guided, Not AST-First}

ATM is adapter-guided for reasons of engineering pragmatism. A universal AST is theoretically attractive, because it appears to provide a unified semantic layer across languages. In practice, however, a multi-agent repository contains not only program code, but also JSON, Markdown, generated artifacts, registries, task ledgers, asset manifests, and domain-specific configuration. Requiring every governance action to first be converted into a single AST would undermine feasibility, both at adoption time and during long-term maintenance.

The adapter-guided approach instead lets each domain supply just enough conflict abstraction for the broker to act: a TypeScript adapter can expose function enclosures, a JSON adapter can expose record keys, a numeric adapter can expose scalar fields, and an atom-map adapter can expose edge or member keys. The broker does not need to understand the full semantics of every domain; it needs to know which mutations share a conflict surface and whether a deterministic merge path exists between them. This design trades global completeness for adoptability and auditability.

\subsection{When Adapter-Guided Fails}

Adapter-guided coordination can fail or degrade under seven conditions, each of which delimits the admission layer rather than refuting it.

\begin{enumerate}
\item \textbf{Adapter-capability gap.} When an adapter cannot identify the actual write surface, the broker can only fall back to a whole-file lock or a validator-level fallback.
\item \textbf{Enclosure gap.} When a patch region cannot be wrapped inside a stable syntactic unit, the broker cannot form a reliable virtual atom.
\item \textbf{Incomplete claim forwarding.} When an active intent is not correctly forwarded at admission time, late enforcement of the B-12 kind can still surface downstream.
\item \textbf{Irreducible human review.} Split suggestions can lower the review burden for a domain owner, but they cannot replace that owner's judgment about semantic partitioning.
\item \textbf{Adversarial or misdeclaring agent.} ATM's positive admission guarantee under the declared model relies on agents and adapters declaring \texttt{\seqsplit{readAtoms}}, \texttt{\seqsplit{writeAtoms}}, ConflictKeys, and shared surfaces conservatively. If an agent or its controlled adapter deliberately conceals a read dependency, under-reports a write surface, or declares an incorrect ConflictKey, the broker may issue an over-optimistic \texttt{\seqsplit{parallel-safe}} verdict, and the admission layer itself cannot detect that behavioral deviation at the pre-write stage. This creates both denial-of-service and silent-corruption risks.

A follow-up framework evidence commit on \texttt{\seqsplit{main}}, \texttt{\seqsplit{7a88af7d3db0be6d7e1b4c59f46706eabc5808a2}}, adds a reviewer-facing last-verified manifest and a small adversarial-adapter-containment evidence package. The package exercises five deterministic local cases covering malformed scope declarations, conservative shared-surface blocking, validator-backed dependency detection, and CAS-mismatch recovery. These cases support only a narrow containment claim: selected malformed or adversarial adapter-boundary conditions can be routed into fail-closed or replay-oriented recovery paths. They do not prove adversarial adapter soundness, completeness, or protection against every possible under-declaration or malicious declaration.

This paper does not claim that the stronger mitigations--adapter signing, capability audits, sandboxing, and broad adversarial injection--have been implemented in the current version; those remain future work.

\item \textbf{Very-large-scale repositories and the single-broker bottleneck.} Topology A assumes that a single broker process serializes every admission decision within one governance domain. In very large repositories, deployments with hundreds of concurrent agents, or extremely high-frequency batch waves, this broker can become a throughput bottleneck. Shard-based or federated broker designs therefore remain future work, as discussed under Topology D in §6.4.
\item \textbf{Non-textual or nondeterministic artifacts.} For binary assets, build outputs, dependency lockfiles with nondeterministic ordering, and generative outputs, the current adapter set does not provide full coverage. ATM's strategy for such artifacts is to fall back to a whole-file lock or to exclude them from the governance scope.
\end{enumerate}

What these seven failure modes share is a common shape: each names a condition under which the admission layer would otherwise have to over-parallelize on incomplete information. ATM's conservative admission posture is therefore to grant fine-grained admission only when the required evidence is available, and to use a fail-closed path when it is not. With respect to adversarial agents and scale pressure, this paper reserves auditable hooks and an evidence chain at the admission layer, rather than claiming that complete adversarial robustness is already in place.

\subsection{Open Questions and Future Work}

Subsequent research can be organized around four main axes.

First, cross-language atom identity still requires a semantic bridge across adapters, for example a shared CID between an API route, a schema contract, and a generated client/server pair.

Second, active-intent forwarding currently remains partly an apply-phase fallback. Future work should push it toward an admission-time default path where owner-map coverage and active-registry visibility allow, so that late enforcement is progressively reduced.

Third, a liveness proof needs to be developed together with a scheduling policy, so that conservative recovery routing after fail-closed containment does not cause starvation inside a high-contention repository.

Fourth, CID migration still requires a machine-verifiable version-negotiation and backfill path.

Several provider-shaped extensions follow naturally from the adjacent systems reviewed above and would compose with ATM's current admission layer without altering its deterministic core. A Tree-sitter-style or parser-backed atomization provider could supply language-specific structural anchors while preserving the adapter contract. A DeliveryLog-style read-set provider could reconstruct \texttt{\seqsplit{readAtoms}} from CLI traces, file-open events, LSP references, test traces, tool-call logs, or agent context citations. A CodeTeam-style design-sketch provider could compile file ownership, public interfaces, dependency edges, and scheduling constraints into ATM task contracts, ConflictKeys, or virtual-atom refinement hints. A structured \texttt{\seqsplit{AdmissionFailureReason}} provider could improve agent-side repair by returning the blocking layer, conflicting CID or ConflictKey, preserved patch envelope, selected recovery route, and validator transcript. A cross-agent review-signature provider could record independent AI review as evidence-closure metadata; the current prototype evidence path remains advisory and does not replace deterministic validators, tests, or human responsibility. Finally, notification-guided post-admission repair could consume broker \texttt{\seqsplit{block}}, \texttt{\seqsplit{serial}}, and fail-closed verdicts as repair triggers. These providers are future work unless explicitly identified as prototype evidence in Section 4 or Appendix A; each turns a strength of adjacent work into an extension point rather than a competitive overlap.

A federated broker spanning clones, hosts, or PR branches is deliberately deferred to Topology D in §6.4. This keeps the single-domain core claim separate from its distributed extensions.

Methodologically, this paper does not promote every engineering accompaniment into a main contribution, but several directions remain important to ATM's completeness. One is an \texttt{\seqsplit{atom-police}}-style governance aide, which flags insufficient atom coverage, owner-map drift, or validator gaps. Another is a Team Agents runtime layer that assigns local roles such as Implementer, Reviewer, Validator, Knowledge Scout, Evidence Collector, or Steward-facing writer while preserving ATM's authority boundaries. A third is the gradual institutionalization of provider-specific Agent SDKs and their skill and CLI wrappers, which absorb multi-vendor agent onboarding, knowledge accumulation, and tool-calling error suppression. What these directions share is an engineering-completeness role rather than a primary novelty claim; they are therefore better positioned as future work and implementation implications than as core contributions of the present paper.

A mature Team Agents runtime would need vendor-neutral provider contracts, permission-broker contracts, observability events, role-level provider selection, and adopter-repository integration configuration. Under such a design, different roles could use different model providers or execution surfaces, while write authority, evidence authority, closure authority, and shared-mutation admission remain governed by ATM. This paper does not claim that all OpenAI, Azure OpenAI, Claude Code, Gemini, Microsoft Foundry, local-model, or custom enterprise-gateway bridges are complete. Such bridges should be treated as future provider adapters unless a specific implementation surface or artifact is listed in Appendix A.

The governing rule is that a worker is not the governance owner. A worker or subagent may read, propose, validate, summarize, or review, but it may not bypass the broker, self-close a task, or write final evidence outside the appropriate governance path. Similarly, a Review Agent can provide an advisory or policy-gated review signature, but it does not replace deterministic validators, the neutral steward, or human responsibility. Should Team Agents be lifted into a method-level contribution in a later paper, the more defensible framing is a role-separated evaluation protocol rather than additional performance evidence. For now, this paper treats that route only as supporting material and does not combine it with the ATM benchmark main result.

ATM and CoAgent \textcolor{atmLinkBlue}{\hyperlink{Item.47}{[11]}} can also form a complementary pipeline. ATM can perform preventive admission at the code-region or artifact-region layer. When an intent is serialized but the downstream tool chain still produces side effects that could not be declared in advance, a CoAgent-style MTPO repair path can take over the reactive recovery. Similarly, when a future system needs to suppress contextual drift between agents earlier than admission, a context-drift synchronization layer can be stacked in front of the broker. This direction is not yet fully documented in the literature; for related failure-mode analyses, see Pan et al. \textcolor{atmLinkBlue}{\hyperlink{Item.42}{[6]}}. Future systems therefore need not choose between preventive, advisory, and synchronization layers; they can divide labor across layers.

SyncMind / SyncBench can serve as a future external replay source for this direction \textcolor{atmLinkBlue}{\hyperlink{Item.79}{[43]}}, but only after out-of-sync recovery instances have been converted into a pre-write intent workload inside a single governance domain. It is not a direct comparator for the present AdmissionBench, nor a source of denominators for this paper's current main result.

\subsection{Deployment Topologies and Future Work}

The admission mechanism described in this paper applies inside a single workspace or filesystem domain, under the single-governance-domain boundary stated in §3.7 and §4.7. This restriction is not a claim that larger topologies are impossible; it is the visibility boundary of the current broker process. We therefore sketch three concrete deployment topologies, ordered from those already supported by field evidence to natural extensions:

\begin{itemize}
\item \textbf{Topology A} --- a single workstation hosting multi-vendor co-writing; all field evidence in §4 belongs to this configuration.
\item \textbf{Topology B} --- a shared on-prem server hosting multi-vendor AI co-writing with remote human prompt input; this is a deployment-only extension that requires no change to ATM's core admission algorithm.
\item \textbf{Topology C} --- a local pre-push admission bridge, validated internally.
\end{itemize}

What these three topologies share is a common core assumption: a single broker process and a single registry own visibility over every concurrent intent within one filesystem domain. Their coverage of development-collaboration scenarios extends from small to large, but the admission algorithm itself remains invariant across them. A more distant distributed extension---Topology D---is treated separately as a distributed extension rather than as part of the current single-domain claim.

\subsubsection*{Topology A: Multi-Vendor Co-Writing on a Single Workstation}

All field evidence in §4 belongs to this topology. On a single developer workstation, multiple vendors' LLM agents write concurrently into the same worktree, the broker acts as a single in-process arbiter, and the registry is a single repo-local write-broker registry file. POS2, the close-orchestration evidence, and the B-12 cross-vendor end-to-end evidence were all completed under this topology.

\subsubsection*{Topology B: Multi-Vendor AI Co-Writing on a Shared On-Prem Server with Remote Human Prompt Input}

The natural extension of Topology A places multiple vendor LLM agents on a shared on-prem enterprise server. All AI inference and writing occur on a single filesystem on that server, while human users submit prompts, task orders, or edit suggestions remotely to the agents running there. The relative position of the broker and the agents is \textbf{identical to Topology A}: the broker is a local in-process arbiter, and every agent is a local process. The only difference is that the physical location of this {\char34}local{\char34} environment has moved from a developer workstation to a shared server.

ATM's admission software requires \textbf{no architectural change} in this topology, but three engineering elements outside the ATM scope must be provisioned alongside it. On-prem LLM inference must be available, for example through Anthropic Enterprise, vLLM, or self-hosted Ollama. A remote prompt-submission interface must be wired through SSH, a web UI, an IDE remote, or a chat API. Multi-tenant isolation must be enforced at the server boundary. What these three elements share is that they belong to the surrounding infrastructure of the on-prem AI center rather than to ATM itself. ATM's role is to serve as the admission-governance layer for concurrent writes by multi-vendor agents inside that center.

\subsubsection*{Topology C: A Local Pre-Push Admission Bridge}

The third topology extends broker admission from the moment of in-workspace write to the moment just before push. Exploiting the decoupling between the broker's mutation request and the proposal source, the pre-push stage constructs the remote delta obtained through fetch as a virtual remote-side mutation request and submits it together with the local branch delta from the merge base into the existing admission pipeline. The common ancestor, the format adapter, the composer, the steward apply, and the refinement loop are all identical to those in Topology A. The admission algorithm, the formal model, and the §3.6 format-adapter design require no modification; the new work is purely a Git-to-broker integration bridge.

The trigger point is placed at push rather than at commit. The former is the natural governance boundary at which local work is about to become shared work. The latter is a private local operation, and a per-commit trigger would slow the edit-and-test loop while adding noise on WIP commits. The independent contribution of this topology is therefore restricted to two areas: format-adapter merging of structured data, and the automation gap for AI agents in conflict-marker scenarios. For pure code merges, standard pull-and-rebase workflows already suffice, and this topology does not claim to replace them.

The corresponding bridge has completed internal validation, with artifacts archived. The validated deliverables are:

\begin{itemize}
\item the \texttt{\seqsplit{atm git admit}} CLI;
\item a pre-push hook;
\item steward dry-run and apply paths;
\item a push-fail fallback;
\item fixture coverage.
\end{itemize}

The MVP mechanics, phased implementation summary, Non-Goals boundary, and acceptance criteria are detailed in Appendix A.4 and in the plan document \texttt{\seqsplit{docs/ai\_atomic\_framework/git-boundary-admission/git-boundary-admission-plan.md}}.

\subsubsection*{Shared Assumptions and Division of Labor across the Three Topologies}

The three topologies share one core assumption: a single broker process and a single registry hold authoritative visibility into every concurrent intent within one filesystem domain. They differ only in the physical location of that domain---a personal workstation, a shared server, or the developer's local Git hook---and in the admission trigger point: live write for Topologies A and B, pre-push for Topology C. They are not mutually exclusive. A developer may co-write locally under Topology A, run the Topology C pre-push admission bridge before pushing, and then land the push on the primary environment of a Topology B shared on-prem server.

With respect to current evidence, the three topologies stand at different maturity layers:

\begin{itemize}
\item \textbf{Topology A} --- field evidence from §4 under same-worktree, same-governance-domain use.
\item \textbf{Topology C} --- internal validation, including a completed bridge and archived artifacts.
\item \textbf{Topology B} --- deployment-only extension, requiring no change to ATM's core admission algorithm.
\end{itemize}

\subsubsection*{A More Distant Extension: Topology D Cross-Machine Patch Synchronization}

If, beyond the Git PR mechanism, patches from multiple remote developers are synchronized directly to a central broker, the design enters the territory of a distributed broker. The basis for marking this case as {\char34}out of scope{\char34} deserves a specific account rather than a one-line invocation of CAP.

Why not simply extend the broker with Raft or Paxos. Classical leader-based consensus algorithms---Raft, Paxos variants, and Multi-Paxos---can in principle solve the problem of multiple broker replicas agreeing on admission order. That direction is technically feasible. For ATM, however, introducing distributed consensus is not merely a matter of swapping the broker's backing store; it requires taking on seven new engineering burdens at once:

\begin{enumerate}
\item \textbf{Federated active-registry replication.} The Active Registry of Definition 6 is currently a broker-local in-memory structure. A distributed version requires a replication protocol, a stale-read tolerance strategy, and read-your-writes semantics.
\item \textbf{Cross-machine ConflictKey equivalence.} If two machines hold adapters at different \texttt{\seqsplit{schema\_version}} values, admission comparisons must first resolve schema reconciliation. §3.7 and §6.3 mark CID schema migration as an open problem.
\item \textbf{Lease and fencing token mechanism.} A distributed broker requires leases and fencing tokens to prevent split-brain scenarios in which two leaders simultaneously admit overlapping intents.
\item \textbf{Distributed apply ordering on the steward side.} The neutral steward can no longer apply only to a single worktree. It must handle partial apply, cross-node rollback, and bounded staleness.
\item \textbf{Liveness and starvation under network partition.} §3.7 and §6.3 mark liveness for conservative admission as open under a single broker, and network partitions make the question harder still.
\item \textbf{Distributed audit of the evidence chain.} A cross-node verdict log requires causal ordering and replayability, rather than a purely local evidence trail.
\item \textbf{Operational complexity.} Quorum loss, network partitions, and stale replica recovery each require their own runbook and fail-closed behavior specification.
\end{enumerate}

The compromise adopted in this paper. Each of the seven items above is an independent research subtopic, and the single-domain core claim of this paper deliberately does not assert that they have been resolved. In practical deployment, Topology C, the pre-push admission bridge, covers an important subset of cross-machine collaboration needs: remote developers can perform Topology A co-writing inside their own worktrees, complete admission with Topology C before pushing, and then rely on the Git, PR, and merge substrate \textcolor{atmLinkBlue}{\hyperlink{Item.45}{[9, 22, 29]}} for cross-clone convergence.

Topology D is therefore positioned as a future direction for cases in which the combination of Topology A, Topology C, and the Git PR workflow is insufficient for high-frequency cross-machine patch synchronization. It is not a deferred deliverable of this paper. The direction is technically feasible, but its engineering scale is significantly larger than that of Topologies A, B, or C. Within the scope of this paper, it is treated only as a research extension direction, to be unfolded separately in future work on federated brokers, bounded staleness, and admission-time consensus protocols.

\section{Conclusion}

This paper argues that what multi-agent software engineering still lacks is not a stronger generator or a later repairer, but an admission layer situated before any governed shared write: a layer that binds task intent, repository scope, admission verdicts, validators, and evidence obligations into a single governance path. AI-Atomic-Framework (ATM) answers this need by translating the otherwise opaque risk of shared writes into locatable, adjudicable, and replayable governance units through adapter-guided semantic atoms, the atom map, virtual atoms, and the CID broker. When the evidence is sufficient, bounded-region admission can proceed. When the evidence is insufficient, or when a conflict cannot be conservatively adjudicated, the broker fails closed. When composition or refinement is required, the write path is rerouted through a composer-routed merge or a refinement loop.

The purpose of the evidence chain in this paper is not to claim that ATM has achieved large-scale comparative superiority, but to show that this governance path is deployable and auditable inside a single controlled filesystem, worktree, or service domain. Deterministic fixtures, self-hosting records, adopter-side observations, same-file boundary cases, and the AdmissionBench paper profile frozen at 20 scenarios and 42 comparisons together support the feasibility, auditability, and bounded recoverability of single-domain pre-write admission. They do not establish a comprehensive cross-organizational or cross-topology advantage claim. The remaining boundary is explicit: distributed governance across clones or PR branches remains future work; end-to-end semantic guarantees under a misdeclaring adapter remain future work; and larger-scale comparative evaluation remains future work.

If shared repository mutation is revealed only after the write---through a Git merge, a failing test, or human code review---the signal that returns to the system is typically late, coarse, and difficult to replay. The conclusion of this paper is therefore direct. For multi-agent co-writing, pre-write admission should be treated as an independent first-class governance problem, not as a byproduct of a downstream version-control flow. ATM's contribution is to condense this problem into a specification-to-evidence governance substrate that is implementable, traceable, and explicit about its own boundary.

\section*{Acknowledgements}

The author used large language model assistants during both framework development and manuscript preparation, including for language editing, structural feedback, literature discovery, code-generation assistance, implementation review, and multi-agent assistant orchestration during development. This use was intentionally human-in-the-loop: assistants proposed, drafted, checked, or compared alternatives, but did not hold final authority over design, evidence, benchmark acceptance, or claims.

The author made and reviewed the conceptual and architectural decisions of ATM, determined which evidence to include or exclude, calibrated the claim boundaries, and interpreted the benchmark and field evidence. The author assumes full responsibility for the manuscript. A detailed transparency statement---including vendor channels, role separation, human decision points, audit boundaries, and explicit non-claims regarding evidence---is provided in Appendix B.

\section*{Appendix A. Evidence Artifact Map}

\subsection*{A.1 Evidence Artifact Index and Release Alignment}

This appendix lists the recommended entry points for paper-citable evidence. The specific artifact names and commits should always defer to the actual files in the repository. To avoid duplicating supplementary indices, Appendix A retains only family-level anchors, artifact roots, and access channels.

\Needspace{5\baselineskip}\TableLead{Table A.1 --- Compact Evidence Artifact Index.}

\begingroup
\tiny
\setlength{\tabcolsep}{1pt}
\arrayrulecolor{black!55}
\rowcolors{2}{gray!8}{white}
\begin{longtable}{@{}L{0.17\textwidth}L{0.22\textwidth}L{0.39\textwidth}L{0.17\textwidth}@{}}
\toprule
Evidence family & Role in paper & Canonical anchor or artifact root & Access channel \\
\midrule
AdmissionBench v0.1 baseline + v0.2 paper profile & Paper-facing admission-result evidence, ablation, policy views, and audit lineage & \texttt{\seqsplit{main@ab8753b7daf0a3c4cd8b4483fe24d519ff2590bd}} -> \texttt{\seqsplit{artifacts/generated/atm-admission-bench/20260625-paper/}}; baseline under \texttt{\seqsplit{artifacts/generated/atm-admission-bench/20260625/}} & Public ATM repository and supplementary release \\
OperationalBench official + supplementary runs & Runtime transparency for admission overhead, recovery routing, $N = 50$ contention, and multi-seed stability & \texttt{\seqsplit{main@c0250009a53b28e887344e71ea675637c97290b0}} -> \texttt{\seqsplit{artifacts/generated/atm-operational-bench/20260627/}}, \texttt{\seqsplit{20260627-extended/}}, and \texttt{\seqsplit{multi-seed-stability-20260627-20260629.\{md,json\}}} & Public ATM repository and supplementary release \\
Public-source and structured-artifact evidence & Framework-mainline support, FastAPI public-source snapshot governance, structured artifact admission, and dual-live public-source conflict demonstration & Framework evidence freeze \texttt{\seqsplit{main@f57dbfe0bdfdf9f939e35400ec346501f4ccb2f3}}; Phase A/B/C artifact roots listed in supplementary artifact index & Public or paper-safe supplementary artifacts, depending on host-repository content \\
Field-collision evidence & POS2 same-file admission, B-12 apply-phase enforcement, BLOCK fail-closed split suggestion, and close-orchestration evidence & \texttt{\seqsplit{docs/ai\_atomic\_framework/broker-collision-evidence/}} and de-identified supplementary packets & Mixed: public summaries and private source packets on request \\
Adoption and self-hosting evidence & npc-brain adopter recoverability, self-hosting forensics, Wave Mode replay, CID stability, and governance incident summaries & Sections 4.2--4.6, Appendix A.5, and supplementary de-identified evidence packets & Mixed: public summaries, de-identified supplementary release, private ledgers on request \\
Evidence-closure and adapter-trust prototypes & Cross-agent review signatures, reviewer-facing verification manifest, and selected adversarial-adapter-containment evidence & \texttt{\seqsplit{artifacts/generated/cross-agent-review-signature/20260628/}}; \texttt{\seqsplit{artifacts/verification/last-verified.json}}; \texttt{\seqsplit{artifacts/adversarial-adapter-containment/20260628/}} & Public ATM repository \\
\bottomrule
\end{longtable}
\endgroup

The full artifact-level index, including individual \texttt{\seqsplit{summary.json}}, \texttt{\seqsplit{results.jsonl}}, command logs, and hash manifests, is provided in the supplementary artifact index. The body and Appendix A retain only the anchors needed to cite the paper's claims.

\subsection*{A.2 Implementation and Commit Provenance}

The primary implementation families of ATM include broker decision, virtual-atom refinement (legacy implementation name: AGR), neutral steward, freeze, patch-envelope, conflict-matrix, format adapters, Wave Mode, and CID verification. The body of the paper deliberately avoids listing large numbers of task IDs. Instead, the appendix keeps only a compact claim-to-verification-route map in §A.4.1, while the full source-path-level verification map is maintained in the repository and supplementary material. This avoids stacking three overlapping indices: the artifact index in A.1, an implementation-status table, and the verification map in A.4.1.

\subsection*{A.3 CID Schema Migration Candidate Paths}

CID schema migration can be approached along three paths.

\begin{enumerate}
\item \textbf{Flag-day migration.} A migration window is locked at the repository level, all active intents are cleared, and CIDs are recomputed afterward. Trade-off: simple but disruptive.
\item \textbf{Dual-read with single-write.} The broker recognizes both v1 and v2 CIDs simultaneously, but new intents are written only in v2. Trade-off: smooth but implementation-heavy.
\item \textbf{Compatibility map.} A signed migration table declares the equivalence between old and new CIDs. Trade-off: traceable but dependent on trusting the migration table.
\end{enumerate}

This paper does not yet select a final option.

\subsection*{A.4 Implementation Verification Map and Topology C Bridge Detail}

This appendix summarizes the implemented capability groups, their verification routes, and their source-code citation anchor in the open-source repository (\texttt{\seqsplit{https://github.com/eaglhuang/AI-Atomic-Framework}}). Detailed source paths, per-command outputs, artifact hashes, and row-level verification records are maintained in the repository and supplementary verification map rather than repeated in the paper body.

\begin{quote}
\textbf{Release-tag pinning rule.} The framework snapshot used for source-code cross-checking in this paper is release tag \texttt{\seqsplit{v0.9.0-alpha.1}} (commit \texttt{\seqsplit{0b31aa8683b44b3a78206132a0bf90a0fde73d1c}}). Readers should treat this tag as the primary source-code citation point in order to avoid line-number drift introduced by the evolution of the \texttt{\seqsplit{main}} branch.
\end{quote}

\subsubsection*{A.4.1 Verification Map}

\begin{table}[H]
\TableLead{Table A.2 --- Compact Verification Map.}
\raggedright
\scriptsize
\setlength{\tabcolsep}{3pt}
\arrayrulecolor{black!55}
\rowcolors{2}{gray!8}{white}
\begin{tabular}{@{}L{0.22\textwidth}L{0.25\textwidth}L{0.47\textwidth}@{}}
\toprule
Verification group & What it verifies & Representative command or artifact \\
\midrule
Core broker and admission path & Broker decision logic, seven-layer gate, ConflictKey behavior, CAS base-hash guarded apply, steward routing & Broker decision, CAS, and steward test commands listed in the supplementary verification map \\
Atom identity and adapter support & Atom-to-CID validation, TypeScript/Python reference paths, format adapters, virtual-atom refinement path & Atom-to-CID and AGR validation commands listed in the supplementary verification map \\
AdmissionBench result chain & v0.1 baseline, blind export, v0.2 paper profile, report rendering, and audit replay & AdmissionBench smoke, paper-profile, and report-generation commands listed in the supplementary verification map \\
OperationalBench result chain & Official runtime-overhead run, $N = 50$ extended run, multi-seed stability, and validation of operational artifacts & OperationalBench paper-profile, extended, and multi-seed commands listed in the supplementary verification notes \\
Public-source and structured-artifact tracks & FastAPI public-source snapshot governance, structured non-code artifact admission, and dual-live public-source conflict evidence & Phase A/B/C artifact roots and paper-safe summaries listed in Table A.1 and the supplementary artifact index \\
Evidence-closure and adapter-trust prototypes & Cross-agent review-signature evidence, reviewer-facing last-verified manifest, and adversarial adapter containment package & \texttt{\seqsplit{artifacts/generated/cross-agent-review-signature/20260628/}}; \texttt{\seqsplit{artifacts/verification/last-verified.json}}; \texttt{\seqsplit{artifacts/adversarial-adapter-containment/20260628/}} \\
Field and adoption evidence & POS2, B-12, BLOCK, close-orchestration, npc-brain cohort, self-hosting forensics, Wave Mode, and CID stability & De-identified supplementary packets and paper-safe summaries; raw private ledgers available on request subject to Appendix A access rules \\
Topology C pre-push bridge & Local pre-push admission bridge, dry-run path, push-fail fallback, and Git-to-broker \texttt{\seqsplit{MutationRequest}} construction & Dry-run bridge command, Topology C plan document, and archived bridge artifacts \\
\bottomrule
\end{tabular}
\end{table}

The compact verification map is intended as a navigation layer. Detailed source paths, per-command outputs, artifact hashes, and row-level verification records are maintained in the repository and supplementary verification map rather than repeated in the paper body.

\subsubsection*{A.4.2 Topology C MVP Mechanics and Implementation Stages}

The Topology C bridge command executes the following sequence before a Git push. All steps correspond to existing components of the §3.4--§3.6 admission pipeline; no new algorithm is introduced.

\begin{enumerate}
\item Fetch remote metadata and compute the local merge base against the tracked branch.
\item Construct the local and remote \texttt{\seqsplit{MutationRequest}} sides from the local and remote diffs.
\item For structured files, parse ConflictKeys with the existing format adapters (§3.6).
\item For files without a structured adapter, fall back to text-range ConflictKeys.
\item Submit both sides to broker admission (§3.4--§3.5).
\item If admission passes, allow the push; if admission is blocked, report the conflict and suggest a rebase, merge, or steward path.
\item If the verdict is composer-routed, produce a deterministic merge plan and optionally apply it via the steward to the working tree, with no auto-commit by default.
\end{enumerate}

The Topology C bridge narrows the gap between ATM admission and ordinary Git collaboration. Before push admission, the bridge fetches the remote branch by default, computes the local merge base, constructs the local-versus-remote delta as local and remote \texttt{\seqsplit{MutationRequest}} sides, and submits the synthesized intents to the existing broker pipeline. If the remote has advanced, the bridge does not assume that the local intent is invalid. It reports whether the local change can still be admitted, should be routed to deterministic composer / steward handling, or should first be rebased, split, or rearbitrated against the refreshed remote base.

This behavior mirrors Git's conservative non-fast-forward boundary while adding ATM's adapter-guided admission layer before the push. The companion push-failure recovery path reruns admission after fetch and classifies the recovery mode as steward apply, rebase, retry-after-no-op, or inspect. Rebase is therefore a supported recovery path, not a universal requirement: some cases can be replayed or queued without semantic rebase, while overlapping or insufficiently evidenced cases remain fail-closed to direct apply.

\subsubsection*{A.4.3 Topology C Implementation Stages (Completed on 2026-06-26)}

\Needspace{5\baselineskip}\TableLead{Table A.3 --- Topology C Implementation Stages.}

\begingroup
\tiny
\setlength{\tabcolsep}{1pt}
\arrayrulecolor{black!55}
\rowcolors{2}{gray!8}{white}
\begin{longtable}{@{}L{0.22\textwidth}L{0.25\textwidth}L{0.47\textwidth}@{}}
\toprule
Stage & Internal work package & Purpose \\
\midrule
\texttt{\seqsplit{S0}} & Architecture lock-in & Lock down the contract and the architecture. \\
\texttt{\seqsplit{S1}} & Ingestion + adapter bridge & Git-diff ingestion, adapter bridge, and CLI admission. \\
\texttt{\seqsplit{S2}} & Hook + steward path & Hook installation, evidence capture, and steward dry-run and apply. \\
\texttt{\seqsplit{S3}} & Coverage + fallback & Fixture coverage, push-fail fallback, and policy and audit. \\
\texttt{\seqsplit{S4}} & Docs + self-hosting evidence & Documentation, self-hosting evidence, and paper-ready evidence. \\
\bottomrule
\end{longtable}
\endgroup

\subsubsection*{A.4.4 Artifact Manifest Snapshot}

A compact public artifact snapshot is retained below for citation convenience. The full public artifact manifest, including row-level paths, command outputs, and hash manifests, is maintained in the supplementary artifact index.

\begin{table}[H]
\TableLead{Table A.4 --- Public Artifact Manifest Snapshot.}
\raggedright
\tiny
\setlength{\tabcolsep}{2pt}
\arrayrulecolor{black!55}
\rowcolors{2}{gray!8}{white}
\begin{tabular}{@{}L{0.17\textwidth}L{0.22\textwidth}L{0.39\textwidth}L{0.17\textwidth}@{}}
\toprule
Artifact & Commit or hash anchor & Role & Access status \\
\midrule
\texttt{\seqsplit{artifacts/generated/atm-admission-bench/20260625/}} & Generator commit \texttt{\seqsplit{3eec69a73a04112e2af8d3630c32138c37143eab}} & v0.1 baseline smoke artifacts & Public release bundle \\
\texttt{\seqsplit{artifacts/blind-bench/20260625/}} & Generator commit \texttt{\seqsplit{3eec69a73a04112e2af8d3630c32138c37143eab}} & Label-retained blind-audit intake & Public release bundle \\
\texttt{\seqsplit{artifacts/generated/atm-admission-bench/20260625-paper/}} & ATM evidence commit \texttt{\seqsplit{ab8753b7daf0a3c4cd8b4483fe24d519ff2590bd}} & v0.2 paper-profile result bundle & Public release bundle \\
\texttt{\seqsplit{artifacts/generated/atm-operational-bench/20260627/}}, \texttt{\seqsplit{artifacts/generated/atm-operational-bench/20260627-extended/}}, and \texttt{\seqsplit{artifacts/generated/atm-operational-bench/multi-seed-stability-20260627-20260629.\{md,json\}}} & ATM evidence commit \texttt{\seqsplit{c0250009a53b28e887344e71ea675637c97290b0}} & OperationalBench official run, higher-contention supplement, and multi-seed stability evidence & Public release bundle \\
\texttt{\seqsplit{docs/reviews/ATM-AdmissionBench-audit.md}} and \texttt{\seqsplit{artifacts/audit/audit-findings.json}} & Frozen audit-input commit \texttt{\seqsplit{3eec69a73a04112e2af8d3630c32138c37143eab}} & Role-separated audit evidence & Public release bundle \\
\texttt{\seqsplit{docs/ai\_atomic\_framework/broker-collision-evidence/}} & 3KLife planning-repository evidence paths & POS2, B-12, and BLOCK field evidence & De-identified or path-level summary in supplementary release \\
npc-brain adoption notes and ledgers & Retained by adopter-side planning records & Cohort recovery and validator-catch reconstruction & Summary-only in paper; raw event ledger is not redistributed by default \\
\bottomrule
\end{tabular}
\end{table}

\subsubsection*{A.4.5 Topology C Non-Goals (MVP)}

\begin{itemize}
\item No per-commit mandatory gate: per-commit overhead is unnecessary because push is the governance boundary.
\item No background daemon or cache: the first version is built around a synchronous hook; caching is listed as a later optimization.
\item No cross-machine broker RPC: this belongs to Topology D in §6.4, requires distributed consensus, and is out of scope for this paper.
\item No fully automated rebase engine: composer-routed merge runs only on bounded-disjoint structured files.
\item The steward apply path does not auto-commit by default: the final commit decision remains with the human or an upper-layer agent.
\item The bridge does not claim to solve every semantic-layer Git conflict; for pure code merges, standard pull-rebase workflow already suffices.
\item The bridge does not resolve the race between two remote developers submitting PRs simultaneously; that case continues to be governed by Git's non-fast-forward rule.
\end{itemize}

\subsubsection*{A.4.6 Topology C Acceptance Conditions}

The bridge is considered acceptance-ready when each of the following conditions holds:

\begin{itemize}
\item The ATM Git-admission CLI can evaluate the local-versus-remote delta before push.
\item The pre-push hook invokes that command and produces clear operator output.
\item Same-file, bounded-disjoint structured edits are routed through the existing broker and composer semantics.
\item Overlapping or insufficiently evidenced cases take a fail-closed path before push and emit auditable evidence.
\item The post-push-fail fallback can explain and replay the same admission path.
\item The evidence can be archived to support the paper's claims without requiring a new envelope schema.
\end{itemize}

The complete contract and design record are kept in \texttt{\seqsplit{docs/ai\_atomic\_framework/git-boundary-admission/git-boundary-admission-plan.md}}.

\subsection*{A.5 Condensed Incident Table}

The table below condenses the incident evidence of \texttt{\seqsplit{TASK-CID-0040}} through \texttt{\seqsplit{TASK-CID-0045}} into three governance-failure categories. The detailed closure packets, repair commits, and forensic-report paths remain authoritative in Appendix A.1 and in the original incident archive.

\Needspace{5\baselineskip}\TableLead{Table A.5 --- Condensed Incident Evidence.}

\begingroup
\tiny
\setlength{\tabcolsep}{2pt}
\arrayrulecolor{black!55}
\rowcolors{2}{gray!8}{white}
\begin{longtable}{@{}L{0.22\textwidth}L{0.25\textwidth}L{0.47\textwidth}@{}}
\toprule
Incident cluster & Mechanism exercised & Outcome and interpretation \\
\midrule
\texttt{\seqsplit{TASK-CID-0040}} claim-displaced-by-import & Claim-ledger consistency and in-progress claim collision detection. & The import flow once overwrote an in-progress claim; the divergence was subsequently detected via an event-ledger mismatch and repaired. The case illustrates that governance state must carry a replayable claim-and-repair trace. \\
\texttt{\seqsplit{TASK-CID-0041}} out-of-scope delivery requiring waiver & Scope gate, closure-packet waiver, and late-enforcement traceability. & Admission did not fully intercept the scope drift at write time; instead the drift was later registered by a closure packet with an explicit waiver. The case supports the paper's honest disclosure of the enforcement boundary, rather than packaging every incident as a clean positive outcome. \\
\texttt{\seqsplit{TASK-CID-0043}} / \texttt{\seqsplit{0044}} / \texttt{\seqsplit{0045}} plan-mirror sync failures & Sole-serialization invariant and planning-mirror to target-ledger closeout consistency. & Closeout drift appeared between the planning side and the target ledger, requiring repair commits to backfill the closure packets. The case illustrates that the broker and steward together must be the sole formal closeout authority. \\
\bottomrule
\end{longtable}
\endgroup

\section*{Appendix B. Authoring Transparency Statement}

\subsection*{B.1 Use of AI-Assisted Writing Tools}

This manuscript was prepared as an instance of the multi-vendor LLM co-synthesis workflow described in this paper. During the three-week npc-brain alignment window from 2026-05-19 to 2026-06-07, the 3KLife project accumulated 320 governed commits, jointly written by 15 LLM agents from different vendor and editor channels. The detailed channel identifiers are retained in the evidence archive rather than expanded in the body. Evidence of these manuscript-side admissions is captured within the self-hosting forensics window reported in Section 4.2 and the multi-vendor self-hosting and Wave Mode discussion in Section 4.6; the manuscript-preparation workload is not separated into a distinct experimental cohort.

\subsection*{B.2 Division of Responsibility}

\begin{itemize}
\item \textbf{AI assistants (multiple vendors).} Drafting, prose refinement, reorganization, citation formatting, candidate literature discovery, structural critique, table layout, and consistency checks across sections.
\item \textbf{Human author.} Research direction, framework design, atom / atom map / CID / virtual atom model, broker and steward architecture, implementation decisions, evidence interpretation, all technical claims, and final acceptance of every section and table.
\end{itemize}

Every paragraph, table, definition, and conclusion in this manuscript was reviewed and accepted by the human author. AI-assistant outputs that conflicted with the author's technical judgment were revised or discarded prior to inclusion.

\subsection*{B.3 Non-Claims}

The use of AI assistants in manuscript preparation is disclosed for transparency and is \textbf{not counted as additional experimental evidence for the framework's effectiveness}. ATM's evaluation (Section 4) stands on its archived fixture runs, field collision artifacts, external adopter records, and self-hosting governance metrics, independent of the authoring process. In particular: (i) the 12-scenario fixture design and the 3 archived deterministic MVP runs (B-02, B-08, B-13) are not influenced by manuscript-side admissions; (ii) the POS2, B-12, and BLOCK same-file collision evidence in Section 4.5 originates from framework-side and adopter-side workloads, not from manuscript drafting; (iii) the npc-brain adoption cohort (N = 37) reported in Section 4.3 is an external adopter sample and is not co-mingled with manuscript-preparation activity.

Prototype cross-agent review signatures, where present, are treated as advisory evidence-closure artifacts. They do not replace deterministic validators, human responsibility, or the archived benchmark evidence used for the paper's main claims, and they are not presented here as proof of semantic correctness.

\subsection*{B.4 Reproducibility Note}

Readers wishing to inspect the manuscript-side admission evidence should use three public ATM repository anchors. Source-mechanism paths are pinned to commit \texttt{\seqsplit{0b31aa8683b44b3a78206132a0bf90a0fde73d1c}} (release tag \texttt{\seqsplit{v0.9.0-alpha.1}}). The AdmissionBench v0.1 baseline and v0.2 paper-profile artifacts are pinned to the later evidence-landing commit \texttt{\seqsplit{ab8753b7daf0a3c4cd8b4483fe24d519ff2590bd}}, while the OperationalBench official and supplementary artifacts are pinned to \texttt{\seqsplit{main@c0250009a53b28e887344e71ea675637c97290b0}}, under the paths listed in Appendix A.1 and Appendix C. Internal task ledgers, vendor account identifiers, and adopter-side records are not redistributed verbatim; de-identified evidence chains, verdict logs, and validator traces are made available on request, subject to the access conditions stated in Appendix A.1.

\section*{Appendix C. Supplementary Data Release and DOI Reservation}

\begin{quote}
\textbf{Pre-submission notice on placeholder identifiers.} The Zenodo DOI (\texttt{\seqsplit{10.5281/zenodo.XXXXXXX}}) and arXiv id are placeholders until release. Until the issued identifiers are available, cite the frozen Git anchors in this appendix instead.
\end{quote}

\subsection*{C.1 Purpose}

This appendix consolidates the supplementary release plan, DOI reservation status, version correspondence, and citation convention. It identifies which anchors are frozen, which become effective only after release, and which remain pending future backfill.

\subsection*{C.2 Release Anchor Summary}

This manuscript separates source, benchmark-evidence, and provenance anchors. The source release tag is the stable code citation point; benchmark evidence landed later on \texttt{\seqsplit{main}} so reviewers can inspect the generated artifact bundles directly.

\Needspace{20\baselineskip}\TableLead{Table C.1 --- Release Anchor Summary.}

\begingroup
\tiny
\setlength{\tabcolsep}{2pt}
\arrayrulecolor{black!55}
\rowcolors{2}{gray!8}{white}
\begin{longtable}{@{}L{0.18\textwidth}L{0.27\textwidth}L{0.18\textwidth}L{0.31\textwidth}@{}}
\toprule
Anchor category & Identifier & Status & Use \\
\midrule
Source release tag (ATM repository) & \texttt{\seqsplit{v0.9.0-alpha.1}} & \textbf{published} & Frozen reference for the source paths and line numbers cited by this manuscript; points to commit \texttt{\seqsplit{0b31aa8683b44b3a78206132a0bf90a0fde73d1c}}. \\
Source release commit hash & \texttt{\seqsplit{0b31aa8683b44b3a78206132a0bf90a0fde73d1c}} & \textbf{published} & The immutable commit that the tag points to, serving as the basis for archive integrity. \\
AdmissionBench evidence landing commit & \texttt{\seqsplit{main@ab8753b7daf0a3c4cd8b4483fe24d519ff2590bd}} & \textbf{published} & Later public commit that first lands the v0.1 baseline and v0.2 paper-profile artifact bundles under \texttt{\seqsplit{artifacts/generated/atm-admission-bench/20260625/}}, \texttt{\seqsplit{artifacts/blind-bench/20260625/}}, and \texttt{\seqsplit{artifacts/generated/atm-admission-bench/20260625-paper/}}; use this anchor for benchmark-number verification. \\
AdmissionBench generator provenance commit & \texttt{\seqsplit{3eec69a73a04112e2af8d3630c32138c37143eab}} & \textbf{published; manifest-recorded} & \texttt{\seqsplit{baseCommit}} and \texttt{\seqsplit{generatorCommit}} recorded in the AdmissionBench manifests; use as generator provenance, not as the public artifact-landing citation point. \\
OperationalBench evidence landing commit & \texttt{\seqsplit{main@c0250009a53b28e887344e71ea675637c97290b0}} & \textbf{published} & Later public commit that lands the OperationalBench official 2026-06-27 paper run (\texttt{\seqsplit{20260627}}), the extended 2026-06-27 higher-contention supplement (\texttt{\seqsplit{20260627-extended}}), and the multi-seed stability artifacts under \texttt{\seqsplit{artifacts/generated/atm-operational-bench/}}; use this anchor for operational-overhead and recovery-routing verification. \\
Supplementary data archive & Zenodo deposit (preparing) & \textbf{pending DOI}; released in sync with arXiv submission & De-identified evidence chains, verdict logs, validator traces, AdmissionBench baseline artifacts, AdmissionBench paper-profile artifacts, and the 12-scenario fixture replay bundle. \\
Supplementary data DOI & reserved (pending Zenodo issuance) & \textbf{placeholder}: \texttt{\seqsplit{10.5281/zenodo.XXXXXXX}} & Replace with the issued DOI after release; until then, cite the frozen Git anchors. \\
Manuscript itself & arXiv submission (pending) & \textbf{pending arXiv id} & Replace with the issued arXiv identifier after release. \\
\bottomrule
\end{longtable}
\endgroup

Version-correspondence convention. This manuscript v3.1 corresponds to source release \texttt{\seqsplit{v0.9.0-alpha.1}}. Later \texttt{\seqsplit{main}}-branch evolution does not update the manuscript automatically: any source-version change must be issued through a new paper revision, with a new row added here so that the citation chain remains explicit.

\subsection*{C.3 Supplementary Data Release Contents}

The supplementary archive groups evidence into four classes: benchmark artifacts (AdmissionBench and OperationalBench runs, summaries, tables, manifests, and hash manifests); mechanism replay artifacts (fixtures, runner outputs, broker decision traces, and structured-artifact admission results); field, adoption, and self-hosting evidence (de-identified POS2/B-12/BLOCK packets, close-orchestration evidence, npc-brain cohort summaries, Wave Mode replay, and CID stability evidence); and methodology or prototype evidence (role-separated audit materials, review-signature prototypes, reviewer-facing verification manifests, and adversarial-adapter-containment artifacts). Private task ledgers, personal names, vendor account identifiers, adopter-internal project details, and raw commercial project material are not redistributed; where release is not possible, the archive provides paper-safe summaries, de-identified traces, or access-on-request instructions.

\subsection*{C.4 Access Conditions}

Benchmark and mechanism-replay artifacts are public through the ATM repository and the supplementary archive. Field, adoption, self-hosting, and methodology evidence is released in de-identified form when it contains adopter-side or internal-project material. Original task ledgers, closure packets, and internal project records remain in the private repository \texttt{\seqsplit{eaglhuang/3KLife}}; review access may be granted on request for a named purpose, without a commitment to automatic authorization or indefinite retention.

\subsection*{C.5 Citation Convention}

When citing the evidence of this manuscript, use only the already-frozen anchors below.

\begin{itemize}
\item For source-mechanism claims, cite the ATM source release \texttt{\seqsplit{v0.9.0-alpha.1}} and commit \texttt{\seqsplit{0b31aa8683b44b3a78206132a0bf90a0fde73d1c}}.
\item For AdmissionBench evidence, cite \texttt{\seqsplit{main@ab8753b7daf0a3c4cd8b4483fe24d519ff2590bd}} together with the \texttt{\seqsplit{artifacts/generated/atm-admission-bench/20260625-paper/}} bundle and the matching \texttt{\seqsplit{20260625}} baseline artifacts.
\item For OperationalBench evidence, cite \texttt{\seqsplit{main@c0250009a53b28e887344e71ea675637c97290b0}} together with the \texttt{\seqsplit{20260627}}, \texttt{\seqsplit{20260627-extended}}, and \texttt{\seqsplit{multi-seed-stability-20260627-20260629}} artifact bundles.
\end{itemize}

The DOI of the supplementary data and the arXiv id of the manuscript remain placeholders and should not be used in formal citations.

After the synchronized arXiv and Zenodo release, this appendix should be updated with the issued DOI and arXiv id as a citation-anchor update rather than a change to the paper's technical content.

\section*{References}

\begin{enumerate}
\item \label{ref:1} Pugachev, Sergey. 2025. {\char34}CodeCRDT: Observation-Driven Coordination for Multi-Agent LLM Code Generation.{\char34} arXiv:2510.18893 [cs.DC]. \url{https://doi.org/10.48550/arXiv.2510.18893}.
\item \label{ref:2} Acharya, Vivek. 2026. {\char34}Semantic Consensus: Process-Aware Conflict Detection and Resolution for Enterprise Multi-Agent LLM Systems.{\char34} arXiv:2604.16339 [cs.AI]. \url{https://doi.org/10.48550/arXiv.2604.16339}.
\item \label{ref:3} Liu, Mengyang, Taozhi Chen, Zhenhua Xu, Xue Jiang, and Yihong Dong. 2026. {\char34}Multi-agent Collaboration with State Management.{\char34} arXiv:2605.20563 [cs.MA]. \url{https://doi.org/10.48550/arXiv.2605.20563}.
\item \label{ref:4} Qian, Kaiyang, Xinmin Fang, and Zhengxiong Li. 2026. {\char34}MPAC: A Multi-Principal Agent Coordination Protocol for Interoperable Multi-Agent Collaboration.{\char34} arXiv:2604.09744 [cs.MA]. \url{https://doi.org/10.48550/arXiv.2604.09744}.
\item \label{ref:5} Zhou, Weixing, Zhiyou Wang, Zeshun Peng, Hetian Chen, Yanfeng Zhang, and Ge Yu. 2026. {\char34}ATCC: Adaptive Concurrency Control for Unforeseen Agentic Transactions.{\char34} arXiv:2603.13906 [cs.DB]. \url{https://doi.org/10.48550/arXiv.2603.13906}.
\item \label{ref:6} Pan, Melissa Z., Mert Cemri, Lakshya A. Agrawal, Shuyi Yang, Bhavya Chopra, Rishabh Tiwari, Kurt Keutzer, Aditya Parameswaran, Kannan Ramchandran, Dan Klein, Joseph E. Gonzalez, Matei Zaharia, and Ion Stoica. 2025. {\char34}Why Do Multiagent Systems Fail?{\char34} In *ICLR 2025 Workshop on Building Trust in Language Models and Applications*. \url{https://openreview.net/forum?id=wM521FqPvI}.
\item \label{ref:7} Sartori, Camilo Chacon. 2026. {\char34}The Specification Gap: Coordination Failure Under Partial Knowledge in Code Agents.{\char34} arXiv:2603.24284 [cs.SE]. \url{https://doi.org/10.48550/arXiv.2603.24284}.
\item \label{ref:8} Ellis, Clarence A., and Simon J. Gibbs. 1989. {\char34}Concurrency Control in Groupware Systems.{\char34} In *Proceedings of the 1989 ACM SIGMOD International Conference on Management of Data*, 399--407. New York: ACM Press. \url{https://doi.org/10.1145/67544.66963}.
\item \label{ref:9} Shapiro, Marc, Nuno Preguica, Carlos Baquero, and Marek Zawirski. 2011. {\char34}Conflict-Free Replicated Data Types.{\char34} In *Stabilization, Safety, and Security of Distributed Systems: 13th International Symposium, SSS 2011*, Lecture Notes in Computer Science 6976, 386-400. Berlin: Springer. \url{https://doi.org/10.1007/978-3-642-24550-3_29}.
\item \label{ref:10} Kung, H. T., and John T. Robinson. 1981. {\char34}On Optimistic Methods for Concurrency Control.{\char34} *ACM Transactions on Database Systems* 6 (2): 213-226. \url{https://doi.org/10.1145/319566.319567}.
\item \label{ref:11} Lyu, Hongtao, Dingyan Zhang, Mingyu Wu, Xingda Wei, and Haibo Chen. 2026. {\char34}CoAgent: Concurrency Control for Multi-Agent Systems.{\char34} arXiv:2606.15376 [cs.DC]. \url{https://doi.org/10.48550/arXiv.2606.15376}.
\item \label{ref:12} Geng, Jiayi, and Graham Neubig. 2026. {\char34}Effective Strategies for Asynchronous Software Engineering Agents.{\char34} arXiv:2603.21489 [cs.CL]. \url{https://doi.org/10.48550/arXiv.2603.21489}.
\item \label{ref:13} Zhang, Qingyu, Junzhe Li, Jiayi Lin, Changhua Luo, and Chenxiong Qian. 2026. {\char34}Rover: Context-aware Conflict Resolution with LLM.{\char34} arXiv:2605.17279 [cs.SE]. \url{https://doi.org/10.48550/arXiv.2605.17279}.
\item \label{ref:14} Ogenrwot, Daniel, and John Businge. 2026. {\char34}AgenticFlict: A Large-Scale Dataset of Merge Conflicts in AI Coding Agent Pull Requests on GitHub.{\char34} arXiv:2604.03551 [cs.SE]. \url{https://doi.org/10.48550/arXiv.2604.03551}.
\item \label{ref:15} Wang, Yifei, Ruiyin Li, Peng Liang, Qiong Feng, Zengyang Li, Mojtaba Shahin, and Arif Ali Khan. 2026. {\char34}CodeTeam: An LLM-Powered Multi-Agent Framework for Repository-Level Code Generation.{\char34} arXiv:2606.22082 [cs.SE]. \url{https://doi.org/10.48550/arXiv.2606.22082}.
\item \label{ref:16} Khan, Sajjad. 2026. {\char34}S-Bus: Automatic Read-Set Reconstruction for Multi-Agent LLM State Coordination.{\char34} arXiv:2605.17076 [cs.LG]. \url{https://doi.org/10.48550/arXiv.2605.17076}.
\item \label{ref:17} Huang, Beichen, Ran Cheng, and Kay Chen Tan. 2025. {\char34}EvoGit: Decentralized Code Evolution via Git-Based Multi-Agent Collaboration.{\char34} arXiv:2506.02049 [cs.SE]. \url{https://doi.org/10.48550/arXiv.2506.02049}.
\item \label{ref:18} Li, Yang, Siqi Ping, Xiyu Chen, Xiaojian Qi, Zigan Wang, Ye Luo, and Xiaowei Zhang. 2025. {\char34}AgentGit: A Version Control Framework for Reliable and Scalable LLM-Powered Multi-Agent Systems.{\char34} arXiv:2511.00628 [cs.SE]. \url{https://doi.org/10.48550/arXiv.2511.00628}.
\item \label{ref:19} Jimenez, Carlos E., John Yang, Alexander Wettig, Shunyu Yao, Kexin Pei, Ofir Press, and Karthik Narasimhan. 2023. {\char34}SWE-bench: Can Language Models Resolve Real-World GitHub Issues?{\char34} arXiv:2310.06770 [cs.CL]. \url{https://doi.org/10.48550/arXiv.2310.06770}.
\item \label{ref:20} Wu, Qingyun, Gagan Bansal, Jieyu Zhang, Yiran Wu, Beibin Li, Erkang Zhu, Li Jiang, Xiaoyun Zhang, Shaokun Zhang, Jiale Liu, Ahmed Hassan Awadallah, Ryen W. White, Doug Burger, and Chi Wang. 2023. {\char34}AutoGen: Enabling Next-Gen LLM Applications via Multi-Agent Conversation.{\char34} arXiv:2308.08155 [cs.AI]. \url{https://doi.org/10.48550/arXiv.2308.08155}.
\item \label{ref:21} Adya, Atul. 1999. {\char34}Weak Consistency: A Generalized Theory and Optimistic Implementations for Distributed Transactions.{\char34} PhD thesis, Massachusetts Institute of Technology. \url{https://hdl.handle.net/1721.1/149899}.
\item \label{ref:22} Lloyd, Wyatt, Michael J. Freedman, Michael Kaminsky, and David G. Andersen. 2011. {\char34}Don't Settle for Eventual: Scalable Causal Consistency for Wide-Area Storage with COPS.{\char34} In *Proceedings of the 23rd ACM Symposium on Operating Systems Principles*, 401-416. \url{https://doi.org/10.1145/2043556.2043593}.
\item \label{ref:23} Liu, Tianyang, Canwen Xu, and Julian McAuley. 2024. {\char34}RepoBench: Benchmarking Repository-Level Code Auto-Completion Systems.{\char34} In *Proceedings of the 12th International Conference on Learning Representations (ICLR 2024)*. \url{https://doi.org/10.48550/arXiv.2306.03091}.
\item \label{ref:24} Ding, Yangruibo, Zijian Wang, Wasi Uddin Ahmad, Hantian Ding, Ming Tan, Nihal Jain, Murali Krishna Ramanathan, Ramesh Nallapati, Parminder Bhatia, Dan Roth, and Bing Xiang. 2023. {\char34}CrossCodeEval: A Diverse and Multilingual Benchmark for Cross-File Code Completion.{\char34} In *Advances in Neural Information Processing Systems 36*. arXiv:2310.11248. \url{https://doi.org/10.48550/arXiv.2310.11248}.
\item \label{ref:25} Li, Wei, Xin Zhang, Zhongxin Guo, Shaoguang Mao, Wen Luo, Guangyue Peng, Yangyu Huang, Houfeng Wang, and Scarlett Li. 2025. {\char34}FEA-Bench: A Benchmark for Evaluating Repository-Level Code Generation for Feature Implementation.{\char34} In *Proceedings of the 63rd Annual Meeting of the Association for Computational Linguistics*, 17160--17176. \url{https://doi.org/10.48550/arXiv.2503.06680}.
\item \label{ref:26} Zan, Daoguang, Ailun Yu, Wei Liu, Dong Chen, Bo Shen, Wei Li, Yafen Yao, Yongshun Gong, Xiaolin Chen, Bei Guan, Zhiguang Yang, Yongji Wang, Qianxiang Wang, and Lizhen Cui. 2024. {\char34}CodeS: Natural Language to Code Repository via Multi-Layer Sketch.{\char34} arXiv:2403.16443 [cs.LG]. \url{https://doi.org/10.48550/arXiv.2403.16443}.
\item \label{ref:27} Ding, Jingzhe, Shengda Long, Changxin Pu, Huan Zhou, Hongwan Gao, Xiang Gao, Chao He, Yue Hou, Fei Hu, Zhaojian Li, Weiran Shi, Zaiyuan Wang, Daoguang Zan, Chenchen Zhang, Xiaoxu Zhang, Qizhi Chen, Xianfu Cheng, Bo Deng, Qingshui Gu, Kai Hua, Juntao Lin, Pai Liu, Mingchen Li, Xuanguang Pan, Zifan Peng, Yujia Qin, Yong Shan, Zhewen Tan, Weihao Xie, Zihan Wang, Yishuo Yuan, Jiayu Zhang, Enduo Zhao, Yunfei Zhao, He Zhu, Chenyang Zou, Ming Ding, Jianpeng Jiao, Jiaheng Liu, Minghao Liu, Qian Liu, Chongyao Tao, Jian Yang, Tong Yang, Zhaoxiang Zhang, Xinjie Chen, Wenhao Huang, and Ge Zhang. 2025. {\char34}NL2Repo-Bench: Towards Long-Horizon Repository Generation Evaluation of Coding Agents.{\char34} arXiv:2512.12730 [cs.SE]. \url{https://doi.org/10.48550/arXiv.2512.12730}.
\item \label{ref:28} Sun, Chengzheng, Xiaohua Jia, Yanchun Zhang, Yun Yang, and David Chen. 1998. {\char34}Achieving Convergence, Causality Preservation, and Intention Preservation in Real-Time Cooperative Editing Systems.{\char34} *ACM Transactions on Computer-Human Interaction* 5 (1): 63-108. \url{https://doi.org/10.1145/274444.274447}.
\item \label{ref:29} Chacon, Scott, and Ben Straub. 2014. *Pro Git*, 2nd ed. Apress / Open Source. \url{https://git-scm.com/book}.
\item \label{ref:30} Bernstein, Philip A., Vassos Hadzilacos, and Nathan Goodman. 1987. *Concurrency Control and Recovery in Database Systems*. Reading, MA: Addison-Wesley. \url{https://www.microsoft.com/en-us/research/people/philbe/book/}.
\item \label{ref:31} Hou, Xinyi, Yanjie Zhao, Yue Liu, Zhou Yang, Kailong Wang, Li Li, Xiapu Luo, David Lo, John Grundy, and Haoyu Wang. 2024. {\char34}Large Language Models for Software Engineering: A Systematic Literature Review.{\char34} *ACM Transactions on Software Engineering and Methodology* 33 (8): 1-79. \url{https://doi.org/10.1145/3695988}.
\item \label{ref:32} Zhao, Wenting, Nan Jiang, Celine Lee, Justin T. Chiu, Claire Cardie, Matthias Gallé, and Alexander M. Rush. 2025. {\char34}Commit0: Library Generation from Scratch.{\char34} In *Proceedings of the 13th International Conference on Learning Representations (ICLR)*. arXiv:2412.01769 [cs.SE]. \url{https://doi.org/10.48550/arXiv.2412.01769}.
\item \label{ref:33} Zhou, Qixing, Jiacheng Zhang, Haiyang Wang, Rui Hao, Jiahe Wang, Minghao Han, Yuxue Yang, Shuzhe Wu, Feiyang Pan, Lue Fan, Dandan Tu, and Zhaoxiang Zhang. 2026. {\char34}FeatureBench: Benchmarking Agentic Coding for Complex Feature Development.{\char34} arXiv:2602.10975 [cs.SE]. \url{https://doi.org/10.48550/arXiv.2602.10975}.
\item \label{ref:34} Ni, Ziyi, Huacan Wang, Shuo Zhang, Shuo Lu, Ziyang He, Wang You, Zhenheng Tang, Yuntao Du, Bill Sun, Hongzhang Liu, Sen Hu, Ronghao Chen, Bo Li, Xin Li, Chen Hu, Binxing Jiao, Daxin Jiang, and Pin Lyu. 2025. {\char34}GitTaskBench: A Benchmark for Code Agents Solving Real-World Tasks Through Code Repository Leveraging.{\char34} arXiv:2508.18993 [cs.SE]. \url{https://doi.org/10.48550/arXiv.2508.18993}.
\item \label{ref:35} Yang, John, Carlos E. Jimenez, Alexander Wettig, Kilian Lieret, Shunyu Yao, Karthik Narasimhan, and Ofir Press. 2024. {\char34}SWE-agent: Agent-Computer Interfaces Enable Automated Software Engineering.{\char34} In *Advances in Neural Information Processing Systems 37*. arXiv:2405.15793. \url{https://doi.org/10.48550/arXiv.2405.15793}.
\item \label{ref:36} Wang, Haoyu, Christopher M. Poskitt, and Jun Sun. 2025. {\char34}AgentSpec: Customizable Runtime Enforcement for Safe and Reliable LLM Agents.{\char34} arXiv:2503.18666 [cs.AI]. \url{https://doi.org/10.48550/arXiv.2503.18666}.
\item \label{ref:37} Zhao, Wei, Zhe Li, Peixin Zhang, and Jun Sun. 2026. {\char34}ClawGuard: A Runtime Security Framework for Tool-Augmented LLM Agents Against Indirect Prompt Injection.{\char34} arXiv:2604.11790 [cs.CR]. \url{https://doi.org/10.48550/arXiv.2604.11790}.
\item \label{ref:38} Winston, Cailin, Claris Winston, and René Just. 2026. {\char34}Solver-Aided Verification of Policy Compliance in Tool-Augmented LLM Agents.{\char34} arXiv:2603.20449 [cs.SE]. \url{https://doi.org/10.48550/arXiv.2603.20449}.
\item \label{ref:39} Sousa, Marcelo, Isil Dillig, and Shuvendu K. Lahiri. 2018. {\char34}Verifying Semantic Conflict-Freedom in Three-Way Program Merges.{\char34} arXiv:1802.06551 [cs.PL]. \url{https://doi.org/10.48550/arXiv.1802.06551}.
\item \label{ref:40} Cavalcanti, Guilherme, Paulo Borba, Leonardo dos Anjos, and Jonatas Clementino. 2024. {\char34}Semistructured Merge with Language-Specific Syntactic Separators.{\char34} arXiv:2407.18888 [cs.SE]. \url{https://doi.org/10.48550/arXiv.2407.18888}.
\item \label{ref:41} Mohammadi, Bardia, Nearchos Potamitis, Lars Klein, Akhil Arora, and Laurent Bindschaedler. 2026. {\char34}Atomix: Timely, Transactional Tool Use for Reliable Agentic Workflows.{\char34} arXiv:2602.14849 [cs.LG]. \url{https://doi.org/10.48550/arXiv.2602.14849}.
\item \label{ref:42} Chen, Zheng, Hanqing Liu, Duling Xu, Dong Dong, Jialin Li, Bangzheng Pu, and Jidong Zhai. 2026. {\char34}Cordon: Semantic Transactions for Tool-Using LLM Agents.{\char34} arXiv:2606.17573 [cs.OS]. \url{https://doi.org/10.48550/arXiv.2606.17573}.
\item \label{ref:43} Guo, Xuehang, Xingyao Wang, Yangyi Chen, Sha Li, Chi Han, Manling Li, and Heng Ji. 2025. {\char34}SyncMind: Measuring Agent Out-of-Sync Recovery in Collaborative Software Engineering.{\char34} arXiv:2502.06994 [cs.SE]. \url{https://doi.org/10.48550/arXiv.2502.06994}.
\item \label{ref:44} Mao, Zhenyu, Jacky Keung, Fengji Zhang, Shuo Liu, Yifei Wang, and Jialong Li. 2025. {\char34}Towards Engineering Multi-Agent LLMs: A Protocol-Driven Approach.{\char34} arXiv:2510.12120 [cs.SE]. \url{https://doi.org/10.48550/arXiv.2510.12120}.
\item \label{ref:45} Hou, Bo, Xin Tan, Kai Zheng, Fang Liu, Yinghao Zhu, and Li Zhang. 2025. {\char34}LLM-Driven Collaborative Model for Untangling Commits via Explicit and Implicit Dependency Reasoning.{\char34} arXiv:2507.16395 [cs.AI]. \url{https://doi.org/10.48550/arXiv.2507.16395}.
\end{enumerate}

\end{document}